\documentclass[journal,11pt,onecolumn,draftclsnofoot]{IEEEtran}
\usepackage{cite}
\usepackage{graphicx,color,epsfig,rotating}
\usepackage{amsfonts,amsmath,amssymb,bbm}
\usepackage{setspace}
\usepackage{algorithm}
\usepackage[algo2e]{algorithm2e} 
\usepackage{subcaption}
\usepackage{mdwtab}
\usepackage{placeins}
\usepackage{psfrag, graphicx}
\usepackage[latin1]{inputenc}
\usepackage{multirow}
\usepackage{stfloats}
\usepackage{tabularx} 
\usepackage{multirow} % for multirow cells
\usepackage{array} 
\usepackage{booktabs} 
\usepackage{url}
\usepackage{bm}
\usepackage{pstricks}
\usepackage{booktabs}
\usepackage{makecell}
\usepackage[dvipsnames]{xcolor}  % color text
\usepackage{geometry}
\usepackage{float}
\usepackage{soul}

\usepackage{booktabs}
\allowdisplaybreaks
\newtheorem{theorem}{Theorem}
\newtheorem{corollary}{Corollary}
\newtheorem{definition}{Definition}
\newtheorem{lemma}{Lemma}

\newtheorem{remark}{Remark}

\newtheorem{example}{Example}

\SetKwInput{KwData}{Input}
\SetKwInput{KwResult}{Output}
\allowdisplaybreaks 
\graphicspath{{./images/}}
\long\def\comment#1{}

\makeatletter

\newcommand{\thickhline}{%
    \noalign {\ifnum 0=`}\fi \hrule height 1pt
    \futurelet \reserved@a \@xhline
}
\newcolumntype{"}{@{\hskip\tabcolsep\vrule width 1pt\hskip\tabcolsep}}
\makeatother

\def \Ucmn{{\Uc^{(m,n)}}}

\def \inputsum{{\sum_{(u,v)\in  \Kc}\Wuv }}

\def \Kcu{{\Kc_{[U]}}}

\def \Zuvall{{\{Z_{u,v}\}_{(u,v)\in  \Kc}}}
\def \Wuvall{{\{W_{u,v}\}_{(u,v)\in  \Kc}}}

\def \Zuv{{Z_{u,v}}}
\def \Wuv{{W_{u,v}}}
\def \Xuv{{X_{u,v}}}
\def \uvinkc{{(u,v)\in \Kc}}
\def \uvinkcu{{(u,v)\in \Kc_u}}
\def \uvin{{(u,v)\in}}
\def \rzsigma{{R_{Z_{\Sigma}}}}
\def \rzsigmastar{{R_{Z_{\Sigma}}^*}}
\def \rx{{R_X}} 
\def \ry{{R_Y}}

\def \rxstar{{R_X^*}} 
\def \rystar{{R_Y^*}}

\def  \zsigma{{Z_{\Sigma}}}
\def  \lzsigma{{L_{Z_{\Sigma}}}}

\def \Scbar{{\overline{\Sc}}}

\def  \Ecmn{{\Ec_{m,n}}}
\def \Tcb{{\bm{\Tc}}}
\def \Scb{{\bm{\Sc}}}

\let\tbf\textbf
\let\tit\textit

\let\mbb\mathbb

\let \bsl\backslash
\let\trm\textrm

\newcommand{\homo}{homogeneous\xspace}

\newcommand{\secty}{security\xspace}

\newcommand{\collusers}{colluding users\xspace}
\newcommand{\collusersets}{colluding user sets\xspace}
\newcommand{\colluserset}{colluding user set\xspace}

\newcommand{\CollUserSet}{Colluding User Set\xspace}

\newcommand{\iptsm}{input sum\xspace}

\newcommand{\reqs}{requirements\xspace}

\newcommand{\pis}{protected input set\xspace}

\newcommand{\PIS}{Protected Input Set\xspace}
\newcommand{\piss}{protected input sets\xspace}

\newcommand{\het}{heterogeneous\xspace}

\newcommand{\genn}{generation\xspace}

\newcommand{\eg}{e.g.\xspace}
\newcommand{\utr}{user-to-relay\xspace}
\newcommand{\rts}{relay-to-server\xspace}
\newcommand{\ie}{i.e.\xspace}
\newcommand{\msg}{message\xspace}
\newcommand{\msgs}{messages\xspace}

\newcommand{\Hie}{Hierarchical\xspace}
\newcommand{\hie}{hierarchical\xspace}
\newcommand{\rsec}{relay security\xspace}
\newcommand{\Rsec}{Relay security\xspace}
\newcommand{\RSec}{Relay Security\xspace}
\newcommand{\ssec}{server security\xspace}
\newcommand{\Ssec}{Server security\xspace}
\newcommand{\SSec}{Server Security\xspace}
\newcommand{\Msp}{More specifically\xspace}
\newcommand{\Ip}{In particular\xspace}
\newcommand{\af}{as follows\xspace}
\newcommand{\resp}{respectively\xspace}
\newcommand{\iid}{i.i.d.\xspace}

\newcommand{\Thm}{Theorem\xspace}

\newcommand{\info}{information\xspace}

\newcommand{\Fig}{Figure\xspace}

\newcommand{\agg}{aggregation\xspace}
\newcommand{\secagg}{secure aggregation\xspace}

\newcommand{\diff}{different\xspace}

\newcommand{\indep}{independent\xspace}

\newcommand{\indepce}{independence\xspace}

\newcommand{\param}{parameter\xspace}

\newcommand{\indiv}{individual\xspace}

\newcommand{\comm}{communication\xspace}

\newcommand{\achvb}{achievable\xspace}

\newcommand{\distn}{distribution\xspace}
\newcommand{\Distn}{Distribution\xspace}

\newfont{\bbb}{msbm10 scaled 700}

\newfont{\bb}{msbm10 scaled 1100}

\newcommand{\vth}{{$v^{\rm th}$ }}

\newcommand{\Ac}{{\cal A}}
\newcommand{\Bc}{{\cal B}}
\newcommand{\Cc}{{\cal C}}

\newcommand{\Ec}{{\cal E}}

\newcommand{\Kc}{{\cal K}}

\newcommand{\Pc}{{\cal P}}
\newcommand{\Qc}{{\cal Q}}
\newcommand{\Rc}{{\cal R}}
\newcommand{\Sc}{{\cal S}}
\newcommand{\Tc}{{\cal T}}
\newcommand{\Uc}{{\cal U}}
\newcommand{\Wc}{{\cal W}}

\newcommand{\Zc}{{\cal Z}}

\newcommand{\eqdef}{\stackrel{\Delta}{=}}

\newcommand{\be}{\begin{equation}}
\newcommand{\ee}{\end{equation}}
\newcommand{\bea}{\begin{eqnarray}}
\newcommand{\eea}{\end{eqnarray}}
\newcommand{\bgsubeq}{\begin{subequations}}
\newcommand{\edsubeq}{\end{subequations}}

\begin{document}
\title{Hierarchical Secure Aggregation with Heterogeneous Security Constraints and Arbitrary User Collusion}
\author{
Zhou Li,~\IEEEmembership{Member,~IEEE},
Xiang Zhang,~\IEEEmembership{Member,~IEEE},
Jiawen Lv,~\IEEEmembership{Member,~IEEE},
Jihao Fan,~\IEEEmembership{Member,~IEEE},
Haiqiang Chen,~\IEEEmembership{Member,~IEEE},
and Giuseppe Caire,~\IEEEmembership{Fellow,~IEEE}

\thanks{Part of this work~\cite{li2025collusionresilienthierarchicalsecureaggregation} was presented at the 2025 IEEE Information Theory Workshop, Sydney, Australia.}

\thanks{Zhou Li, Jiawen Lv, and Haiqiang Chen are with the School of Computer, Electronics and Information, 
Guangxi University, Nanning 530004, China (e-mail: lizhou@gxu.edu.cn, lvljw1001@gmail.com, and haiqiang@gxu.edu.cn).}

\thanks{X. Zhang and G. Caire are with the Department of Electrical Engineering and Computer Science, Technical University of Berlin, 10623 Berlin, Germany (e-mail: \{xiang.zhang, caire\}\@tu-berlin.de).
}

\thanks{Jihao Fan is with School of Cyber Science and Engineering, Nanjing University of Science and Technology, Nanjing 210094, China and also with Laboratory for Advanced Computing and Intelligence Engineering, Wuxi 214083, China (e-mail: jihao.fan@outlook.com).
}

\thanks{Corresponding author: Xiang Zhang.}
}

\maketitle

\begin{abstract} 
In hierarchical secure aggregation (HSA), a server communicates with clustered users through an intermediate layer of relays to compute the sum of users' inputs under two security requirements -- server security and relay
security. Server security requires that the server learns nothing beyond the desired sum even when colluding with a subset of users, while relay security requires that each relay remains oblivious to the users' inputs under collusion. Existing work on HSA enforces homogeneous  security where \tit{all} inputs must be protected against \tit{any} subset of potential colluding users with sizes up to a predefined threshold. 
Such a \homo formulation cannot capture scenarios with \tit{\het} \secty \reqs where \diff users may demand various levels of protection.

In this paper, we study hierarchical secure aggregation (HSA) with heterogeneous security requirements and arbitrary user collusion. Specifically, we consider scenarios where the inputs of certain groups of users must remain information-theoretically secure against inference by the server or any relay, even if the server or any relay colludes with an arbitrary subset of other users. Under server security, the server learns nothing about these protected inputs beyond the prescribed aggregate sum, despite any such collusion. Under relay security, each relay similarly obtains no information about the protected inputs under the same collusion model.

We characterize the optimal communication rates achievable across all layers for all parameter regimes. Furthermore, we study the minimum source keys required at the users to ensure security. For this source key requirement, we provide tight characterizations in two broad regimes determined by the security and collusion constraints, and establish a general information-theoretic lower bound together with a bounded-gap achievable scheme for the remaining regime. Our results reveal how heterogeneous security constraints fundamentally affect optimal source keys allocation, and strictly generalize existing results on hierarchical secure aggregation.

\end{abstract}

\section{Introduction}
\label{sec:intro}

Federated learning (FL) is a distributed learning framework that enables multiple users to collaboratively train a shared model while keeping their data locally~\cite{mcmahan2017communication,konecny2016federated,kairouz2021advances,yang2018applied}. In a standard FL system, each user computes local model updates and periodically transmits them to an aggregation server, which combines the updates to form a global model. Although raw data are not shared, model updates themselves can leak sensitive information, as adversaries can perform inference attacks such as model inversion or gradient leakage~\cite{Zhu2020DLG,geiping2020inverting}. In particular, an honest-but-curious server, or a server colluding with some users, may exploit repeated aggregation rounds to infer private information~\cite{melis2019exploiting}. These vulnerabilities highlight that FL alone does not provide formal privacy guarantees~\cite{bouacida2021vulnerabilities,mothukuri2021survey}, motivating the use of secure aggregation (SA) techniques that allow the server to compute only a prescribed aggregate (e.g., the sum) while keeping individual updates hidden~\cite{bonawitz2017practical,bonawitz2016practical}. Existing SA schemes are predominantly cryptographic, providing security only against computationally bounded adversaries.

Beyond cryptographic approaches, \emph{information-theoretic secure aggregation} has been studied to characterize the fundamental limits under perfect security against computationally unbounded adversaries~\cite{9834981,zhao2023secure,zhang2024optimal,Zhang_Li_Wan_DSA,Li_Zhang_GroupwiseDSA,Li_Zhang_WeaklyDSA}. This line of work focuses on determining the optimal communication rates and source key rates needed to securely compute the aggregate. For example, Zhao and Sun~\cite{zhao2023secure} fully characterized one-shot SA with collusion of up to \(T\) users, showing that perfect security can be achieved without loss in communication efficiency, at the cost of a source key rate that scales linearly with the collusion size. Subsequent works have extended this analysis to scenarios including user dropout and collusion~\cite{9834981,so2022lightsecagg,jahani2022swiftagg,jahani2023swiftagg+}, groupwise keys~\cite{zhao2023secure,wan2024information,wan2024capacity,Li_Zhang_GroupwiseDSA}, decentralized secure aggregation~\cite{Zhang_Li_Wan_DSA,Li_Zhang_GroupwiseDSA,Li_Zhang_WeaklyDSA}, user selection~\cite{zhao2022mds,zhao2023optimal,zhao2024secure}, oblivious servers~\cite{sun2023secure}, malicious users~\cite{karakocc2021secure}, vector secure aggregation~\cite{yuan2025vectorlinearsecureaggregation}, 
and decentralized secure aggregation~\cite{Zhang_Li_Wan_DSA,Li_Zhang_GroupwiseDSA,Li_Zhang_WeaklyDSA}.
% hierarchical secure aggregation~\cite{zhang2024optimal,10806947,egger2024privateaggregationhierarchicalwireless,zhang2025fundamental,11195652,egger2023private,lu2024capacity}

\begin{figure}[h]
    \centering
    \includegraphics[width=0.4\textwidth]{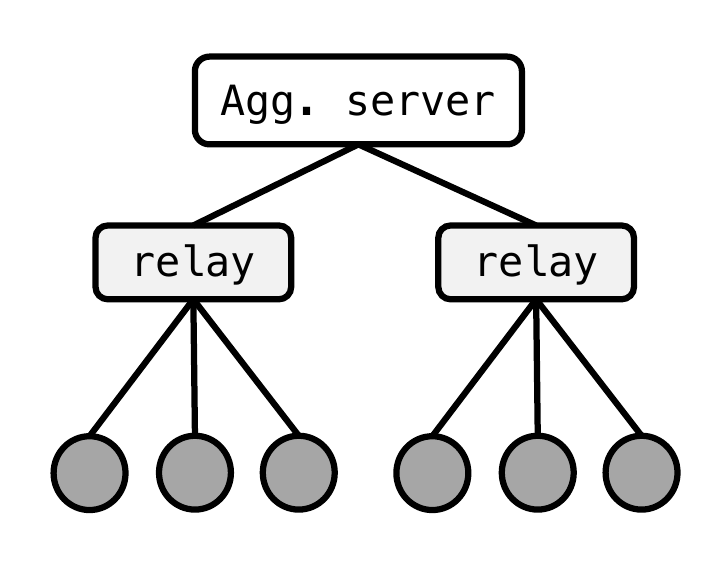}
    \caption{\small Client-edge-cloud architecture in FL. Shaded circles represent users.}
    \label{fig:network model11}
\end{figure}

Despite these advances, existing works on information-theoretic SA~\cite{9834981,so2022lightsecagg,jahani2022swiftagg,jahani2023swiftagg+,zhao2023secure,wan2024information,wan2024capacity,zhao2022mds,zhao2023optimal,li2023weakly,sun2023secure} mainly focus on classical client-server architectures or overlook the randomness consumption required for key generation in hierarchical networks. This motivates the study of \emph{how network topology fundamentally impacts the design of SA protocols, considering both communication and key generation efficiency}.

We consider \emph{hierarchical secure aggregation (HSA)}~\cite{10806947,egger2024privateaggregationhierarchicalwireless,zhang2025fundamental,11195652,egger2023private,lu2024capacity} in a three-layer network consisting of a server, $U\ge 2$ relays, and $UV$ users, where each relay is associated with a disjoint cluster of $V$ users (Fig.~\ref{fig:network model11}). Each user holds an \emph{input} representing its local model and a secret \emph{key} unknown to both the relays and the server. The server aims to recover the sum of all users' inputs subject to two security constraints:

\begin{enumerate}
    \item \emph{Server security}: the server learns nothing beyond the sum, even when colluding with up to $T$ users.
    \item \emph{Relay security}: each relay learns nothing about its users' inputs, even if colluding with up to $T$ users.
\end{enumerate}

One main motivation of this work is to relax the conventional uniform security requirement and investigate its impact on randomness consumption. Specifically, we consider \emph{heterogeneous security constraints}, under which only prescribed subsets of users are required to be protected against adversarial collusion. These subsets, referred to as \emph{security sets} $\bm{\mathcal{S}}$, consist of groups of users $\mathcal{S}_m\in \bm{\mathcal{S}}$ whose inputs must remain information-theoretically secure. Adversarial behavior is captured by a collection of \emph{collusion sets} $\bm{\mathcal{T}}$, where each collusion set $\mathcal{T}_n \in \bm{\mathcal{T}}$ specifies a subset of users that may jointly collude with a relay or the server. Security is enforced only for the specified pairs $(\mathcal{S}_m, \mathcal{T}_n)$, rather than uniformly for all users, leading to a heterogeneous security constraints and arbitrary user collusion model.

The second motivation is to understand the interaction between heterogeneous security constraints and arbitrary collusion patterns. We also aim to study how this combined structure affects the randomness cost in HSA. In contrast to typical HSA with uniform protection, the heterogeneous security constraints setting makes the dependence on collusion patterns explicit and nontrivial. We characterize this dependence by jointly considering the network topology, security sets, and collusion sets. The user-to-relay and relay-to-server communication rates, denoted by $R_X$ and $R_Y$, respectively, represent the number of transmitted bits required per 1-bit input.
Let $R_{Z_\Sigma}$ denote the source key rate, i.e., the total source key required per input bit to generate the users' keys. This formulation provides a principled framework for quantifying the communication efficiency and randomness consumption, and for designing communication- and key-efficient hierarchical secure aggregation protocols under heterogeneous security constraints.

A key challenge in HSA with heterogeneous security constraints lies in the intricate interplay among security sets $\bm{\mathcal{S}}$, collusion sets $\bm{\mathcal{T}}$, and the hierarchical aggregation structure. Each $(\mathcal{S}_m, \mathcal{T}_n)$ pair may span multiple clusters without following any layered or nested structure, creating complex combinatorial interactions dictated by the most adversarial combinations. Users may participate in different numbers of security sets or lie outside all sets, so the allocation of randomness (keys) must be carefully coordinated: some users require no keys, while others contribute portions of keys to satisfy multiple security sets simultaneously, making optimal key assignment a nontrivial optimization problem. In addition, HSA must guarantee security at multiple layers: server security ensures the server learns no information beyond the aggregate, while relay security ensures each relay remains oblivious to protected inputs under the same collusion model. These distinct and potentially competing constraints significantly complicate both the achievability and converse analyses, rendering HSA with heterogeneous security substantially more challenging than HSA with uniform security.

\subsection{Summary of Contributions}
The main contributions of this paper are summarized as follows:

\begin{itemize}
    \item We formulate an information-theoretic model for HSA with heterogeneous security constraints and arbitrary user collusion in a three-layer network, capturing arbitrary security sets and collusion sets under an honest-but-curious threat model while enforcing both relay and server security.
    \item We propose HSA schemes that achieve the optimal communication rates $R_X$ and $R_Y$, and determine the minimum source key rate $R_{Z_\Sigma}$ via a linear program that accounts for the combinatorial interactions between security sets and collusion sets. In broad regimes, the achievable schemes match the converse bounds exactly, providing tight characterizations of the optimal key rate.
    \item For remaining regimes, we derive a general lower bound on the source key rate and construct explicit achievable schemes whose performance is within a bounded gap of this bound. These results identify the dominant security-collusion constraints that govern randomness consumption and illustrate how heterogeneity in cluster structure and security requirements affects optimal key allocation.
\end{itemize}

\tit{Notation.} Throughout the paper, the following notation is used: $[m:n] \eqdef  \{m,m+1,\cdots,n\}$ if $m\le n$ and $[m:n]=\emptyset$ if $m>n$. $[1:n]$ is written as $[n]$ for brevity. Calligraphic letters (\eg, $\Ac,\Bc$) denote sets. $\Ac \backslash \Bc \eqdef \{x\in \Ac: x\notin \Bc\}$.
Given a set of random variables $X_1,\cdots,X_m$, denote $ X_{\Sc}\eqdef \{X_i\}_{i\in \Sc}$ and $X_{1:m}\eqdef \{X_1,\cdots,X_m\}$.
$\binom{\Ac}{n}\eqdef \{\Sc \subseteq \Ac: |\Sc|=n\} $ denotes the collection of all $n$-subsets of $\Ac$ and $\binom{\Ac}{0}=\emptyset$. $\Pc(\Ac) \eqdef \cup_{n=0}^{|\Ac|} \binom{\Ac}{n}  $ denotes the power set of $\Ac$. For two tuples $(i,j)$ and $(i^\prime,j^\prime)$, we say $(i,j)<(i^\prime,j^\prime)$ if either $i< i^\prime$ or $i= i^\prime,j< j^\prime$.

% \tit{Notation.} Throughout the paper, the following notation is used: $[m:n] \eqdef  \{m,m+1,\cdots,n\}$ if $m\le n$ and $[m:n]=\emptyset$ if $m>n$. $[1:n]$ is written as $[n]$ for brevity. Calligraphic letters (\eg, $\Ac,\Bc$) denote sets. $\Ac \backslash \Bc \eqdef \{x\in \Ac: x\notin \Bc\}$.
% %Given a set of random variables $X_1,\cdots,X_m$, denote $ X_{\Sc}\eqdef \{X_i\}_{i\in \Sc}$ and $X_{1:m}\eqdef \{X_1,\cdots,X_m\}$.
% %$\binom{\Ac}{n}\eqdef \{\Sc \subseteq \Ac: |\Sc|=n\} $ denotes the collection of all $n$-subsets of $\Ac$ and $\binom{\Ac}{0}=\emptyset$. $\Pc(\Ac) \eqdef \cup_{n=0}^{|\Ac|} \binom{\Ac}{n}  $ denotes the power set of $\Ac$. 
% For two tuples $(i,j)$ and $(i^\prime,j^\prime)$, we say $(i,j)<(i^\prime,j^\prime)$ if either $i< i^\prime$ or $i= i^\prime,j< j^\prime$. 

%%%%%%%%%%%%%%%%%%%%%%%%%%%%%
%%%%%%%%%%%%%%%%%%%%%%%%%%%%%

\section{Problem Statement}
\label{sec:problem statement}

\subsection{Network Model}
\label{subsec:network model}
\begin{figure}[h]
    \centering
    \includegraphics[width=0.48\textwidth]{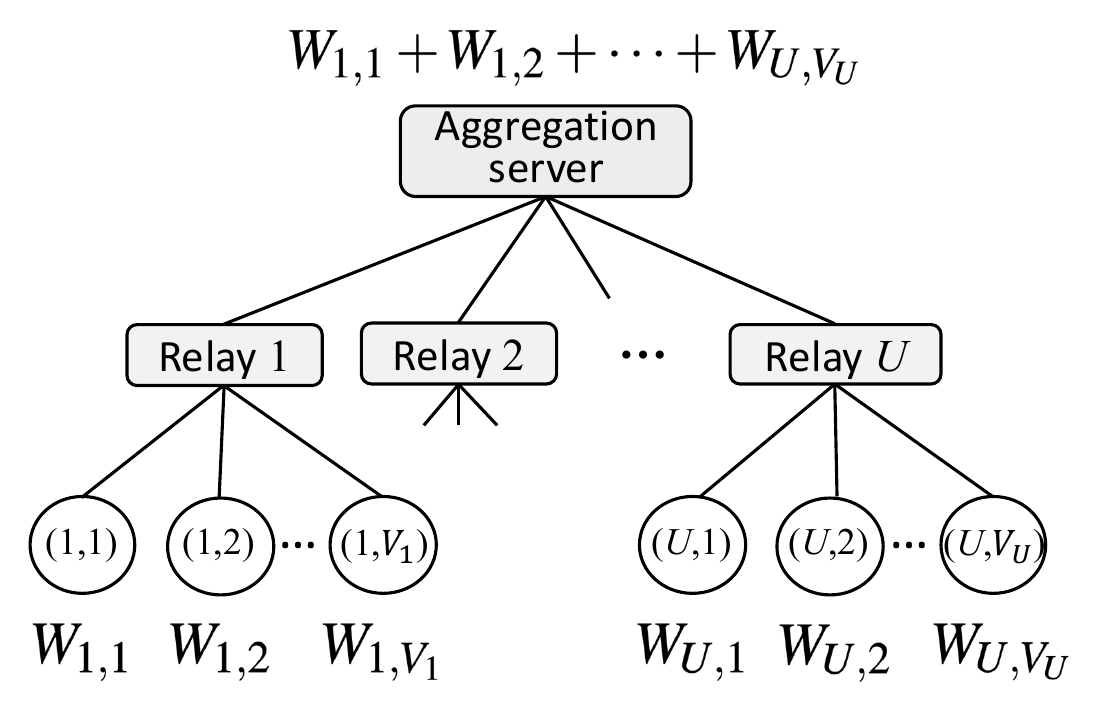}
    \caption{\small \Hie network with $U$ relays where each relay is associated with a disjoint cluster of  users. The \agg server  aims to compute the sum of inputs $W_{1,1}+W_{1,2}+\cdots+W_{U,V_U}    $ of all users.}
    \label{fig:network model}
\end{figure}

Consider \secagg in a 3-layer \hie network consisting of an \agg server, $U\ge 2$ relays and $K\ge U$ users as shown in Fig.~\ref{fig:network model}.
Each relay is associated with a disjoint subset of users. Specifically,
Relay $u\in [U]$ is connected to a cluster of $V_u$ users denoted by $\mathcal{K}_{\{u\}} \eqdef \{(u,v)\}_{v\in [V_u]} = \{(u,1), \cdots, (u,V_u)\}$,  where $(u,v)$ denotes the \vth user in cluster $u$.
Let $ \Kc_{\Uc} \eqdef \cup_{u\in\Uc} \mathcal{K}_{\{u\}} $ denote the users  contained in the clusters in $\Uc\subseteq [U]$.
For ease of notation, we write $ \Kc =\Kcu= \cup_{u\in [U]}\mathcal{K}_{\{u\}}$ as the set of all $K=|\mathcal{K}|=\sum_{u=1}^UV_u$ users.
Each user $(u,v)$ has an \emph{input} vector $W_{u,v}$  
consisting of $L$ \iid uniform symbols from some finite field $\mathbb{F}_q$, and the inputs of  the  users are mutually \indep, \ie,
\begin{align}
H(W_{u,v}) & = L \trm{ (in $q$-ary units), } \forall (u,v) \in \Kc, 
\label{h2} \\
 H\left(\{\Wuv\}_{(u,v)\in\Kc}   \right) & = \sum_{(u,v)\in\Kc } H\left(\Wuv \right).
 \label{eq:input independence}
\end{align} 
$W_{u,v}$ is an abstraction of  the local models/gradients in FL. 
To protect the inputs,  each user also possesses a individual key $\Zuv$--referred to as an \emph{\indiv key}--containing $L_Z$ symbols from $\mbb{F}_q$,  which is  unknown to the server and relays. 
For ease of notation, let
$
\Wc_{\Kc}= \Wuvall, \Zc_\Kc =  \Zuvall 
$ 
denote the set of all inputs and \indiv keys, \resp.
The inputs are assumed to be \indep of the keys, \ie, 
\be 
\label{ind}
H\left( \Wc_\Kc,  \Zc_\Kc     \right)
= \sum_{\uvinkc} H\left(\Wuv\right) + 
H\left( \Zc_\Kc\right).
\ee
Moreover, the \indiv keys $Z_{1,1}, \cdots, Z_{U,V_U}$ can be arbitrarily correlated\footnote{Another line of research~\cite{zhao2023secure,wan2024information,wan2024capacity} focuses on the use of groupwise keys, where each subset of users of a given size shares an independent key. In this setting, the individual keys exhibit certain correlations that allow them to be generated via pairwise key agreement protocols, thereby eliminating the need for a designated key distribution server.}, which necessitates the use of a trusted third-party entity that  generates  and distributes the \indiv keys to each user. 
We assume  that the \indiv keys are generated from  a \emph{source key} variable $\zsigma$ that contains $\lzsigma$ symbols, such that 
\be
\label{total rand}
H\left(\Zc_\Kc | \zsigma \right) =0.
\ee

Similar to~\cite{zhang2024optimal}, a two-hop \comm protocol is used. In the first hop, User $(u,v)$ sends a \msg  $\Xuv$ to Relay $u$, where $\Xuv$--containing $L_X$ symbols--is a deterministic function of $\Wuv$ and $\Zuv$,  \ie, 
\be
\label{messageX}
H\left(\Xuv |\Wuv, \Zuv  \right)=0, \;\forall (u,v) \in \Kc.
\ee 
In the second hop, Relay $u$ generates a \msg $Y_u$ (containing $L_Y$ symbols) and sends it to the server.  $Y_u$ is a deterministic function of the received \msgs from the associated user, \ie,  
\be
\label{messageY}
H\left(Y_u | \{\Xuv\}_{\uvinkcu} \right)=0, \;\forall u\in[U].
\ee 
From the messages $Y_1, \cdots, Y_U$, the server should correctly recover the \iptsm $\inputsum$ :
\be
\label{corr}
\mathrm{[Recovery]}\quad 
H\Big( \sum_{\uvinkc} \Wuv \Big|Y_{1:U} \Big)=0
\ee

%%%%%%%%%%%%
\subsection{Security Model}
\label{subsec:security model}
Existing studies of \hie \secagg (HSA)~\cite{zhang2024optimal, zhang2025fundamental} adopt a \emph{uniform} security model: Neither the server nor any relay is allowed to infer any information about the \emph{entire} input set $\Wc_\Kc$ beyond what they are supposed to learn--namely, the input sum for the server and  the inputs of the colluding  users. This is imposed under a threshold-based collusion model, where  the \agg scheme must remain secure against collusion  between the server/relay and \emph{any subset of up to $T$} users for some predefine threshold  $ T<K$. 
In this paper, we consider a more general security model which allows for
\emph{nonuniform} security guarantees across users. Specifically, we require the protection (of the inputs  of) of a series of subsets  of users $\Sc_1,\cdots, \Sc_M\subseteq \Kc $ against a series of \collusers $\Tc_1, \cdots, \Tc_N \subset \Kc$, where $\Sc_m, m\in[M]$ can be any nonempty subset of $\Kc$, and $\Tc_n, n\in[N]$ is not necessarily restricted to lie in $\cup_{t=0}^T\binom{\Kc}{t}$.
\Msp, our security model is detailed \af. 

\begin{definition}[\PIS]
\label{def:protected input set}
A \emph{\pis} $\Sc_m \subseteq \Kc$ is a set of users whose inputs should be protected from the server and each relay under potential collusion. Together, we consider $M$ \diff \piss 
\be
\label{eq: def S=(S1,...,SM)}
\mathrm{[Protected\;Input\;Sets]} \quad
\Scb \eqdef \left \{\Sc_1,\cdots, \Sc_M\right\}, 
\ee 
where the corresponding inputs in each $\Wc_{\Sc_m},\forall m\in[M] $ should be protected simultaneously. 
\end{definition}

\begin{definition}[\CollUserSet]
\label{def:colluding user set}
A \emph{\colluserset} $\Tc_n \subset \Kc  $ is a subset of users that can potentially collude with the server and the relays during \agg. We  consider $N$ \diff \collusersets
\be
\label{eq: def T=(T1,...,TN)}
\mathrm{[Colluding\;User\;Sets]} \quad
\Tcb\eqdef \left\{\Tc_1,\cdots, \Tc_N\right\}, 
\ee
where security should be guaranteed for every $\Tc_n, \forall n\in[N]$.
\end{definition}

Both $\Scb$ and  $\Tcb$ are assumed to be monotone--that is,  if $\Sc \in \bm{\Sc}, \Tc \in \Tcb$, then any subset of $\Sc$  and $\Tc$ must also belong to $\bm{\Sc}$ and $\Tcb$, \resp. This is because protecting the inputs in $\Sc$ against the users in $\Tc$ automatically implies the protection of the inputs in any subset of $\Sc$ against any subset of $\Tc$. %\bl{As a result, in what follows, we exclude all subsets of $\Sc$ (or $\Tc$) whenever $\Sc \in \Scb$ ($\Tc \in \Tcb$), so that $\Scb$ ($\Tcb$) contains only those sets that are not subsets of any other entry.}

Two types of security constraints must be satisfied, namely \ssec and \rsec, which are defined as follows.

\tbf{\SSec}.
\Ssec requires that the server infer nothing about the inputs   $W_{\Sc_m}$ for any $m \in [M]$ beyond the desired input sum (and the \info
available from the colluding users $\Tc_n$ if there is an nonempty intersection between $\Sc_m$ and $\Tc_n$), even if the server colludes with the users in any $\Tc_n, n\in[N]$. 
\Msp, the \ssec constraint can be expressed as
\be
\label{serversecurity}
\mathrm{[Server\;Security]} \quad
I\left(Y_{1:U} ; \Wc_{\Sc_m}   \Big| \sum\nolimits_{(u,v)\in\Kc} W_{u,v},  \{\Wuv, \Zuv\}_{\uvin\Tc_n}   \right)=0,\;  \forall m \in [M],\forall  n \in [N],
\ee
which necessitates the statistical \indepce between the \rts \msgs $Y_1, \cdots, Y_U$ and the input in any $\Sc_m$, when conditioned on the desired input sum and  the contents accessible from the colluding users.  
The conditioning on the input sum reflects the fact that the server is supposed to recover it, and hence it should be regarded as known information when evaluating security.

\tbf{\RSec}. 
\Rsec requires that each relay infer nothing about the inputs $\Wc_{\Sc_m}$ for any  $m\in[M]$ besides what is known from collusion, even if it  colludes with users in any $\Tc_n, \forall n\in[N]$. Specially, \rsec can be expressed  as
\be
\label{relaysecurity}
\mathrm{[Relay\;Security]} \quad
I\left(\left\{X_{u,v}\right\}_{v\in[V_u]}; \Wc_{\Sc_m} \Big| \{ W_{i,j}, Z_{i,j} \}_{(i,j)\in\mathcal{T}_n} \right) = 0, \;\forall u\in[U], \forall m \in [M], \forall n \in [N] 
\ee
which ensures \indepce between the inputs $\Wc_{\Sc_m}$ and the \utr \msgs $X_{u,1}, \cdots, X_{u,V_u}$ received by Relay $u$. 
Note that the relays are merely \comm facilitators and does not perform partial input \agg. Hence, they should be remain oblivious to  the inputs.

% \if0
% \Tosumrz, a comparison of the security models is given in Table~\ref{table:comparison of security models}.  
% \begin{table}[h]
%     \centering
%     \caption{Comparison of security models.}
%     \begin{tabular}{|c|p{9cm}|}
%      \hline
%       Security model  & \hspace{3.5cm} Description  \\
%       \hline
%       Uniform (\cite{zhang2024optimal, zhang2025fundamental})   & All inputs should be protected given any subset of up  to $T$ \collusers ($\Scb=\{ \Kc\}, \Tcb= \cup_{t=0}^T \binom{\Kc}{t}   $).     \\
%      \hline  
%      Nonuniform (Ours) & Inputs in every $\Sc \in \{\Sc_1, \cdots, \Sc_M\} $  should be protected given any \collusers $\Tc \in \{\Tc_1, \cdots, \Tc_N\}$. The choice of  $\Scb$ and $\Tcb$ can be arbitrary, providing flexibility in security levels across users.    \\
%      \hline
%     \end{tabular}
%     \label{table:comparison of security models}
% \end{table}
% \fi

\begin{remark}
\emph{The server and relay security constraints in (\ref{serversecurity}) and (\ref{relaysecurity}) must hold for every pair $(\Sc_m, \Tc_n), m \in [M], n \in [N]$. 
It should be emphasized that the security constraints do not require protecting the inputs in the union $\cup_{m=1}^M \Sc_m$ against the colluding users $\cup_{n=1}^N \Tc_n$, which would impose a stronger condition than our definition.
The specific choices of the \piss $\{\Sc_1,  \cdots, \Sc_M\}$ and the \collusersets $\{\Tc_1,  \cdots, \Tc_N\}$ can be flexibly selected depending on the security demands of the users. For example, if we choose $\Scb=\Pc({\Kc})$, and $\Tcb=\cup_{t=0}^T \binom{\Kc}{t}$, then (\ref{serversecurity}) and (\ref{relaysecurity}) specializes to the conventional HSA security model with collusion threshold $T$.
}
\end{remark}

%%%%%%%%%
\tbf{Performance Metrics.} 
We study both the \comm and secret key \genn efficiency of the proposed hierarchical secure aggregation with heterogeneous security constraints and arbitrary user collusion problem. \Ip, the \comm rates $R_X$ and $R_Y$ characterize the \comm efficiency over the two network hops, \ie,
how many symbols that each message $X_{u,v}$ and $Y_u$ contains (normalized by the input size $L$) \resp, \ie, 
\be 
\label{eq:def of Rx,Ry}
\rx  \eqdef \frac{L_X}{L}, \; \ry\eqdef \frac{L_Y}{L}
\ee 
The \emph{source key rate} $\rzsigma$ characterizes how many \indep symbols that the source key variable $\zsigma$ contains (normalized by $L$), \ie, 
\be 
\label{eq:def source key rate}
\rzsigma  \eqdef \frac{\lzsigma}{L}
\ee 
It is directly related to the \comm overhead incurred by the  key \distn process as illustrated in Remark~\ref{remark:key dist overhead}.

\begin{remark}[Key \Distn Overhead]
\label{remark:key dist overhead}
Because the keys of  the users can be arbitrarily correlated, a trusted third-party entity is needed  to 
\end{remark}

A rate tuple $(\rx,\ry, \rzsigma)$ is said to be achievable if there exists a \secagg scheme, \ie, a design of the source and \indiv keys $\zsigma, \{\Zuv\}_{\uvinkc}$, the \utr \msgs $\{X_{u ,v}\}_{\uvinkc}$, and the \rts \msgs $Y_{1:U}$, with rates $\rx,\ry$ and $\rzsigma$ for which the recovery constraint (\ref{corr}) and the security constraints (\ref{serversecurity}), (\ref{relaysecurity}) are satisfied. We aim to find the optimal rate region $\Rc^*$, which is defined as the closure of all achievable rate tuples. Let $\rxstar, \rystar$, and $\rzsigmastar$ denote the individually minimal values of $\rx,\ry$, and $\rzsigma$ \resp.

\subsection{Auxiliary Definitions}
\label{subsec:auxiliary defs} 
In this section, we introduce several definitions that are necessary for presenting the main  results of the paper.

Some users are imposed to be protected by some key variable even if they do not explicitly belong to any $\mathcal{S}_m$ set. Such implicit security sets are specified below. %For two sets $\mathcal{A}, \mathcal{B}$, the set difference $\mathcal{A} \setminus \mathcal{B}$ is defined as the set of elements that belong to $\mathcal{A}$ but not to $\mathcal{B}$. For any two users $(u_1,v_1)$ and $(u_2,v_2)$, we define $(u_1,v_1)<(u_2,v_2)$, if $u_1 < u_2$, or $u_1=u_2$ and $v_1<v_2$.

\begin{definition}[Security relay sets $\mathcal{U}^{(m,n)}$ and $\mathcal{F}^{(m,n)}$]
\label{def:secrelay11}
For any pair of user sets $\mathcal{S}_m$ and $\mathcal{T}_n$, define
\begin{align}
&\mathcal{U}^{(m,n)}
\eqdef \big\{ u \in [U] :
 |\mathcal{S}_m \cap \mathcal{K}_{\{u\}}| \neq 0,  |(\mathcal{S}_m \cup \mathcal{T}_n)\cap \mathcal{K}_{\{u\}}| = V_u
\big\}, \label{def:Uset}\\
&\mathcal{F}^{(m,n)}
\eqdef \big\{ u \in [U] :
 |\mathcal{S}_m \cap \mathcal{K}_{\{u\}}| = 0, |\mathcal{T}_n \cap \mathcal{K}_{\{u\}}| = V_u
\big\}. \label{def:Fset}
\end{align}
\end{definition}
Intuitively, the sets $\mathcal{U}^{(m,n)}$ and $\mathcal{F}^{(m,n)}$ classify relays according to their roles in protecting the inputs of the security set $\mathcal{S}_m$ under a collusion set $\mathcal{T}_n$.

A relay $u \in \mathcal{U}^{(m,n)}$ if it is connected to at least one user in $\mathcal{S}_m$ and all users connected to this relay are contained in $\mathcal{S}_m \cup \mathcal{T}_n$. From the server's perspective, the inputs of users in $\mathcal{K}_{\{u\}} \setminus \mathcal{S}_m$ are already known, as they belong to the collusion set $\mathcal{T}_n$. Therefore, the only information that needs protection at relay $u$ comes from users in $\mathcal{S}_m \cap \mathcal{K}_{\{u\}}$. Since no intrinsic randomness is available to mask these protected inputs, these users must consume at least one symbol of independent key. Such relays directly observe protected inputs and play a critical role in the security analysis.

In contrast, a relay $u\in \mathcal{F}^{(m,n)}$ if it is connected exclusively to users in $\mathcal{T}_n$ and has no connection to any user in $\mathcal{S}_m$. These relays do not observe any protected inputs, and all information passing through them is already known to the server. Hence, no key is required, and they can be regarded as fully compromised in the security analysis.

Finally, for a relay
$u \in [U]\setminus\big(\mathcal{U}^{(m,n)} \cup \mathcal{F}^{(m,n)}\big),$
if a relay is connected to at least one user in $\mathcal{S}_m$, it is also connected to at least one user outside $\mathcal{S}_m \cup \mathcal{T}_n$. In this case, although protected inputs are present, the inputs $W_{u,v}$ of users outside $\mathcal{S}_m \cup \mathcal{T}_n$ are unknown to the server and can effectively serve as implicit keys to mask the protected inputs. Therefore, such relays do not require additional independent keys to ensure server security. Relays without any connection to $\mathcal{S}_m$ do not observe protected inputs and also require no key.

We will use the following example to explain the definitions.
\begin{example}\label{ex1}
Consider $U=3$ and $V_1=V_2=V_3=2$, there are 6 users $(1,1),(1,2),(2,1),(2,2),$ $(3,1),(3,2)$. The security input sets are $\Scb=(\mathcal{S}_1,\cdots,\mathcal{S}_{6})=(\emptyset, \{(1,1)\}, \{(1,2)\}, \{(2,1)\}, \{(2,2)\},$ $\{(1,1),$ $(2,1)\})$, and the colluding user sets are $\Tcb=(\mathcal{T}_1,\cdots,\mathcal{T}_{8})=(\emptyset, \{(1,2)\}, \{(2,2)\}, \{(3,1)\}, \{(1,2),(2,2)\},\{(1,2),(3,1)\},\{(2,2),(3,1)\},$ $\{(1,2),(2,2),(3,1)\})$.
\end{example}

Consider the pair $(\mathcal{S}_6=\{(1,1),(2,1)\}, \mathcal{T}_8=\{(1,2),(2,2),(3,1)\})$. When $u=1$, the user set associated with relay~1 is $\mathcal{K}_{\{1\}}=\{(1,1),(1,2)\}$. Since $|\mathcal{S}_6\cap\mathcal{K}_{\{1\}}|=|\{(1,1),(2,1)\}\cap\{(1,1),(1,2)\}|=1\neq 0$, and $|(\mathcal{S}_6\cup\mathcal{T}_8)\cap\mathcal{K}_{\{1\}}|=|(\{(1,1),(2,1)\}\cup\{(1,2),(2,2),(3,1)\})\cap\{(1,1),(1,2)\}|=2=V_1$, hence $1\in\mathcal{U}^{(6,8)}$. 
When $u=2$, we have $|\mathcal{S}_6\cap\mathcal{K}_{\{2\}}|=|\{(1,1),(2,1)\}\cap\{(2,1),(2,2)\}|=|\{(2,1)\}|\neq 0$, and $|(\mathcal{S}_6\cup\mathcal{T}_8)\cap\mathcal{K}_{\{2\}}|=|(\{(1,1),(2,1)\}\cup\{(1,2),(2,2),(3,1)\})\cap\{(2,1),(2,2)\}|=|\{(2,1),(2,2)\}|=2=V_2$, then $2\in \mathcal{U}^{(6,8)}$.
When $u=3$, we have $|\mathcal{S}_6\cap\mathcal{K}_3|=|\{(1,1),(2,1)\}\cap\{(3,1),(3,2)\}|= 0$, the condition is not satisfied, then $3\notin \mathcal{U}^{(6,8)}$. So, $\mathcal{U}^{(6,8)}=\{1,2\}$. 
For any $m,n$, $\mathcal{F}^{(m,n)}=\emptyset$. Intuitively, $h^{(m,n)}_u = 1$ indicates that at least $L$ key symbols are needed to protect the inputs of users in cluster $u$. Consequently, a total of $|\mathcal{U}^{(m,n)}|$ key symbols are required to secure the inputs of all users in the cluster set $\mathcal{U}^{(m,n)}$.

\begin{definition}[Implicit Security Input Set $\mathcal{S}_I$]\label{def:imp}
Let $\mathcal{K}_{[U]}$ denote the set of all $K$ users.

\smallskip
\noindent\textbf{Implicit relay security input set.}
Under the relay security constraint, the \emph{implicit relay security input set} (IRSIS) is defined as
\begin{equation}
\mathcal{S}_{I_1} \triangleq 
\Big\{
\mathcal{K}_{[U]} \setminus \big( (\mathcal{S}_m \cap \mathcal{K}_{\{u\}}) \cup \mathcal{T}_n \big)
:\;
\big| (\mathcal{S}_m \cap \mathcal{K}_{\{u\}}) \cup \mathcal{T}_n \big| = K-1,
\;\forall u\in[U],\, m\in[M],\, n\in[N]
\Big\}
\setminus \{\mathcal{S}_i\}_{i\in[M]} .
\end{equation}

\smallskip
\noindent\textbf{Implicit server security input set.}
Under the server security constraint, the \emph{implicit server security input set} (ISSIS) is defined as
\begin{equation}
\mathcal{S}_{I_2} \triangleq
\Big\{
\mathcal{K}_{[U]} \setminus \big( \mathcal{K}_{\mathcal{U}^{(m,n)}} \cup \mathcal{T}_n \big)
:\;
\big| \mathcal{K}_{\mathcal{U}^{(m,n)}} \cup \mathcal{T}_n \big| = K-1,
\;\forall m\in[M],\, n\in[N]
\Big\}
\setminus \{\mathcal{S}_i\}_{i\in[M]} .
\end{equation}

\smallskip
\noindent
The \emph{implicit security input set} is then defined as
\begin{equation}
\mathcal{S}_I \triangleq \mathcal{S}_{I_1} \cup \mathcal{S}_{I_2}.
\end{equation}
\end{definition}
Intuitively, an implicit security input set consists of inputs that are not explicitly required to be protected by the security constraints, but must nonetheless remain secure in order to preserve the correctness of the aggregation.

For Example~\ref{ex1}, we enumerate all $u,m,n$ and observe that there exist no indices $(u,m,n)$ such that 
$|(\mathcal{S}_m \cap \mathcal{K}_{\{u\}}) \cup \mathcal{T}_n| = K-1$, which implies that 
$\mathcal{S}_{I_1} = \emptyset$.

Next, we enumerate all $(m,n)$ and find that for $(m,n) = (6,8)$, 
$|\mathcal{K}_{\mathcal{U}^{(6,8)}} \cup \mathcal{T}_8| 
= |\{(1,1),(1,2),(2,1),(2,2)\} \cup \{(1,2),(2,2),(3,1)\}| 
= 5 = K-1$. 
Therefore, the corresponding implicit server security input set is 
$\mathcal{S}_{I_2} = \{(3,2)\}$, and hence 
$\mathcal{S}_I = \mathcal{S}_{I_1} \cup \mathcal{S}_{I_2} = \{(3,2)\}$.

This example shows that although User $(3,2)$ does not belong to any explicit security input set, 
there exist implicit security constraints that still require this user to be protected by key variables. 
Intuitively, when the server colludes with Users $(1,2)$, $(2,2)$, and $(3,1)$, 
the requirement of keeping Users $(1,1)$ and $(2,1)$ secure forces User $(3,2)$ to remain secure as well, 
which will later be formalized through entropy arguments; see~(\ref{corollary1_eqX}) and~(\ref{corollary1_eqY}).

\begin{definition}[Total Security Input Set $\overline{\mathcal{S}}$] \label{def:tot} 
The union of explicit and implicit security input sets is defined as the total security input set, 
\begin{eqnarray}
\overline{\mathcal{S}}\eqdef \cup_{m\in[M]} \mathcal{S}_m 
\cup\mathcal{S}_I.
\end{eqnarray}
\end{definition}

For Example \ref{ex1}, we have $\overline{\mathcal{S}}=\{(1,1),(1,2),(2,1),$ $(2,2),(3,2)\}$.

\begin{definition}[$\mathcal{A}_{u,m,n}$ and $\mathcal{E}_{m,n}$] \label{def:totset1}
For any $u\in [U], m\in [M], n\in [N]$, the intersection of $(\mathcal{S}_m\cap\mathcal{K}_{\{u\}})\cup\mathcal{T}_n$ and $\overline{\mathcal{S}}$ is defined as 
%For each pair of security input set $\mathcal{S}_m$ and colluding user set $\mathcal{T}_n$,  
%its overlap with the total security input set $\overline{\mathcal{S}}$ is denoted as
\begin{eqnarray}
\mathcal{A}_{u,m,n}\eqdef ((\mathcal{S}_m\cap\mathcal{K}_{\{u\}})\cup\mathcal{T}_n)\cap\overline{\mathcal{S}} 
\end{eqnarray}
and its maximum cardinality is denoted as 
\be
\label{eq:def a*}
a^*\eqdef \max_{u\in [U],m\in[M], n \in [N]} |\mathcal{A}_{u,m,n}|. 
\ee 
The intersection of $\mathcal{K}_{\mathcal{U}^{(m,n)}}\cup\mathcal{T}_n$ and $\overline{\mathcal{S}}$ is defined as 
\begin{eqnarray}
\mathcal{E}_{m,n}\eqdef \mathcal{K}_{\mathcal{U}^{(m,n)}}\cup\mathcal{T}_n\cap\overline{\mathcal{S}} 
\end{eqnarray}
and its maximum cardinality is denoted as 
\be
\label{eq:def e*}
e^*\eqdef \max_{m\in[M], n \in [N]} |\mathcal{E}_{m,n}|. 
\ee
\end{definition}
For Example \ref{ex1}, $\mathcal{A}_{1,6,8}= ((\{(1,1),(2,1)\}\cap\{(1,1),(1,2)\})\cup\{(1,2),(2,2),(3,1)\})\cap\{(1,1),(1,2),(2,1),$ $(2,2),$ $(3,2)\} = \{(1,1),(1,2),(2,2)\} $, 
$\mathcal{A}_{2,6,8}= ((\{(1,1),(2,1)\}\cap\{(2,1),(2,2)\})\cup\{(1,2),(2,2),(3,1)\})$ $\cap\{(1,1),$ $(1,2),(2,1),$ $(2,2),(3,2)\} = \{(1,1),(2,1),(2,2)\} $. Hence, $a^*=3$.

$\mathcal{K}_{\mathcal{U}^{(6,8)}}=\mathcal{K}_{\{1,2\}}=\{(1,1),(1,2),(2,1),(2,2)\}$. $\mathcal{E}_{6,8}=\mathcal{K}_{\mathcal{U}^{(6,8)}}\cup\mathcal{T}_8\cap\overline{\mathcal{S}}=\{(1,1),(1,2),(2,1),(2,2)\}$. Hence, $e^*=4$.

\begin{definition}[Maximal Summation cardinality of $\mathcal{U}^{(m,n)}%\setminus\mathcal{F}^{(m,n)}
$ and $\mathcal{T}_n\cap\overline{\mathcal{S}}$] \label{def:totset2}
For each pair of security input set $\mathcal{S}_m$ and colluding user set $\mathcal{T}_n$, the 
maximal summation cardinality of $\mathcal{U}^{(m,n)}%\setminus\mathcal{F}^{(m,n)}
$ and $\mathcal{T}_n\cap\overline{\mathcal{S}}$ is denoted as
\be
\label{eq:def d*}
d^* \eqdef \max_{m\in[M], n \in [N]}(|\mathcal{U}^{(m,n)}|+ |\mathcal{T}_n\cap\overline{\mathcal{S}}|).
\ee 
\end{definition}
For Example \ref{ex1}, $|\mathcal{U}^{(6,8)} 
|=|\{1,2\}|=2$, $|\mathcal{T}_8\cap\overline{\mathcal{S}}|=|\{(1,2),(2,2),(3,1)\}\cap\{(1,1),(1,2),(2,1),$ $(2,2),(3,2)\}|=2$, and $|\mathcal{U}^{(6,8)} 
|+ |\mathcal{T}_8\cap\overline{\mathcal{S}}|=4$. Hence, $d^*=4$.

\begin{definition}[Union of Maximum $\mathcal{A}_{u,m,n}$ and $\mathcal{E}_{m,n}$] \label{def:uni2}
Find all $u,m,n$ such that $|\mathcal{A}_{u,m,n}|=|\overline{\mathcal{S}}|$ 
and denote the union of the corresponding $\mathcal{S}_m\cap\mathcal{K}_{\{u\}}, \mathcal{T}_n$ sets as $\mathcal{Q}_1$.
\begin{eqnarray}
\mathcal{Q}_1 \eqdef \cup_{u,m,n: |\mathcal{A}_{u,m,n}| = |\overline{\mathcal{S}}|} (\mathcal{S}_m\cap\mathcal{K}_{\{u\}})\cup\mathcal{T}_n. %\mathcal{A}_{u,m,n}.
\end{eqnarray}
Find all $m,n$ such that $|\mathcal{E}_{m,n}|=|\overline{\mathcal{S}}|$ denote 
 the union of $\mathcal{K}_{\mathcal{U}^{(m,n)}}\cup\mathcal{T}_n$
as $\mathcal{Q}_{2}$
\begin{eqnarray}
    \mathcal{Q}_2\eqdef \cup_{m,n:|\mathcal{E}_{(m,n)}|=|\overline{\mathcal{S}}|}\mathcal{K}_{\mathcal{U}^{(m,n)}}\cup\mathcal{T}_n.
\end{eqnarray}
And define the set $\mathcal{Q}$ as the union of $\mathcal{Q}_1$ and $\mathcal{Q}_2$:
\begin{eqnarray}
\mathcal{Q} \eqdef \mathcal{Q}_1 \cup \mathcal{Q}_2 .
\end{eqnarray}
\end{definition}

For Example~\ref{ex1}, we have $|\overline{\mathcal{S}}| = 5$. Since there exist no $u, m, n$ such that 
$|\mathcal{A}_{u,m,n}| = 5$ or $|\mathcal{E}_{m,n}| = 5$, it follows that 
$\mathcal{Q}_1 = \emptyset$ and $\mathcal{Q}_2 = \emptyset$. Therefore, $\mathcal{Q} = \emptyset$.

\section{Main Results}
\label{sec:main results}
The main results of this paper are presented in three theorems. Each theorem addresses a set of disjoint conditions, determined by the relevant quantities defined in Section~\ref{subsec:auxiliary defs}. Specifically, we enumerate these conditions and their corresponding cases--which together span the entire \param regime--as follows:

\textbf{Condition 1}: Case 1)  $a^*=K$; or Case 2) $a^*\le K-1$ and $|\Kc_{\Ucmn}\cup  \Tc_n|=K$; or Case 3) $a^*\le K-1$, $|\Kc_{\Ucmn}\cup  \Tc_n|\le K-1$, and $\max\{a^*,e^*\}\le |\Scbar|-1$, 
    or Case 4) $a^*\le K-1$,
    $|\Kc_{\Ucmn}\cup  \Tc_n|\le K-1$, $\max\{a^*,e^*\}=|\Scbar|$,   and $|\Qc|\le K-1$.

{{\textbf{Condition 2}:}} $a^*\le K-1$,
    $|\Kc_{\Ucmn}\cup  \Tc_n|\le K-1$, $\max\{a^*,e^*\}=|\Scbar|$, $|\Qc|=K$, and $e^*<a^*=|\Scbar|$.

{{\textbf{Condition 3}:}} $a^*\le K-1$,
    $|\Kc_{\Ucmn}\cup  \Tc_n|\le K-1$, $\max\{a^*,e^*\}=|\Scbar|$, $|\Qc|=K$, and $a^*\le e^*=|\Scbar|$.

\begin{figure}[ht]
\begin{center}
\includegraphics[width=.7\textwidth]{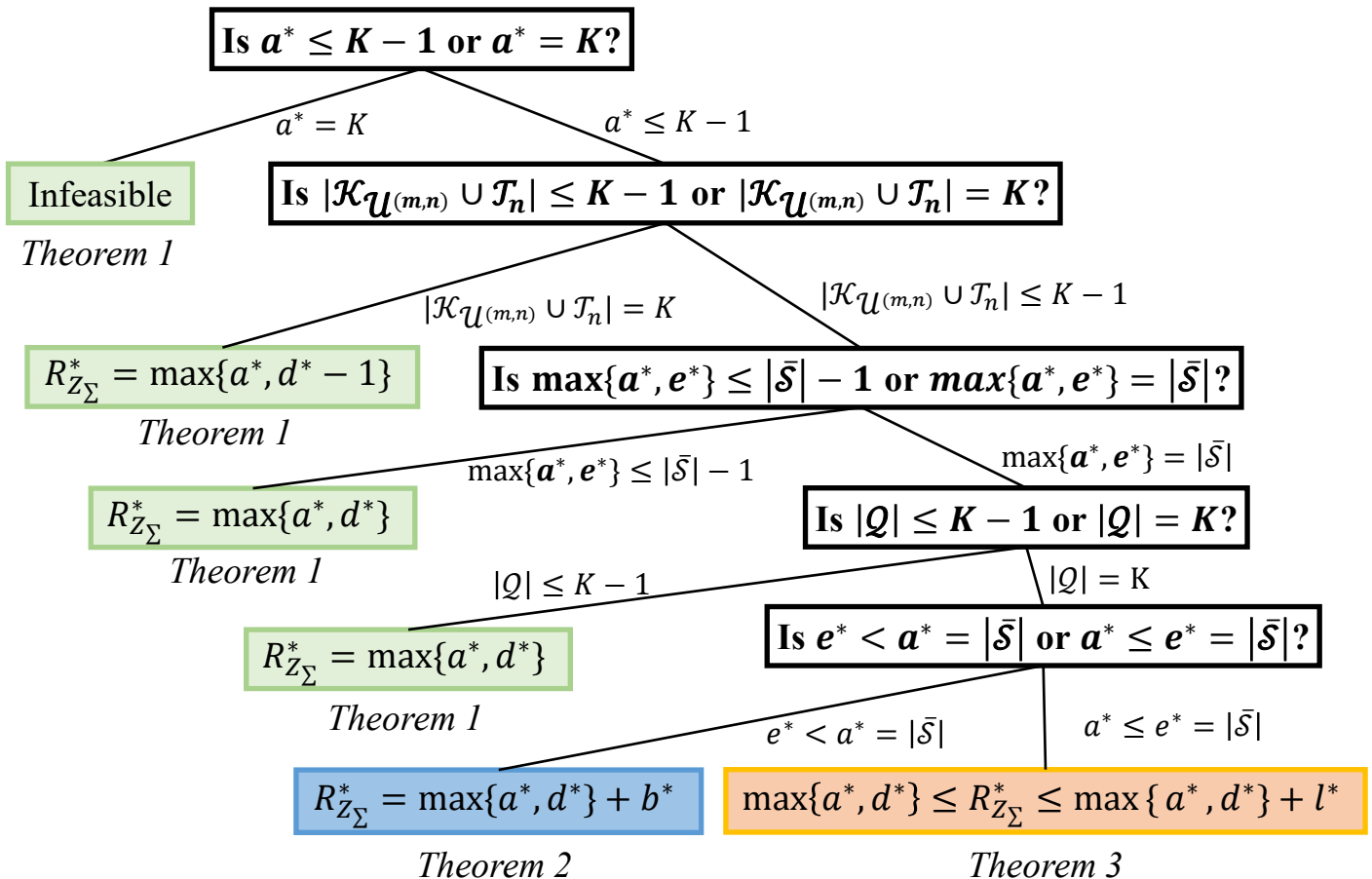}
\caption{Summary of results based on parameter regimes. The cases covered by \Thm~\ref{thm:1}, \Thm~\ref{thm:2} and \Thm~\ref{thm:3} are marked by green, blue and orange boxes, \resp.
}
\label{fig:tree graph summary of results}
\end{center}
\end{figure}

For Conditions 1 and 2, we characterize the optimal source key rate as shown in Theorems~\ref{thm:1} and \ref{thm:2}. For Condition 3, we derive a lower bound on the optimal source key rate and propose an achievable scheme that attains it within a constant additive gap, as shown in \Thm~\ref{thm:3}.
A summary of  the results is given in \Fig\ref{fig:tree graph summary of results}.

% \if0
% \begin{table}[htbp]
% \centering
% \caption{Summary of Results  } % title
% \label{tab:mytable}  
% \begin{tabular}{|c|c|c|c|}
% \hline
% Condition & Case & Covered by  & \Skr optimality\\
% \thickhline  
% % First row: multirow in col1 and col2
% \multirow{2}{*}{Cond. 1} &
% 1)     & Infeasible & N/A  \\
% \cline{2-4}
%  & 2), 3), 4) & \Thm~\ref{thm:1} & Optimal ($\rzsigmastar$ is an integer)   \\
% \cline{1-4}
% Cond. 2 &  &\Thm~\ref{thm:2} & Optimal ($\rzsigmastar$ may contain a fractional part) \\
% \cline{1-4}
% Cond. 3 &  &\Thm~\ref{thm:3} &  Constant additive gap\xz{is this accurate?} \\
% \hline
% \end{tabular}
% \end{table}
% \fi

\begin{theorem}
\label{thm:1}
Given any security input sets $ \Scb=  \{\mathcal{S}_m\}_{m =1}^M$ and collusion sets $\Tcb=\{\mathcal{T}_n\}_{n =1}^N$, the proposed HSA problem is infeasible if $a^*=K$ (Condition 1, Case 1)). For the remaining cases in Condition 1, the optimal rate region is given by
\be 
\Rc^* = \left\{ (\rx, \ry, \rzsigma )|    \rx \ge 1, \ry\ge 1, \rzsigma \ge \rzsigmastar\right \},
\ee 
where
\be 
\label{eq: rzsigma, thm 1}
\rzsigmastar=
\Bigg\{ 
\begin{array}{ll}
 \max \{a^*, d^*-1 \} ,    & \trm{Condition 1, Case 2)}  \\
 \max \{a^*, d^* \} ,    & \trm{Condition 1, Case 3), 4)} 
\end{array}
\ee 
\end{theorem}

% \begin{remark}\label{rem:thm1_implications}
% Theorem~\ref{thm:1} implies that, whenever the problem is feasible, the optimal total key rate \(R_{Z_\Sigma}^*\) is an \emph{integer} under all cases in Condition~1. This integrality follows directly from the combinatorial structure of the quantities \(a^*\) and \(d^*\), which characterize the fundamental security bottlenecks of the system. Specifically,
% \begin{itemize}
%   \item \(a^*\) captures the worst-case number of protected users that can be exposed at a single relay under some pair \((\mathcal{S}_m,\mathcal{T}_n)\);
%   \item \(d^*-1\) (or \(d^*\)) represents the minimum number of keys required to protect users whose exposures result from the joint observations of multiple relays after server-side aggregation.
% \end{itemize}
% \end{remark}

% \begin{remark}\label{rem:thm1_implications}
% Theorem~\ref{thm:1} implies that, whenever the problem is feasible, the optimal total key rate $R_{Z_\Sigma}^*$ is an integer under Condition~1. This integrality comes directly from the combinatorial structure of $a^*$ and $d^*$, which capture the fundamental security bottlenecks: $a^*$ is the worst-case number of protected users exposed at a single relay, and $d^*-1$ (or $d^*$) is the minimum number of keys needed to protect users exposed via joint observations across relays.
% \end{remark}

\begin{remark}\label{rem:thm1_implications}
Theorem~\ref{thm:1} implies that, whenever the problem is feasible, the optimal total key rate $R_{Z_\Sigma}^*$ is an integer under Condition~1. This integrality comes directly from the combinatorial structure of $a^*$ and $d^*$, which capture the fundamental security bottlenecks: $a^*$ characterizes the worst-case number of protected users exposed at a single relay, while $d^*-1$ (or $d^*$) captures the minimum number of keys required to protect users exposed to the server via joint observations across relays. Hence, the optimal total key rate equals $\max\{a^*,\,d^*\}$ (or $\max\{a^*,\,d^*-1\}$), determined directly by the security and collusion sets.
\end{remark}

% \begin{remark}\label{rem:thm1_implications}
% Theorem~\ref{thm:1} implies that, whenever the problem is feasible, the optimal total key rate $R_{Z_\Sigma}^*$ is an integer under Condition~1. This integrality comes directly from the combinatorial structure of $a^*$ and $d^*$, which capture the fundamental security bottlenecks: $a^*$ characterizes the worst-case number of protected users exposed at a single relay, while $d^*-1$ (or $d^*$) captures the minimum number of keys required to protect users exposed to the server via joint observations across relays.
% \end{remark}

% \begin{remark}
% Since $a^*$ and $d^*$ count critical user groups that must be masked, the required total key rate is integer-valued and equals $\max\{a^*,\,d^*\}$ (or $\max\{a^*,\,d^*-1\}$) under Condition~1, determined directly by the security and collusion sets.
% \end{remark}

\begin{theorem}
\label{thm:2}
Under Condition 2, the optimal rate regime of hierarchical secure aggregation with heterogeneous security constraints and arbitrary user collusion is given by
\be 
\Rc^* = \left\{ (\rx, \ry, \rzsigma )|    \rx \ge 1, \ry\ge 1, \rzsigma \ge \rzsigmastar\right \},
\ee 
where
\be 
\label{eq: rzsigma, thm 2}
\rzsigmastar = \max\{a^*, d^*\} + b^*,
\ee
and $b^*$ is the optimal objective value of the following linear program (LP) with  variables $b_{u,v}$:
\begin{align}
\label{eq: LP, thm 2}
 & \min~~~ \max_{\substack{u',m,n:| \Ac_{u',m,n}  |=|\Scbar|  }}  \sum_{(u,v)\in \Tc_n  \bsl \Scbar  }b_{u,v}\\
& \quad  \; \mathrm{s.t.} \quad
\sum_{(u,v) \in \Kc \bsl ((\mathcal{S}_m \cap \mathcal{K}_{\{u'\}}) \cup \mathcal{T}_n)} b_{u,v}\ge 1,  ~~ \forall u'\in[U],m\in[M],n\in[n] \trm{ s.t. } | \Ac_{u',m,n}  |=|\Scbar| \\
 & \hspace{2.8cm} b_{u,v}\ge 0,\;~~~~~~~~~~~~~~~~~~~ \forall (u,v)\in \Kc\bsl \Scbar 
\end{align}
\end{theorem}

%\xz{The crux of this theorem is to provide an  \emph{intuitive} explanation to the LP--the meaning of its variables, objective, and constraints that correspond to the relay/server security constraints.}

\begin{remark}
Theorem~\ref{thm:2} characterizes the optimal key rate under Condition~2, which is more involved than Condition~1. The additional term $b^*$ accounts for the extra key symbols that must be assigned to the users in $\mathcal{T}_n \setminus \overline{\mathcal{S}}$ in order to satisfy the correctness constraints specified by Lemma~\ref{lemma2}. The LP~\eqref{eq: LP, thm 2} determines the minimal total number of such additional symbols while ensuring that all users are protected and the correctness constraints are fulfilled.

\end{remark}

\begin{theorem}
\label{thm:3}
Under Condition 3, the following rate region is \achvb:
\be
\Rc = \left\{ (\rx, \ry, \rzsigma )|    \rx \ge 1, \ry\ge 1, \rzsigma \ge \max\{a^*, d^*\} + l^*    \right \},
\ee 
where $l^*$ is the optimal objective value of the following LP with variables $l_{u,v}$: 
\begin{align}
\label{eq: LP, thm 3}
 &~~~~~~~~~ \min~ \sum_{(u, v) \in \mathcal{K} \setminus \overline{\mathcal{S}}} l_{u,v}\\
& \quad  \; \mathrm{s.t.} \quad
\sum_{(u,v) \in \mathcal{K} \bsl ((\mathcal{S}_m \cap \mathcal{K}_{\{u'\}}) \cup \mathcal{T}_n)} l_{u,v}\ge 1, ~~ \forall u'\in[U],m\in[M],n\in[N] \trm{ s.t. } |\mathcal{A}_{u',m,n}|=|\Scbar| \\
& \hspace{1.3cm}\sum_{(u,v) \in \Kc \bsl \left( \Kc_{\Ucmn} \cup \mathcal{T}_n\right)} l_{u,v}\ge 1, ~~~~~~ \forall m\in[M],n\in[N] \trm{ s.t. } |\Ecmn|=|\Scbar| \\
 & \hspace{2.8cm} l_{u,v}\ge 0,\;~~~~~~~~~~~~~~~~~~~ \forall (u,v)\in \Kcu\bsl \Scbar 
\end{align}
Conversely, for any hierarchical secure aggregation with heterogeneous security constraints and arbitrary user collusion scheme, the  source key rate is lower bounded by
\be
\label{eq: bounds of Rzsigmastar, thm 3}
\rzsigma \ge \max\{a^*, d^*\}.
\ee 
As a result, the optimal source key rate is characterized within a constant additive gap of at most $l^*$, \ie, 
\begin{align}
    \max\{a^*, d^*\} \le \rzsigmastar  \le \max\{a^*, d^*\} + l^*
\end{align}
\end{theorem}
\begin{remark}
Unlike Theorems~1 and~2, which tightly characterize the optimal source key rate, Theorem~3 only provides a lower bound on $R_{Z_\Sigma}^*$ and an achievable scheme that matches this bound up to a finite gap. This gap arises because, in this regime, arbitrary security and collusion sets interact with the hierarchical aggregation structure in a highly coupled manner: multiple adversarial security-collusion pairs simultaneously induce binding constraints, preventing a straightforward match between converse and achievability. Whether the lower bound is tight or improved coding schemes can close the gap remains an open problem.
\end{remark}

We highlight the implications of the theorems as follows:
\begin{enumerate}
    \item For any parameter regime, the optimal communication rates $R_X = R_Y = 1$ are achievable, showing that heterogeneous security constraints and arbitrary user collusion do not incur any communication overhead; secure aggregation can be performed at the same cost as unsecured aggregation in hierarchical networks.

    \item The minimum required source key rate is fundamentally determined by the interaction between security and collusion sets. Heterogeneous security constraints reduce randomness consumption compared to uniform protection, since only users whose inputs must be protected contribute to the key budget. This precisely quantifies the randomness savings enabled by heterogeneous security in hierarchical settings.

    \item Under uniform security constraints, where all users' inputs must be protected against all admissible collusions, the proposed scheme reduces to the hierarchical secure aggregation model in~\cite{zhang2024optimal}. Our results thus strictly generalize existing HSA results by accommodating heterogeneous secure constraints and arbitrary collusion, while recovering known optimal characterizations as special cases.

    \item Secure aggregation with heterogeneous requirements was first studied in star networks by Li et al.~\cite{li2025weakly}. We extend this to hierarchical secure aggregation, showing that heterogeneous security constraints reduce randomness consumption not only in star networks but also in multi-layer aggregation systems with heterogeneous clusters.
\end{enumerate}

\section{Proof of Theorem~\ref{thm:1}} \label{sec:converseex1}
Before presenting the general proof of Theorem~\ref{thm:1}, we illustrate the main idea using Example~\ref{ex1}.

\subsection{Converse Proof of Example~\ref{ex1}} \label{sec:ex1conv}

In general, the converse proof relies on lower bounding the entropy of properly constructed subsets of source keys. Specifically, for any $u',m,n$, it suffices to show that
$H(\{Z_{u,v}\}_{(u,v)\in\mathcal{A}_{u',m,n}})\ge|\mathcal{A}_{u',m,n}|L$ and
$H(\{Z_{u,v}\}_{(u,v)\in\mathcal{E}_{m,n}})\ge(|\mathcal{U}^{(m,n)}|+|\mathcal{T}_n\cap\overline{\mathcal{S}}|)L$
or
$(|\mathcal{U}^{(m,n)}|-1+|\mathcal{T}_n\cap\overline{\mathcal{S}}|)L$.

For Example~\ref{ex1} (see Definition~\ref{def:secrelay11} and Fig.~\ref{fig:example thm 1,U=3,V=2}), since $\max\{a^*,e^*\}=4\le|\overline{\mathcal{S}}|-1=4$, this setting falls into Condition~1, Case~3) of Theorem~\ref{thm:1}, and hence $\rzsigmastar=\max\{a^*,d^*\}=4$. Consider $m=6$ and $n=8$, for which $\mathcal{U}^{(6,8)}=\{1,2\}$ and $\mathcal{T}_8\cap\overline{\mathcal{S}}=\{(1,2),(2,2)\}$. Thus, $\mathcal{K}_{\mathcal{U}^{(6,8)}}=\{(1,1),(1,2),(2,1),(2,2)\}$ and
$|\mathcal{K}_{\mathcal{U}^{(6,8)}}\cup\mathcal{T}_8|
=|\{(1,1),(1,2),(2,1),(2,2),(3,1)\}|=5<K$,
which places this example in Condition~1, Case~3).
\begin{figure}[ht]
    \centering  
    \includegraphics[width=0.45\textwidth]{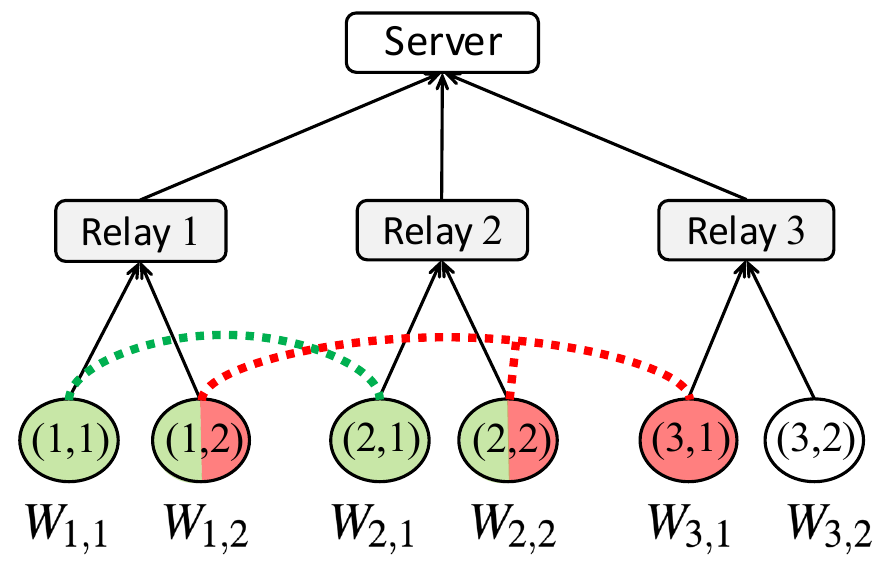}
    \caption{\small Network with $U=3$ relays, each serving $V=2$ users.  Security input sets are groups of users (in green) connected by green dashed links, while collusion sets are groups (in red) connected by red dashed links.   }
    \label{fig:example thm 1,U=3,V=2}
\end{figure} 

By Definition~\ref{def:totset1}, $\mathcal{E}_{m,n}=\mathcal{K}_{\mathcal{U}^{(m,n)}}\cup(\mathcal{T}_n\cap\overline{\mathcal{S}})$. Applying the chain rule of entropy, we obtain
\begin{align}
H(\{Z_{u,v}\}_{(u,v)\in\mathcal{E}_{6,8}})
\geq H(Z_{1,1},Z_{1,2},Z_{2,1},Z_{2,2}\mid Z_{1,2},Z_{2,2})
+H(Z_{1,2},Z_{2,2}) \ge (|\mathcal{U}^{(6,8)}|+2)L = 4L. \label{eq:ex1-entropy}
\end{align}
It therefore suffices to show that
$H(Z_{1,1},Z_{1,2},Z_{2,1},Z_{2,2}\mid Z_{1,2},Z_{2,2})\ge2L$
and
$H(Z_{1,2},Z_{2,2})\ge2L$, which we establish next.

First, consider $H(Z_{1,1},Z_{1,2},Z_{2,1},Z_{2,2}|Z_{1,2},Z_{2,2}) \geq 2L$. The intuitive explanation is that we need 2 key symbols to protect the input message $W_{u,v}$ sent to relays 1 and 2.   
To formalize this idea in terms of entropy, we have
\begin{align}
&H(Z_{1,1},Z_{1,2},Z_{2,1},Z_{2,2}|Z_{1,2},Z_{2,2})\geq H(Z_{1,1},Z_{1,2},Z_{2,1},Z_{2,2}|Z_{1,2},Z_{2,2},Z_{3,1})\\
%&\geq& H(Z_{1,1},Z_{1,2},Z_{2,1},Z_{2,2}|Z_{1,2},Z_{2,2},Z_{3,1},W_{1,1},W_{1,2},W_{2,1},W_{2,2},W_{3,1})\\
\geq&I(Z_{1,1},Z_{1,2},Z_{2,1},Z_{2,2};X_{1,1},X_{1,2},X_{2,1},X_{2,2}|Z_{1,2},Z_{2,2},Z_{3,1},W_{1,1},W_{1,2},W_{2,1},W_{2,2},W_{3,1})~~\\
\overset{(\ref{messageX})}{=}&H(X_{1,1},X_{1,2},X_{2,1},X_{2,2}|Z_{1,2},Z_{2,2},Z_{3,1},W_{1,1},W_{1,2},W_{2,1},W_{2,2},W_{3,1})\label{eq:t1}\\
\overset{(\ref{messageY})}{\geq}&H(Y_1,Y_2|Z_{1,2},Z_{2,2},Z_{3,1},W_{1,1},W_{1,2},W_{2,1},W_{2,2},W_{3,1})\label{geq:t2}\\
=&H(Y_1,Y_2|Z_{1,2},Z_{2,2},Z_{3,1},W_{1,2},W_{2,2},W_{3,1})-I(Y_1,Y_2;W_{1,1},W_{2,1}|Z_{1,2},Z_{2,2},Z_{3,1},W_{1,2},W_{2,2},W_{3,1})\\
\geq&H(Y_1,Y_2|Z_{1,2},Z_{2,2},Z_{3,1},W_{1,2},W_{2,2},W_{3,1})\notag\\
&-I\Big(Y_1,Y_2,Y_3,\sum_{(u,v)\in \mathcal{K}}W_{u,v};W_{1,1},W_{2,1}\big|Z_{1,2},Z_{2,2},Z_{3,1},W_{1,2},W_{2,2},W_{3,1}\Big)\\
=&H(Y_1,Y_2|Z_{1,2},Z_{2,2},Z_{3,1},W_{1,2},W_{2,2},W_{3,1})\notag\\
&-I\Big(\sum_{(u,v)\in \mathcal{K}}W_{u,v};W_{1,1},W_{2,1}\big|Z_{1,2},Z_{2,2},Z_{3,1},W_{1,2},W_{2,2},W_{3,1}\Big)\notag\\
&-I\Big(Y_1,Y_2,Y_3;W_{1,1},W_{1,2}\big|\sum_{(u,v)\in \mathcal{K}} W_{u,v},Z_{1,2},Z_{2,2},Z_{3,1},W_{1,2},W_{2,2},W_{3,1}\Big)\label{eq:t3}\\
\overset{(\ref{ind})(\ref{serversecurity})}{\ge}&2L. \label{eq:t4}
\end{align}
Equation (\ref{eq:t1}) follows from $X_{u,v}$ being a function of $W_{u,v}$ and $Z_{u,v}$ (see (\ref{messageX})), while (\ref{geq:t2}) is derived from $Y_u$ being a function of $X_{u,v}, v \in [V_u]$ (see (\ref{messageY})). In (\ref{eq:t3}), the second term vanishes due to the independence between the inputs and the keys, and the third term is zero due to the server security constraint (\ref{serversecurity}) with $\mathcal{S}_6 = \{(1,1), (2,1)\}$ and $\mathcal{T}_8 = \{(1,2), (2,2), (3,1)\}$. Lastly, in (\ref{eq:t4}), $H(Y_1, Y_2 \mid Z_{1,2}, Z_{2,2}, Z_{3,1}, W_{1,2}, W_{2,2}, W_{3,1}) \geq 2L$ follows from the transmission size constraint, which will be formally established in Lemma \ref{lemma1} (\ref{lemma1_eqX}).

Second, consider $H(Z_{2,1},Z_{2,2}) \geq 2L$.
\begin{eqnarray}
H(Z_{2,1},Z_{2,2})=H(Z_{2,2})+H(Z_{2,1}|Z_{2,2})\geq 2L.\label{eq:t5}
\end{eqnarray}
(\ref{eq:t5}) follows from the transmission size constraint, which will be proved in lemma \ref{lemma3} (\ref{lemma3_eqX}). 

\subsection{Achievable Scheme of Example \ref{ex1}} \label{sec:ex1ach}
We now show that the converse bound $R^*_{Z_\Sigma}=4$ is achievable. 
Suppose each input consists of a single symbol over $\mathbb{F}_5$. Let the source key be $\zsigma=(N_1,N_2,N_3,N_4)$, where $N_1,\ldots,N_4$ are independent and uniformly distributed over $\mathbb{F}_5$. Hence, the source key rate is $\rzsigma=4$, which is integer-valued as characterized by Theorem~\ref{thm:1}.

The individual keys assigned to the users are given by
\begin{align}
\label{eq:Z, example 1ach}
 Z_{1,1} = N_1,\; Z_{1,2} = N_2,\; Z_{2,1} = N_3,\; Z_{2,2} = N_4,\; Z_{3,1} = 0,\; Z_{3,2} = -(N_1+N_2+N_3+N_4).
\end{align}
Note that $Z_{3,1}=0$, since the input $W_{3,1}$ does not require protection under the specified security input sets. Each user $(u,v)$ sends a message $X_{u,v}=W_{u,v}+Z_{u,v}$ to relay $u$. In particular,
\begin{align}
   & X_{1,1}=W_{1,1}+N_1,\; X_{1,2}=W_{1,2}+N_2,\; X_{2,1}=W_{2,1}+N_3, \notag\\
   & X_{2,2}=W_{2,2}+N_4,\; X_{3,1}=W_{3,1},\; X_{3,2}=W_{3,2}-(N_1+N_2+N_3+N_4).
\end{align}

Upon receiving the messages from its associated users, relay $u$ forwards $Y_u=X_{u,1}+X_{u,2}$ to the server. In particular,
\begin{align}
\label{eq: Y msgs, example 1ach}
Y_1 & = W_{1,1}+W_{1,2}+N_1+N_2, ~~Y_2  = W_{2,1}+W_{2,2}+N_3+N_4, \notag \\
Y_3 & = W_{3,1}+W_{3,2}-(N_1+N_2+N_3+N_4).
\end{align}
Since each message consists of a single symbol, the achieved communication rates are $\rx=\ry=1$.

\textbf{Input sum recovery.} The server simply adds up all received \msgs to obtain
$Y_1+Y_2+Y_3= \sum_{(u,v)\in \mathcal{K}} \Wuv$.
This simple recovery method owes to the zero-sum property of the \indiv keys in (\ref{eq:Z, example 1ach}).

\textbf{Proof of relay security.} 
For brevity of notation, let $\Cc_{\Tc_n}\eqdef\{W_{u,v}, Z_{u,v}\}_{(u,v)\in \Tc_n}  $ denote the collection of inputs and keys at the colluding users $\Tc_n,n\in[8]$.
Consider the most challenging case where relay 1 colludes with users in $\Tc_8 =\{(1,2), (2,2), (3,1)\}$. Note that the key design in (\ref{eq:Z, example 1ach}) has the property that \emph{any 4 out of the 6 \indiv keys are mutually \indep}. As a result, even colluding with the 3 users in $\Tc_8$ and obtaining their keys $Z_{1,2}, Z_{2,2}$ and $Z_{3,1}$, relay 1  is unable to infer $Z_{1,1}$, which protects the input $W_{1,1}$ in $\Sc_2=\{(1,1)\}$. More formally, 
\begin{align}
& I(X_{1,1}, X_{1,2}; W_{1,1}|\Cc_{\Tc_8}  ) \notag\\
%&= H(X_{1,1}, X_{1,2}|\Cc_{\Tc_8}   ) - H(X_{1,1}, X_{1,2}|\Cc_{\Tc_8} ,W_{1,1}  )\\
 =& H(X_{1,1}|\Cc_{\Tc_8}   ) - H(X_{1,1}|\Cc_{\Tc_8} ,W_{1,1}  )
\label{eq:step0, proof of rsec, exmaple 1}\\
 =& H(X_{1,1}|\Cc_{\Tc_8}   ) - H(Z_{1,1}|\Cc_{\Tc_8} ,W_{1,1}  )
\label{eq:step1, proof of rsec, exmaple 1}\\
  \overset{(\ref{ind})}{=}& H(X_{1,1}|\Cc_{\Tc_8}   ) - H(Z_{1,1}|Z_{1,2}, Z_{2,2},Z_{3,1}  )
\label{eq:step2, proof of rsec, exmaple 1}\\
  \overset{(\ref{eq:Z, example 1ach})}{=}& H(X_{1,1}|\Cc_{\Tc_8}   ) - H(N_1|N_2,N_4,2N_3-2N_1 )
\label{eq:step3, proof of rsec, exmaple 1}\\
 \leq& H(X_{1,1}  ) - H(N_1)
\label{eq:step4, proof of rsec, exmaple 1}\\
%& \le H(X_{1,1}  ) - H(N_1) \\
 \le& 1 - H(N_1)=0
\label{eq:step5, proof of rsec, exmaple 1}
\end{align}
where (\ref{eq:step0, proof of rsec, exmaple 1}) is  because $X_{1,2}=W_{1,2}+Z_{1,2}$ can be recovered from  $ \Cc_{\Tc_8}$ as $(1,2)\in \Tc_8$. In (\ref{eq:step1, proof of rsec, exmaple 1}), we plugged in the \msg $X_{1,1}=W_{1,1}+Z_{1,1}$. (\ref{eq:step2, proof of rsec, exmaple 1}) is due to the \indepce of the inputs and the keys (See (\ref{ind})). (\ref{eq:step4, proof of rsec, exmaple 1}) is because $N_1,\cdots,N_4$ are \iid uniform. (\ref{eq:step5, proof of rsec, exmaple 1}) is because $X_{1,1}$ contains one symbol and uniform \distn maximizes the entropy. Since mutual \info cannot be negative, we  have $I(X_{1,1}, X_{1,2}; W_{1,1}|\Cc_{\Tc_8}  )=0$, proving security for relay 1. The proof is similar for relay 2. Although $W_{3,1}$ is exposed to relay 3 (since $Z_{3,1}=0$), it does not need to be protected anyways.

\textbf{Proof of server security.} Still consider the most challenging case where the server colludes  with users in $\Tc_8$. The inputs in $\Sc_3=\{(1,2)\}, \Sc_5=\{(2,2)\}$ and $\Sc_6=\{(1,1), (2,1)\}$ need to be protected \resp. Take $\Sc_6$ for example. Intuitively, \ssec is guaranteed because (i) each of $W_{1,1}$ and $W_{2,1}$ is protected by an \indep key, and (ii) the only way for the server to cancel out the noise variables $N_1,\cdots, N_4$  is $Y_1+Y_2 +Y_3$, which yields $W_{1,1} + W_{2,1}+W_{3,2} $ (after removing what is already known from $\Cc_{\Tc_8}$). It can be seen that $W_{1,1} + W_{2,1} $ is protected by $W_{3,2}$, therefore leaking no \info about $W_{1,1}, W_{2,1}$ to the server. More formally, denoting the input sum  by $W_{\Sigma} \eqdef \sum_{(u,v)\in \mathcal{K}}\Wuv$, we have
\begin{align}
& I(Y_1, Y_2, Y_3; W_{1,1}, W_{2,1}|\Cc_{\Tc_8}, W_{\Sigma} )\notag \\
=&  H(Y_1, Y_2, Y_3|\Cc_{\Tc_8}, W_{\Sigma} )
- H(Y_1, Y_2, Y_3|\Cc_{\Tc_8}, W_{\Sigma},W_{1,1}, W_{2,1} ) \notag   \\
=&  H(Y_1, Y_2|\Cc_{\Tc_8}, W_{\Sigma} )
- H(Y_1, Y_2|\Cc_{\Tc_8}, W_{\Sigma},W_{1,1}, W_{2,1} )
\label{eq:step0, proof of ssec, example1}\\
\overset{(\ref{ind})}{=}& H(Y_1, Y_2|\Cc_{\Tc_8}, W_{\Sigma} ) - H(N_1+N_2, N_3+N_4|N_2, N_4    )
\label{eq:step1, proof of ssec, example1}\\
%  =&  H(Y_1, Y_2|\Cc_{\Tc_8}, W_{\Sigma} ) - H(N_1,N_3)\label{eq:step2, proof of ssec, example1}\\
%\le &  H(Y_1, Y_2 ) - H(N_1,N_3) \\
 \le &  H(Y_1) + H(Y_2 ) - H(N_1,N_3) \\
 \le & 1+1 - 2=0 
\end{align}
where  (\ref{eq:step0, proof of ssec, example1}) is because $Y_3=W_{\Sigma}-Y_1-Y_2$. In (\ref{eq:step1, proof of ssec, example1}), we plugged in the message design (\ref{eq: Y msgs, example 1ach}) and utilized the \indepce between inputs and keys (See (\ref{ind})). The last step is because each of $Y_1$ and $Y_2$ contains one symbol and uniform \distn maximizes the entropy, and also because $N_1,\cdots,  N_4$ are \iid uniform. As a result, we proved \ssec for $\Sc_6$. For $\Sc_3$ and $\Sc_5$, the proof follows similarly.
\hfill $\lozenge$

\subsection{Infeasibility Proof of $R_{Z_{\Sigma}}$ under Condition 1, Case 1)} \label{sec:inf}
Consider Condition 1, Case 1), where $a^* = K$, i.e., there exists a relay $u'$ such that $|\mathcal{S}_m \cap \mathcal{K}_{\{u'\}} \cup \mathcal{T}_n| = K$. We will show that under this condition, relay security cannot be satisfied if relay $u'$ colludes with the users in $\mathcal{T}_n$.

By the relay security requirement \eqref{relaysecurity}, we have
\begin{eqnarray}
    0&\overset{(\ref{relaysecurity})}{=
    }&I\left(\left\{W_{u,v}\right\}_{(u,v)\in\mathcal{S}_m}; \left\{X_{u',v}\right\}_{v\in[V_{u'}]} \Big| \{ W_{i,j}, Z_{i,j} \}_{(i,j)\in\mathcal{T}_n} \right)\\
    &=&H\left(\left\{W_{u,v}\right\}_{(u,v)\in\mathcal{S}_m} \Big| \{ W_{i,j}, Z_{i,j} \}_{(i,j)\in\mathcal{T}_n} \right)\notag\\
    &&-H\left(\left\{W_{u,v}\right\}_{(u,v)\in\mathcal{S}_m} \Big| \{ W_{i,j}, Z_{i,j} \}_{(i,j)\in\mathcal{T}_n}, \left\{X_{u',v}\right\}_{v\in[V_{u'}]} \right)\\
    &\overset{(\ref{ind})(\ref{messageX})(\ref{messageY})}{=}&H\left(\left\{W_{u,v}\right\}_{(u,v)\in\mathcal{S}_m\setminus\mathcal{T}_n} \Big| \{ W_{i,j}, Z_{i,j} \}_{(i,j)\in\mathcal{T}_n} \right)\notag\\
    &&-H\left(\left\{W_{u,v}\right\}_{(u,v)\in\mathcal{S}_m\setminus\mathcal{T}_n} \Big|\{ W_{i,j}, Z_{i,j}, X_{i,j} \}_{(i,j)\in\mathcal{T}_n}, \{Y_u\}_{u\in[U]\setminus\{u'\}},\left\{X_{u',v}\right\}_{v\in[V_{u'}]},\{Y_{u'}\} \right)\label{eq:contradt1}\\
    &\overset{(\ref{corr})}{\geq}&H\left(\left\{W_{u,v}\right\}_{(u,v)\in\mathcal{S}_m\setminus\mathcal{T}_n} \right)-H\Big(\left\{W_{u,v}\right\}_{(u,v)\in\mathcal{S}_m\setminus\mathcal{T}_n} \Big|\sum_{(u,v)\in \mathcal{S}_m\setminus\mathcal{T}_n}W_{u,v}\Big)\label{eq:contradt2}\\
&=&|\mathcal{S}_m\setminus\mathcal{T}_n|L-(|\mathcal{S}_m\setminus\mathcal{T}_n|-1)L\\
&=&L. \label{eq: infeasible proof11}
\end{eqnarray}

In \eqref{eq:contradt1}, the first term follows from the independence of $W_{u,v}$ and $Z_{u,v}$, while the second term is justified because $X_{i,j}$ is determined by $W_{i,j}$ and $Z_{i,j}$, and $Y_u$ is determined by $\{X_{u,v}\}_{v\in [V_u]}$. Since $\mathcal{S}_m \cap \mathcal{K}_{u'} \cup \mathcal{T}_n = \mathcal{K}_{[U]}$, it follows that $\mathcal{K}_{[U]\setminus \{u'\}} \subset \mathcal{T}_n$, and thus $\{Y_u\}_{u\in[U]\setminus\{u'\}}$ is determined by $\{X_{i,j}\}_{(i,j)\in \mathcal{T}_n}$. 
In \eqref{eq:contradt2}, the second term uses the correctness constraint: the sum $\sum_{(u,v)\in \mathcal{S}_m\setminus \mathcal{T}_n} W_{u,v}$ can be derived from $\sum_{(u,v)\in \mathcal{K}_{[U]}} W_{u,v}$ and $\{W_{u,v}\}_{(u,v)\in \mathcal{T}_n}$. Since $|\mathcal{S}_m \cap \mathcal{K}_{\{u\}} \cup \mathcal{T}_n| = K$, we have $|\mathcal{S}_m \cup \mathcal{T}_n| = K$ and $(\mathcal{K}_{[U]} \setminus \mathcal{T}_n) \subset (\mathcal{S}_m \setminus \mathcal{T}_n)$. 
Therefore, \eqref{eq: infeasible proof11} implies $0 \ge L$, which contradicts the relay security requirement. We conclude that hierarchical secure aggregation with heterogeneous security constraints is infeasible for $a^* = K$, i.e., $\Rc^* = \emptyset$.

\subsection{General Converse Proof under Condition 1, Cases 2), 3), and 4)
} \label{sec:convpfthm1}

We now generalize the previous argument to all parameter settings. To do so, we first present a useful lemma. Intuitively, it states that each message $X_{u,v}$ must carry at least $L$ symbols (the size of the input) even if all other inputs are known. Similarly, each relay message $Y_u$ must contain at least $L$ symbols if at least one input $X_{u,v}$ connected to relay $u$ remains unknown.
\begin{lemma}\label{lemma1}
For any $u\in[U]$, $(u,v)\in \mathcal{K}_{[U]}$, we have 
\begin{eqnarray}
    &&H\left(X_{u,v}\big|\{W_{i,j},Z_{i,j}\}_{(i,j)\in\mathcal{K}_{[U]}\backslash \{(u,v)\}}\right) \geq L, \label{lemma1_eqX}\\
    &&H\left(Y_{u}\big|\{W_{i,j},Z_{i,j}\}_{(i,j)\in\mathcal{K}_{[U]}\backslash \{(u,v)\}}\right) \geq L. \label{lemma1_eqY}
\end{eqnarray}
\end{lemma}

\begin{IEEEproof}  
First, consider (\ref{lemma1_eqX})
\begin{eqnarray}
    && H\left(X_{u,v}\big|\{W_{i,j},Z_{i,j}\}_{(i,j)\in\mathcal{K}_{[U]}\backslash \{(u,v)\}}\right)\notag\\
    &\geq& I\Big(X_{u,v};\sum_{(i,j)\in\mathcal{K}_{[U]}} W_{i,j}\Big|\{W_{i,j},Z_{i,j}\}_{(i,j)\in\mathcal{K}_{[U]}\backslash \{(u,v)\}}\Big)\\
    &=& H\Big(\sum_{(i,j)\in\mathcal{K}_{[U]}} W_{i,j}\Big|\{W_{i,j},Z_{i,j}\}_{(i,j)\in\mathcal{K}_{[U]}\backslash \{(u,v)\}}\Big)\notag \\
    && -H\Big(\sum_{(i,j)\in\mathcal{K}_{[U]}} W_{i,j}\Big|\{W_{i,j},Z_{i,j}\}_{(i,j)\in\mathcal{K}_{[U]}\backslash \{(u,v)\}},X_{u,v}\Big)\\
    &\overset{(\ref{ind})(\ref{messageX})(\ref{messageY})}{=}& H\left( W_{u,v}\right)-H\Big(\sum_{(i,j)\in\mathcal{K}_{[U]}} W_{i,j}\Big|\{W_{i,j},Z_{i,j}\}_{(i,j)\in\mathcal{K}_{[U]}\backslash \{(u,v)\}},X_{u,v},\{Y_u\}_{u\in[U]}\Big)\label{pf_lemma1_t1}\\
    &\overset{(\ref{h2})(\ref{corr})}{=}& L.\label{pf_lemma_t2}\label{eq:e1} 
\end{eqnarray}
where the first term of (\ref{pf_lemma1_t1}) follows from the fact that input $W_{u,v}$ is independent of other inputs and keys $\{W_{i,j},Z_{i,j}\}_{(i,j) \in \mathcal{K}_{[U]}\backslash\{(u,v)\}}$ (see (\ref{ind})) and the second term of (\ref{pf_lemma1_t1}) follows from the fact that $\{X_{i,j}\}_{(i,j)\in\mathcal{K}_{[U]}\backslash \{(u,v)\}}$ is determined by $\{W_{i,j},Z_{i,j}\}_{(i,j)\in\mathcal{K}_{[U]}\backslash \{(u,v)\}}$ (see (\ref{messageX})) and $\{Y_u\}_{u\in[U]}$ is determined by $\{X_{u,v}\}_{(u,v)\in \mathcal{K}_{[U]}}$  (see (\ref{messageY})). In (\ref{pf_lemma_t2}), we use the property that $W_{u,v}$ has $L$ uniform symbols (see (\ref{h2})) and the desired sum $\sum_{(u,v)\in\mathcal{K}_{[U]}} W_{u,v}$ can be decoded with no error from all messages $\{Y_u\}_{u\in[U]}$ (see (\ref{corr})).

Second, consider (\ref{lemma1_eqY})
\begin{eqnarray}
    && H\left(Y_{u}\big|\{W_{i,j},Z_{i,j}\}_{(i,j)\in\mathcal{K}_{[U]}\backslash \{(u,v)\}}\right)\notag\\
    &\geq& I\Big(Y_{u};\sum_{(i,j)\in\mathcal{K}_{[U]}} W_{i,j}\Big|\{W_{i,j},Z_{i,j}\}_{(i,j)\in\mathcal{K}_{[U]}\backslash \{(u,v)\}}\Big)\\
    &=& H\Big(\sum_{(i,j)\in\mathcal{K}_{[U]}} W_{i,j}\Big|\{W_{i,j},Z_{i,j}\}_{(i,j)\in\mathcal{K}_{[U]}\backslash \{(u,v)\}}\Big)\notag\\
    && -H\Big(\sum_{(i,j)\in\mathcal{K}_{[U]}} W_{i,j}\Big|\{W_{i,j},Z_{i,j}\}_{(i,j)\in\mathcal{K}_{[U]}\backslash \{(u,v)\}},Y_{u}\Big)\\
    &\overset{(\ref{ind})(\ref{messageX})(\ref{messageY})}{=}& H\left( W_{u,v}\right)-H\Big(\sum_{(i,j)\in\mathcal{K}_{[U]}} W_{i,j}\Big|\{W_{i,j},Z_{i,j}\}_{(i,j)\in\mathcal{K}_{[U]}\backslash \{(u,v)\}},\{Y_u\}_{u\in[U]}\Big)\label{pf_lemma1_t2}\\
    &\overset{(\ref{h2})(\ref{corr})}{=}& L, \label{eq:e12} 
\end{eqnarray}
where the second term in (\ref{pf_lemma1_t2}) follows from the fact that
(i) $\{X_{i,j}\}_{(i,j)\in\mathcal{K}_{[U]}\setminus{(u,v)}}$ is completely determined by $\{W_{i,j},Z_{i,j}\}_{(i,j)\in\mathcal{K}_{[U]}\setminus{(u,v)}}$ according to (\ref{messageX}), and
(ii) $\{Y_{u'}\}_{u'\in[U]\setminus\{u\}}$ is in turn determined by $\{X_{i,j}\}_{(i,j)\in\mathcal{K}_{[U]}\setminus\{(u,v)\}}$ as specified in (\ref{messageY}).
\end{IEEEproof}

% \hfill\QED

Next, we show that, under the security and correctness constraints, the keys used by users outside the sets 
$(\mathcal{S}_m \cap \mathcal{K}_{\{u'\}}) \cup \mathcal{T}_n$ and 
$\mathcal{K}_{\mathcal{U}^{(m,n)}} \cup \mathcal{T}_n$ 
each contain at least $L$ symbols of entropy, conditioned on the information available to the colluding users.

\begin{lemma} \label{lemma2} 
For any $\mathcal{K}_{\{u'\}}, \mathcal{K}_{\mathcal{U}^{(m,n)}}, \mathcal{S}_m, \mathcal{T}_n$ 
and any $u'\in[U], m\in[M], n\in[N]$, 
where $\mathcal{S}_m \cap \mathcal{T}_n = \emptyset,
|(\mathcal{S}_m \cap \mathcal{K}_{\{u'\}}) \cup \mathcal{T}_n| \le K-1,
|\mathcal{K}_{\mathcal{U}^{(m,n)}} \cup \mathcal{T}_n| \le K-1,$
we have
\begin{align}
    &H\Big(\{Z_{u,v}\}_{(u,v) \in \mathcal{K}_{[U]} \setminus((\mathcal{S}_m\cap\mathcal{K}_{\{u'\}})\cup\mathcal{T}_n)}\;\big|\;\{Z_{u,v}\}_{(u,v)\in\mathcal{T}_n}\Big) \geq L, \label{lemma2_eqX}\\
    &H\Big(\{Z_{u,v}\}_{(u,v) \in \mathcal{K}_{[U]} \setminus(\mathcal{K}_{\mathcal{U}^{(m,n)}}\cup\mathcal{T}_n)}\;\big|\;\{Z_{u,v}\}_{(u,v)\in\mathcal{T}_n}\Big) \geq L. \label{lemma2_eqY}
\end{align}
\end{lemma}

First, consider (\ref{lemma2_eqX})
\begin{eqnarray}
&&H\left(\{Z_{u,v}\}_{(u,v) \in \mathcal{K}_{[U]} \setminus((\mathcal{S}_m\cap\mathcal{K}_{\{u'\}})\cup\mathcal{T}_n)}\big|\{Z_{u,v}\}_{(u,v)\in\mathcal{T}_n}\right) \notag\\
&\geq&I\left(\{Z_{u,v}\}_{(u,v) \in \mathcal{K}_{[U]} \setminus((\mathcal{S}_m\cap\mathcal{K}_{\{u'\}})\cup\mathcal{T}_n)};\{Z_{u,v}\}_{(u,v) \in(\mathcal{S}_m\cap\mathcal{K}_{\{u'\}})}\big|\{Z_{u,v}\}_{(u,v)\in\mathcal{T}_n}\right)\\
&\overset{(\ref{ind})}{=}&I\left(\{Z_{u,v},W_{u,v}\}_{(u,v) \in \mathcal{K}_{[U]} \setminus((\mathcal{S}_m\cap\mathcal{K}_{\{u'\}})\cup\mathcal{T}_n)};\{Z_{u,v},W_{u,v}\}_{(u,v) \in(\mathcal{S}_m\cap\mathcal{K}_{\{u'\}})}\big|\{Z_{u,v},W_{u,v}\}_{(u,v)\in\mathcal{T}_n}\right) \label{eq:tx12}\\
&\overset{(\ref{messageX})}{\geq}&I\left(\{X_{u,v},W_{u,v}\}_{(u,v) \in \mathcal{K}_{[U]} \setminus((\mathcal{S}_m\cap\mathcal{K}_{\{u'\}})\cup\mathcal{T}_n)};\{X_{u,v},W_{u,v}\}_{(u,v)\in(\mathcal{S}_m\cap\mathcal{K}_{\{u'\}})}\big|\{Z_{u,v},W_{u,v}\}_{(u,v)\in\mathcal{T}_n}\right)\\
&\geq&I\left(\{X_{u,v},W_{u,v}\}_{(u,v)\in \mathcal{K}_{[U]} \setminus((\mathcal{S}_m\cap\mathcal{K}_{\{u'\}})\cup\mathcal{T}_n)};\{W_{u,v}\}_{(u,v)\in(\mathcal{S}_m\cap\mathcal{K}_{\{u'\}})}\big|\right.\notag\\
&&\left.\{Z_{u,v},W_{u,v}\}_{(u,v)\in\mathcal{T}_n},\{X_{u,v}\}_{(u,v)\in(\mathcal{S}_m\cap\mathcal{K}_{\{u'\}})}\right)\\
&=&H\left(\{W_{u,v}\}_{(u,v)\in(\mathcal{S}_m\cap\mathcal{K}_{\{u'\}})}\big|\{Z_{u,v},W_{u,v}\}_{(u,v)\in\mathcal{T}_n},\{X_{u,v}\}_{(u,v)\in(\mathcal{S}_m\cap\mathcal{K}_{\{u'\}})}\right)\notag \\
&&-~H\left(\{W_{u,v}\}_{(u,v)\in(\mathcal{S}_m\cap\mathcal{K}_{\{u'\}})}\big|\{Z_{u,v},W_{u,v}\}_{(u,v)\in\mathcal{T}_n},\{X_{u,v}\}_{(u,v)\in(\mathcal{S}_m\cap\mathcal{K}_{\{u'\}})},\right.\notag\\
&&\left.\{X_{u,v},W_{u,v}\}_{(u,v)\in \mathcal{K}_{[U]} \setminus((\mathcal{S}_m\cap\mathcal{K}_{\{u'\}})\cup\mathcal{T}_n)}\right)\\
&=&H\left(\{W_{u,v}\}_{(u,v)\in(\mathcal{S}_m\cap\mathcal{K}_{\{u'\}})}\big|\{Z_{u,v},W_{u,v}\}_{(u,v)\in\mathcal{T}_n}\right)\notag \\
&&-~I\left(\{W_{u,v}\}_{(u,v)\in(\mathcal{S}_m\cap\mathcal{K}_{\{u'\}})};\{X_{u,v}\}_{(u,v)\in(\mathcal{S}_m\cap\mathcal{K}_{\{u'\}})}\big|\{Z_{u,v},W_{u,v}\}_{(u,v)\in\mathcal{T}_n}\right)\notag \\
&&-~H\Big(\{W_{u,v}\}_{(u,v)\in(\mathcal{S}_m\cap\mathcal{K}_{\{u'\}})}\Big|\{Z_{u,v},W_{u,v}\}_{(u,v)\in\mathcal{T}_n},\{X_{u,v}\}_{(u,v)\in\mathcal{K}_{[U]}},\{Y_u\}_{u\in[U]},\sum_{(u,v)\in \mathcal{K}_{[U]}} W_{u,v},\notag\\
&&\{X_{u,v},W_{u,v}\}_{(u,v)\in \mathcal{K}_{[U]} \setminus((\mathcal{S}_m\cap\mathcal{K}_{\{u'\}})\cup\mathcal{T}_n)})\label{pf_lemma2_t1}\\
&\overset{(\ref{relaysecurity})}{\geq}&H\left(\{W_{u,v}\}_{(u,v)\in(\mathcal{S}_m\cap\mathcal{K}_{\{u'\}})\setminus\mathcal{T}_n}\big|\{Z_{u,v},W_{u,v}\}_{(u,v)\in\mathcal{T}_n}\right)\notag \\
&&-~H\Big(\{W_{u,v}\}_{(u,v)\in(\mathcal{S}_m\cap\mathcal{K}_{\{u'\}})\setminus\mathcal{T}_n}\Big|\sum_{(u,v)\in (\mathcal{S}_m\cap\mathcal{K}_{\{u'\}})\setminus\mathcal{T}_n} W_{u,v}\Big)\label{pf_lemma2_t3}\\
&\overset{(\ref{h2})}{=}&|(\mathcal{S}_m\cap\mathcal{K}_{\{u'\}})\setminus\mathcal{T}_n|L-(|(\mathcal{S}_m\cap\mathcal{K}_{\{u'\}})\setminus\mathcal{T}_n|-1)L\\
&=&L.
\end{eqnarray}
In (\ref{pf_lemma2_t1}), the second term equals zero due to the relay security constraint (\ref{relaysecurity}) applied to $\mathcal{S}_m \cap \mathcal{K}_{\{u'\}}$ and $\mathcal{T}_n$. The third term holds because of the following dependencies: $\{X_{u,v}\}_{(u,v) \in \mathcal{T}_n}$ is determined by $\{Z_{u,v}, W_{u,v}\}_{(u,v) \in \mathcal{T}_n}$ (see (\ref{messageX})); $\{X_{u,v}\}_{(u,v) \in \mathcal{K}_{[U]}}$ is determined by $\{X_{u,v}\}_{(u,v) \in \mathcal{T}_n}$, $\{X_{u,v}\}_{(u,v) \in \mathcal{S}_m \cap \mathcal{K}_{\{u'\}}}$, and $\{X_{u,v}\}_{(u,v) \in \mathcal{K}_{[U]} \setminus (\mathcal{S}_m \cap \mathcal{K}_{\{u'\}} \cup \mathcal{T}_n)}$; $\{Y_u\}_{u \in [U]}$ is determined by $\{X_{u,v}\}_{(u,v) \in \mathcal{K}_{[U]}}$ (see (\ref{messageY})); and finally, $\sum_{(u,v) \in \mathcal{K}_{[U]}} W_{u,v}$ is determined by $\{Y_u\}_{u \in [U]}$ (see (\ref{corr})). The second term of (\ref{pf_lemma2_t3}) holds because $\sum_{(u,v)\in (\mathcal{S}_m\cap\mathcal{K}_{\{u'\}})\setminus\mathcal{T}_n} W_{u,v}$ is determined by $\{W_{u,v}\}_{(u,v)\in\mathcal{T}_n}$, $\{W_{u,v}\}_{(u,v)\in \mathcal{K}_{[U]} \setminus((\mathcal{S}_m\cap\mathcal{K}_{\{u'\}})\cup\mathcal{T}_n)}$, and $\sum_{(u,v)\in \mathcal{K}_{[U]}} W_{u,v}$.

Second, consider (\ref{lemma2_eqY})
\begin{align}
&H\left(\{Z_{u,v}\}_{(u,v) \in \mathcal{K}_{[U]} \setminus(\mathcal{K}_{\mathcal{U}^{(m,n)}}\cup\mathcal{T}_n)}\big|\{Z_{u,v}\}_{(u,v)\in\mathcal{T}_n}\right) \notag\\
\geq&I\left(\{Z_{u,v}\}_{(u,v) \in \mathcal{K}_{[U]} \setminus(\mathcal{K}_{\mathcal{U}^{(m,n)}}\cup\mathcal{T}_n)};\{Z_{u,v}\}_{(u,v) \in\mathcal{K}_{\mathcal{U}^{(m,n)}}}\big|\{Z_{u,v}\}_{(u,v)\in\mathcal{T}_n}\right)\\
\overset{(\ref{ind})}{=}&I\left(\{Z_{u,v},W_{u,v}\}_{(u,v) \in \mathcal{K}_{[U]} \setminus(\mathcal{K}_{\mathcal{U}^{(m,n)}}\cup\mathcal{T}_n)};\{Z_{u,v},W_{u,v}\}_{(u,v) \in\mathcal{K}_{\mathcal{U}^{(m,n)}}}\big|\{Z_{u,v},W_{u,v}\}_{(u,v)\in\mathcal{T}_n}\right) \label{eq:tx1}\\
\overset{(\ref{messageX})}{\geq}&I\left(\{X_{u,v},W_{u,v}\}_{(u,v) \in \mathcal{K}_{[U]} \setminus(\mathcal{K}_{\mathcal{U}^{(m,n)}}\cup\mathcal{T}_n)};\{X_{u,v},W_{u,v}\}_{(u,v)\in\mathcal{K}_{\mathcal{U}^{(m,n)}}}\big|\{Z_{u,v},W_{u,v}\}_{(u,v)\in\mathcal{T}_n}\right)\\
\overset{(\ref{messageY})}{\geq}&I\left(\{X_{u,v},W_{u,v}\}_{(u,v) \in \mathcal{K}_{[U]} \setminus(\mathcal{K}_{\mathcal{U}^{(m,n)}}\cup\mathcal{T}_n)};\{Y_u\}_{u\in\mathcal{U}^{(m,n)}},\{W_{u,v}\}_{(u,v)\in\mathcal{K}_{\mathcal{U}^{(m,n)}}}\big|\{Z_{u,v},W_{u,v}\}_{(u,v)\in\mathcal{T}_n}\right)\label{pf_lemma2_t4}\\
\geq&I\left(\{X_{u,v},W_{u,v}\}_{(u,v)\in \mathcal{K}_{[U]} \setminus(\mathcal{K}_{\mathcal{U}^{(m,n)}}\cup\mathcal{T}_n)};\{W_{u,v}\}_{(u,v)\in\mathcal{K}_{\mathcal{U}^{(m,n)}}}|\{Z_{u,v},W_{u,v}\}_{(u,v)\in\mathcal{T}_n},\{Y_u\}_{u\in\mathcal{U}^{(m,n)}}\right)\\
=&H\left(\{W_{u,v}\}_{(u,v)\in\mathcal{K}_{\mathcal{U}^{(m,n)}}}\big|\{Z_{u,v},W_{u,v}\}_{(u,v)\in\mathcal{T}_n},\{Y_u\}_{u\in\mathcal{U}^{(m,n)}}\right)\notag \\
&-~H\left(\{W_{u,v}\}_{(u,v)\in\mathcal{K}_{\mathcal{U}^{(m,n)}}}\big|\{Z_{u,v},W_{u,v}\}_{(u,v)\in\mathcal{T}_n},\{Y_u\}_{u\in\mathcal{U}^{(m,n)}},\{X_{u,v},W_{u,v}\}_{(u,v)\in \mathcal{K}_{[U]} \setminus(\mathcal{K}_{\mathcal{U}^{(m,n)}}\cup\mathcal{T}_n)}\right)\notag\\
\geq&H\Big(\{W_{u,v}\}_{(u,v)\in\mathcal{K}_{\mathcal{U}^{(m,n)}}\setminus\mathcal{T}_n}\Big|\sum_{(u,v)\in\mathcal{K}_{[U]}}W_{u,v},\{Z_{u,v},W_{u,v}\}_{(u,v)\in\mathcal{T}_n},\{Y_u\}_{u\in[U]}\Big)\notag \\
&-~H\Big(\{W_{u,v}\}_{(u,v)\in\mathcal{K}_{\mathcal{U}^{(m,n)}}\setminus\mathcal{T}_n}\Big|\sum_{(u,v)\in \mathcal{K}_{\mathcal{U}^{(m,n)}}\setminus\mathcal{T}_n} W_{u,v}\Big)\label{pf_lemma2_t5}\\
=&H\Big(\{W_{u,v}\}_{(u,v)\in\mathcal{K}_{\mathcal{U}^{(m,n)}}\setminus\mathcal{T}_n}\Big|\sum_{(u,v)\in\mathcal{K}_{[U]}}W_{u,v},\{Z_{u,v},W_{u,v}\}_{(u,v)\in\mathcal{T}_n}\Big)\notag \\
&-~I\Big(\{W_{u,v}\}_{(u,v)\in\mathcal{K}_{\mathcal{U}^{(m,n)}}\setminus\mathcal{T}_n};\{Y_u\}_{u\in[U]}\Big|\sum_{(u,v)\in\mathcal{K}_{[U]}}W_{u,v},\{Z_{u,v},W_{u,v}\}_{(u,v)\in\mathcal{T}_n}\Big)\notag \\
&-~H\Big(\{W_{u,v}\}_{(u,v)\in\mathcal{K}_{\mathcal{U}^{(m,n)}}\setminus\mathcal{T}_n}\Big|\sum_{(u,v)\in \mathcal{K}_{\mathcal{U}^{(m,n)}}\setminus\mathcal{T}_n} W_{u,v}\Big)\label{pf_lemma2_t6}\\
\overset{(\ref{ind})(\ref{h2})(\ref{serversecurity})}{=}&|\mathcal{K}_{\mathcal{U}^{(m,n)}}\setminus\mathcal{T}_n|L-(|\mathcal{K}_{\mathcal{U}^{(m,n)}}\setminus\mathcal{T}_n|-1)L\label{pf_lemma2_t7}\\
=&L.
\end{align}
The second term of (\ref{pf_lemma2_t5}) holds because $\{X_{u,v}\}_{(u,v)\in\mathcal{T}_n}$ is determined by $\{Z_{u,v},W_{u,v}\}_{(u,v)\in\mathcal{T}_n}$ (see (\ref{messageX})), and $\{Y_u\}_{u\in [U]\setminus\mathcal{U}^{(m,n)}}$ is determined by $\{X_{u,v}\}_{(u,v)\in\mathcal{T}_n}$ and $\{X_{u,v}\}_{(u,v)\in \mathcal{K}_{[U]} \setminus(\mathcal{K}_{\mathcal{U}^{(m,n)}}\cup\mathcal{T}_n)}$ (see (\ref{messageY})). Consequently, by (\ref{corr}), $\sum_{(u,v)\in \mathcal{K}_{\mathcal{U}^{(m,n)}}\setminus\mathcal{T}_n} W_{u,v}$ is determined by $\{Y_u\}_{u\in [U]\setminus\mathcal{U}^{(m,n)}}$, $\{W_{u,v}\}_{(u,v)\in \mathcal{K}_{[U]} \setminus(\mathcal{K}_{\mathcal{U}^{(m,n)}}\cup\mathcal{T}_n)}$, $\{Y_u\}_{u\in \mathcal{U}^{(m,n)}}$, and  $\{W_{u,v}\}_{(u,v)\in\mathcal{T}_n}$.
In (\ref{pf_lemma2_t6}), the second term equals zero due to the server security constraint (\ref{serversecurity}) applied to $\mathcal{K}_{\mathcal{U}^{(m,n)}}\setminus\mathcal{T}_n\subset \mathcal{S}_m$ and $\mathcal{T}_n$.

Specializing Lemma \ref{lemma2} to the case where $|(\mathcal{S}_m \cap \mathcal{K}_{\{u\}}) \cup \mathcal{T}_n| = K-1$, we obtain the following corollaries for elements in the implicit security input set $\mathcal{S}_{I}$ (Definition \ref{def:imp}).

\begin{corollary}
Let $u' \in [U]$, $m \in [M]$, $n \in [N]$, and consider $\mathcal{K}_{\{u'\}}, \mathcal{S}_m, \mathcal{T}_n$ such that 
$|(\mathcal{S}_m \cap \mathcal{K}_{\{u'\}}) \cup \mathcal{T}_n| = K-1$. 
Define the remaining index $\{(u,v)\} = \mathcal{K}_{[U]} \setminus ((\mathcal{S}_m \cap \mathcal{K}_{\{u'\}}) \cup \mathcal{T}_n)$. 
Then
\begin{eqnarray}
    H\left(Z_{u,v}\big|\{Z_{i,j}\}_{(i,j)\in\mathcal{T}_n}\right) \overset{(\ref{lemma2_eqX})}{\geq} L. \label{corollary1_eqX}
\end{eqnarray}
\end{corollary}

\begin{corollary}
For $m \in [M]$, $n \in [N]$ with $\mathcal{S}_m \cap \mathcal{T}_n = \emptyset$ and 
$\bigl|\mathcal{K}_{\mathcal{U}^{(m,n)}} \cup \mathcal{T}_n\bigr| = K-1$, 
denote the remaining element $\{(u,v)\} = \mathcal{K}_{[U]}\setminus (\mathcal{K}_{\mathcal{U}^{(m,n)}}\cup\mathcal{T}_n)$. Then
\begin{eqnarray}
    H\left(Z_{u,v}\big|\{Z_{i,j}\}_{(i,j)\in\mathcal{T}_n}\right) \overset{(\ref{lemma2_eqY})}{\geq} L. \label{corollary1_eqY}
\end{eqnarray}
\end{corollary}

We now consider arbitrary sets $\mathcal{S}_m$ and $\mathcal{T}_n$. The total length of the keys used by users in the explicit security input set $\mathcal{S}_m \cap \mathcal{K}_{\{u'\}}$, given the information available to colluding users, must be at least $|\mathcal{S}_m \cap \mathcal{K}_{\{u'\}}| L$. Similarly, for the explicit security input set $\mathcal{K}_{\mathcal{U}^{(m,n)}}$, the required total key length conditioned on colluding users depends on whether $|\mathcal{K}_{\mathcal{U}^{(m,n)}} \cup \mathcal{T}_n| = K$: it is at least $(|\mathcal{U}^{(m,n)}|-1)L$ if the condition holds, and $|\mathcal{U}^{(m,n)}|L$ otherwise.

\begin{lemma} \label{lemma3} 
Let $u' \in [U]$, $m \in [M]$, $n \in [N]$, and consider arbitrary sets $\mathcal{K}_{\{u'\}}, \mathcal{K}_{\mathcal{U}^{(m,n)}}, \mathcal{S}_m, \mathcal{T}_n$ such that $\mathcal{S}_m \cap \mathcal{T}_n = \emptyset$ and $\lvert (\mathcal{S}_m \cap \mathcal{K}_{\{u'\}}) \cup \mathcal{T}_n \rvert \le K-1$. Then
\begin{align}
    H\big(\{Z_{u,v}\}_{(u,v) \in \mathcal{S}_m \cap \mathcal{K}_{\{u'\}}} \,\big|\, \{Z_{u,v}\}_{(u,v)\in \mathcal{T}_n}\big) 
    &\ge |\mathcal{S}_m \cap \mathcal{K}_{\{u'\}}| L, \label{lemma3_eqX}\\[1mm]
    H\big(\{Z_{u,v}\}_{(u,v) \in \mathcal{K}_{\mathcal{U}^{(m,n)}}} \,\big|\, \{Z_{u,v}\}_{(u,v)\in \mathcal{T}_n}\big) 
    &\ge 
    \Bigg\{
    \begin{array}{cl}
        (|\mathcal{U}^{(m,n)}|-1)L, &\mbox{if}~|\mathcal{K}_{\mathcal{U}^{(m,n)}}\cup\mathcal{T}_n|= K,\\
         |\mathcal{U}^{(m,n)}|L, &\mbox{otherwise}.
    \end{array}
    \label{lemma3_eqY}
\end{align}
\end{lemma}

First, consider (\ref{lemma3_eqX})
\begin{eqnarray}
&& H\left(\{Z_{u,v}\}_{(u,v) \in (\mathcal{S}_m\cap\mathcal{K}_{\{u'\}})}\big|\{Z_{u,v}\}_{(u,v)\in\mathcal{T}_n}\right)\\
%&\geq&H\left(\{Z_{u,v}\}_{(u,v) \in (\mathcal{S}_m\cap\mathcal{K}_{\{u'\}})}\big|\{Z_{u,v}\}_{(u,v)\in\mathcal{T}_n},\{W_{u,v}\}_{(u,v) \in (\mathcal{S}_m\cap\mathcal{K}_{\{u'\}}\cup\mathcal{T}_n)}\right) \\
&\geq&I\left(\{Z_{u,v}\}_{(u,v) \in (\mathcal{S}_m\cap\mathcal{K}_{\{u'\}})};\{X_{u,v}\}_{(u,v) \in (\mathcal{S}_m\cap\mathcal{K}_{\{u'\}})}\big|\{Z_{u,v}\}_{(u,v)\in\mathcal{T}_n},\{W_{u,v}\}_{(u,v) \in (\mathcal{S}_m\cap\mathcal{K}_{\{u'\}}\cup\mathcal{T}_n)}\right) ~~~~\\
&\overset{(\ref{messageX})}{=}&H\left(\{X_{u,v}\}_{(u,v) \in (\mathcal{S}_m\cap\mathcal{K}_{\{u'\}})}\big|\{Z_{u,v}\}_{(u,v)\in\mathcal{T}_n},\{W_{u,v}\}_{(u,v) \in (\mathcal{S}_m\cap\mathcal{K}_{\{u'\}}\cup\mathcal{T}_n)}\right) \label{pf_lemma3_t1}\\
&=&H\left(\{X_{u,v}\}_{(u,v) \in (\mathcal{S}_m\cap\mathcal{K}_{\{u'\}})}\big|\{Z_{u,v},W_{u,v}\}_{(u,v)\in\mathcal{T}_n}\right)\notag\\
&&-I\left(\{X_{u,v}\}_{(u,v) \in (\mathcal{S}_m\cap\mathcal{K}_{\{u'\}})};\{W_{u,v}\}_{(u,v) \in (\mathcal{S}_m\cap\mathcal{K}_{\{u'\}})}\big|\{Z_{u,v},W_{u,v}\}_{(u,v)\in\mathcal{T}_n}\right)\\
&\overset{(\ref{lemma1_eqX})}{\geq}&|\mathcal{S}_m\cap\mathcal{K}_{\{u'\}}|L-I\left(\{X_{u,v}\}_{(u,v) \in \mathcal{K}_{\{u\}}};\{W_{u,v}\}_{(u,v) \in (\mathcal{S}_m)}\big|\{Z_{u,v},W_{u,v}\}_{(u,v)\in\mathcal{T}_n}\right)\label{pf_lemma3_t2}\\
&\overset{(\ref{relaysecurity})}{\geq}&|\mathcal{S}_m\cap\mathcal{K}_{\{u'\}}|L. \label{pf_lemma3_t3}
\end{eqnarray}
Where (\ref{pf_lemma3_t1}) follows the fact that $\{X_{u,v}\}_{(u,v) \in (\mathcal{S}_m\cap\mathcal{K}_{\{u'\}})}$ is determined by $\{W_{u,v},Z_{u,v}\}_{(u,v) \in (\mathcal{S}_m\cap\mathcal{K}_{\{u'\}})}$. In (\ref{pf_lemma3_t2}), the first term follows from chain-rule and applying Lemma \ref{lemma1}
(\ref{lemma1_eqX}) to each expanded term; the second term is zero due to relay security constraint with $\mathcal{S}_m$ and $\mathcal{T}_n$ (see (\ref{relaysecurity})). Note that $\{X_{u,v}\}_{(u,v) \in \mathcal{K}_{\{u\}}}$ is the same as $\{X_{u,v}\}_{v \in [V_u]}$.

Second, consider (\ref{lemma3_eqY}), we will show that $H\left(\{Z_{u,v}\}_{(u,v) \in \mathcal{K}_{\mathcal{U}^{(m,n)}}}\big|\{Z_{u,v}\}_{(u,v)\in\mathcal{T}_n}\right)\geq (|\mathcal{U}^{(m,n)}|-1)L$ if $|\mathcal{K}_{\mathcal{U}^{(m,n)}}\cup\mathcal{T}_n|= K$; Otherwise, $H\left(\{Z_{u,v}\}_{(u,v) \in \mathcal{K}_{\mathcal{U}^{(m,n)}}}\big|\{Z_{u,v}\}_{(u,v)\in\mathcal{T}_n}\right)\geq |\mathcal{U}^{(m,n)}|L$ when $|\mathcal{K}_{\mathcal{U}^{(m,n)}}\cup\mathcal{T}_n|\leq K-1$. Let us prove them respectively,
\begin{eqnarray}
&&H\left(\{Z_{u,v}\}_{(u,v) \in \mathcal{K}_{\mathcal{U}^{(m,n)}}}\big|\{Z_{u,v}\}_{(u,v)\in\mathcal{T}_n}\right)\\
&\geq&H\left(\{Z_{u,v}\}_{(u,v) \in \mathcal{K}_{\mathcal{U}^{(m,n)}}}\big|\{Z_{u,v}\}_{(u,v)\in\mathcal{T}_n},\{W_{u,v}\}_{(u,v) \in \mathcal{K}_{\mathcal{U}^{(m,n)}}}\right)\\
&\geq&I\left(\{Z_{u,v}\}_{(u,v) \in \mathcal{K}_{\mathcal{U}^{(m,n)}}};\{Y_u\}_{u\in \mathcal{U}^{(m,n)}}\big|\{Z_{u,v}\}_{(u,v)\in\mathcal{T}_n},\{W_{u,v}\}_{(u,v) \in \mathcal{K}_{\mathcal{U}^{(m,n)}}}\right)\\
&\overset{(\ref{messageX})(\ref{messageY})
}{=}&H\left(\{Y_u\}_{u\in \mathcal{U}^{(m,n)}}\big|\{Z_{u,v}\}_{(u,v)\in\mathcal{T}_n},\{W_{u,v}\}_{(u,v) \in \mathcal{K}_{\mathcal{U}^{(m,n)}}}\right)\\
&\geq&H\left(\{Y_u\}_{u\in \mathcal{U}^{(m,n)}}\big|\{Z_{u,v}\}_{(u,v)\in\mathcal{T}_n},\{W_{u,v}\}_{(u,v) \in \mathcal{K}_{\mathcal{U}^{(m,n)}}\setminus\mathcal{T}_n},\{W_{u,v}\}_{(u,v)\in\mathcal{T}_n}\right)\\
&=&H\left(\{Y_u\}_{u\in \mathcal{U}^{(m,n)}}\big|\{Z_{u,v},W_{u,v}\}_{(u,v)\in\mathcal{T}_n}\right)\notag\\
&&-I\left(\{Y_u\}_{u\in \mathcal{U}^{(m,n)}};\{W_{u,v}\}_{(u,v) \in \mathcal{K}_{\mathcal{U}^{(m,n)}}\setminus\mathcal{T}_n}\big|\{Z_{u,v},W_{u,v}\}_{(u,v)\in\mathcal{T}_n}\right)\\
&\geq&H\left(\{Y_u\}_{u\in \mathcal{U}^{(m,n)}}\big|\{Z_{u,v},W_{u,v}\}_{(u,v)\in\mathcal{T}_n}\right)\notag\\
&&-I\Big(\{Y_u\}_{u\in [U]},\sum_{(u,v)\in\mathcal{K}_{[U]}}W_{u,v};\{W_{u,v}\}_{(u,v) \in \mathcal{S}_{m}}\Big|\{Z_{u,v},W_{u,v}\}_{(u,v)\in\mathcal{T}_n}\Big)\label{pf_lemma3_t4}\\
&=&|\mathcal{U}^{(m,n)}|L-I\Big(\sum_{(u,v)\in\mathcal{K}_{[U]}}W_{u,v};\{W_{u,v}\}_{(u,v) \in \mathcal{S}_m}\Big|\{Z_{u,v},W_{u,v}\}_{(u,v)\in\mathcal{T}_n}\Big)\notag\\
&&-I\Big(\{Y_u\}_{u\in [U]};\{W_{u,v}\}_{(u,v) \in \mathcal{S}_m}\Big|\sum_{(u,v)\in\mathcal{K}_{[U]}}W_{u,v},\{Z_{u,v},W_{u,v}\}_{(u,v)\in\mathcal{T}_n}\Big)\label{pf_lemma3_t5}\\
%&\geq&\mbox{min}(|\mathcal{U}^{(m,n)}|,U-1)L.\\
& \geq& \Bigg\{
    \begin{array}{cl}
        (|\mathcal{U}^{(m,n)}|-1)L, &\mbox{if}~|\mathcal{K}_{\mathcal{U}^{(m,n)}}\cup\mathcal{T}_n|= K,\\
         |\mathcal{U}^{(m,n)}|L, &\mbox{otherwise},
    \end{array}
\end{eqnarray}
where in (\ref{pf_lemma3_t4}), the second term follows the fact that $\mathcal{U}^{(m,n)}\subset [U]$ and $(\mathcal{K}_{\mathcal{U}^{(m,n)}}\setminus\mathcal{T}_n) \subset \mathcal{S}_m$. In (\ref{pf_lemma3_t5}), the first term follows from chain-rule and applying Lemma \ref{lemma1}
(\ref{lemma1_eqY}) to each expanded term; the third term is zero due to server security constraints with $\mathcal{S}_m$ and $\mathcal{T}_n$ (see (\ref{serversecurity})).

When $|\mathcal{K}_{\mathcal{U}^{(m,n)}}\cup\mathcal{T}_n|\leq K-1$, the second term of (\ref{pf_lemma3_t5}) is zero due to $|\mathcal{S}_{m}\cup\mathcal{T}_n|\leq K-1$ and the independence of $W_{u,v}$ and $Z_{u,v}$. When $|\mathcal{K}_{\mathcal{U}^{(m,n)}}\cup\mathcal{T}_n|= K$, the second term of (\ref{pf_lemma3_t5}) is $L$ due to $|\mathcal{S}_{m}\cup\mathcal{T}_n|= K$

% 
% \begin{eqnarray}
% &&H\left(\{Z_{u,v}\}_{(u,v) \in \mathcal{K}_{\mathcal{U}^{(m,n)}}}\big|\{Z_{u,v}\}_{(u,v)\in\mathcal{T}_n}\right)\\
% &\overset{(\ref{pf_lemma3_t4})}{\geq}&|\mathcal{U}^{(m,n)}\setminus\mathcal{F}^{(m,n)}|L\notag\\
% &&-I\left(\sum_{(u,v)\in\mathcal{K}_{[U]}}W_{u,v};\{W_{u,v}\}_{(u,v) \in \mathcal{K}_{\mathcal{U}^{(m,n)}}\setminus\mathcal{T}_n}\big|\{Z_{u,v}\}_{(u,v)\in\mathcal{T}_n},\{W_{u,v}\}_{(u,v)\in\mathcal{T}_n}\right)\notag\\
% &&-I\left(\{Y_u\}_{u\in [U]};\{W_{u,v}\}_{(u,v) \in \mathcal{K}_{\mathcal{U}^{(m,n)}}\setminus\mathcal{T}_n}\big|\sum_{(u,v)\in\mathcal{K}_{[U]}}W_{u,v},\{Z_{u,v}\}_{(u,v)\in\mathcal{T}_n},\{W_{u,v}\}_{(u,v)\in\mathcal{T}_n}\right)\label{pf_lemma3_t6}\\
% &\geq&(U-1)L.
% \end{eqnarray}
% \end{lemma}
% Where (\ref{pf_lemma3_t6}) applies (\ref{pf_lemma3_t4}). In (\ref{pf_lemma3_t6}), the first is $UL$ follows the fact that $|\mathcal{U}^{(m,n)}\setminus\mathcal{F}^{(m,n)}|= U$; the second term is $L$ because $|\mathcal{U}^{(m,n)}|= U$ and $|\mathcal{K}_{\mathcal{U}^{(m,n)}}\cup\mathcal{T}_n|=K$.

\begin{lemma} \label{lemma4} 
Let $m \in [M]$, $n \in [N]$, and consider sets $\mathcal{S}_m, \mathcal{T}_n$ such that $\mathcal{S}_m \cap \mathcal{T}_n = \emptyset$ and $|\mathcal{S}_m \cup \mathcal{T}_n| \le K-1$. Then
\begin{align}
&H\Big(\{Z_{u,v}\}_{(u,v) \in ((\mathcal{S}_m \cap \mathcal{K}_{\{u'\}}) \cup \mathcal{T}_n) \cap \overline{\mathcal{S}}} \,\Big|\, 
\{Z_{u,v}\}_{(u,v) \in \mathcal{T}_n \setminus \overline{\mathcal{S}}}\Big)
\ge |((\mathcal{S}_m \cap \mathcal{K}_{\{u'\}}) \cup \mathcal{T}_n) \cap \overline{\mathcal{S}}| L, \label{lemma4_eqX} \\[1mm]
&H\Big(\{Z_{u,v}\}_{(u,v) \in (\mathcal{K}_{\mathcal{U}^{(m,n)}} \cup \mathcal{T}_n) \cap \overline{\mathcal{S}}} \,\Big|\, 
\{Z_{u,v}\}_{(u,v) \in \mathcal{T}_n \setminus \overline{\mathcal{S}}}\Big)
\ge |\mathcal{T}_n \cap \overline{\mathcal{S}}| L + 
\begin{cases}
(|\mathcal{U}^{(m,n)}|-1)L, & \text{if } |\mathcal{K}_{\mathcal{U}^{(m,n)}} \cup \mathcal{T}_n| = K,\\
|\mathcal{U}^{(m,n)}| L, & \text{otherwise}.
\end{cases} \label{lemma4_eqY}
\end{align}
\end{lemma}
First, consider (\ref{lemma4_eqX})
\begin{eqnarray}
&&H\left(\{Z_{u,v}\}_{(u,v) \in ((\mathcal{S}_m\cap\mathcal{K}_{\{u'\}})\cup\mathcal{T}_n)\cap\overline{\mathcal{S}}}\big|\{Z_{u,v}\}_{(u,v)\in\mathcal{T}_n\setminus \overline{\mathcal{S}}}\right)\\
&=&H\left(\{Z_{u,v}\}_{(u,v) \in (\mathcal{T}_n\cap\overline{\mathcal{S}})}\big|\{Z_{u,v}\}_{(u,v)\in\mathcal{T}_n\setminus \overline{\mathcal{S}}}\right)\notag\\
&&+H\left(\{Z_{u,v}\}_{(u,v) \in (\mathcal{S}_m\cap\mathcal{K}_{\{u'\}})\setminus(\mathcal{T}_n\cap\overline{\mathcal{S}})}\big|\{Z_{u,v}\}_{(u,v)\in\mathcal{T}_n\setminus \overline{\mathcal{S}}},\{Z_{u,v}\}_{(u,v)\in(\mathcal{T}_n\cap \overline{\mathcal{S}})}\right)\\
&=&H\left(\{Z_{u,v}\}_{(u,v) \in \mathcal{T}_n\cap\overline{\mathcal{S}}}\big|\{Z_{u,v}\}_{(u,v)\in\mathcal{T}_n\setminus \overline{\mathcal{S}}}\right)+H\left(\{Z_{u,v}\}_{(u,v) \in (\mathcal{S}_m\cap\mathcal{K}_{\{u'\}})}\big|\{Z_{u,v}\}_{(u,v)\in\mathcal{T}_n}\right)\\
&\geq&|\mathcal{T}_n\cap\overline{\mathcal{S}}|L+|(\mathcal{S}_m\cap\mathcal{K}_{\{u'\}})\setminus\mathcal{T}_n|L\label{pf_lemma4_t1}\\
&=&|((\mathcal{S}_m\cap\mathcal{K}_{\{u'\}})\cup\mathcal{T}_n)\cap\overline{\mathcal{S}}| L.
\end{eqnarray}
where in (\ref{pf_lemma4_t1}), the first term will be proven in (\ref{pf_lemma4_t4}); the second term follows (\ref{lemma3_eqX}) in lemma 3.

 Second, consider (\ref{lemma4_eqY})
\begin{eqnarray}
&&H\left(\{Z_{u,v}\}_{(u,v) \in (\mathcal{K}_{\mathcal{U}^{(m,n)}}\cup\mathcal{T}_n)\cap\overline{\mathcal{S}}}\big|\{Z_{u,v}\}_{(u,v)\in\mathcal{T}_n\setminus\overline{\mathcal{S}}}\right)\\
&=&H\left(\{Z_{u,v}\}_{(u,v) \in \mathcal{T}_n\cap\overline{\mathcal{S}}}\big|\{Z_{u,v}\}_{(u,v)\in\mathcal{T}_n\setminus\overline{\mathcal{S}}}\right)\\
&&+H\left(\{Z_{u,v}\}_{(u,v) \in \mathcal{K}_{\mathcal{U}^{(m,n)}}\setminus(\mathcal{T}_n\cap\overline{\mathcal{S}})}\big|\{Z_{u,v}\}_{(u,v)\in\mathcal{T}_n\setminus\overline{\mathcal{S}}},\{Z_{u,v}\}_{(u,v) \in \mathcal{T}_n\cap\overline{\mathcal{S}}}\right)\\
&=&H\left(\{Z_{u,v}\}_{(u,v) \in \mathcal{T}_n\cap\overline{\mathcal{S}}}\big|\{Z_{u,v}\}_{(u,v)\in\mathcal{T}_n\setminus\overline{\mathcal{S}}}\right)+H\left(\{Z_{u,v}\}_{(u,v) \in \mathcal{K}_{\mathcal{U}^{(m,n)}}}\big|\{Z_{u,v}\}_{(u,v)\in\mathcal{T}_n}\right)\\
& \geq& |\mathcal{T}_n\cap\overline{\mathcal{S}}|L +\Bigg\{
    \begin{array}{cl}
        (|\mathcal{U}^{(m,n)}|-1)L, &\mbox{if}~|\mathcal{K}_{\mathcal{U}^{(m,n)}}\cup\mathcal{T}_n|= K,\\
         |\mathcal{U}^{(m,n)}|L, &\mbox{otherwise}, 
    \end{array}\label{pf_lemma4_t2}
\end{eqnarray}
where in (\ref{pf_lemma4_t2}), the first term will be proven in (\ref{pf_lemma4_t4}); the second term follows (\ref{lemma3_eqY}) in lemma 3.

Next, we will show that $H\left(\{Z_{u,v}\}_{(u,v) \in \mathcal{T}_n\cap\overline{\mathcal{S}}}\big|\{Z_{u,v}\}_{(u,v)\in\mathcal{T}_n\setminus \overline{\mathcal{S}}}\right)\geq |\mathcal{T}_n\cap\overline{\mathcal{S}}|L$ in the first term of (\ref{pf_lemma4_t1}) and (\ref{pf_lemma4_t2}). For each element $(u',v') \in \mathcal{T}_n \cap \overline{\mathcal{S}}$, we have
\begin{eqnarray}
&&H\left(\{Z_{u,v}\}_{(u,v) \in \mathcal{T}_n\cap\overline{\mathcal{S}}}\big|\{Z_{u,v}\}_{(u,v)\in\mathcal{T}_n\setminus \overline{\mathcal{S}}}\right)\notag\\
 &\geq&\sum_{u',v': (u',v')\in \mathcal{T}_n\cap\overline{\mathcal{S}}}H\left(Z_{u',v'}\big|\{Z_{u,v}\}_{(u,v)\in\mathcal{T}_n\setminus \overline{\mathcal{S}}},\{Z_{u,v}\}_{(u,v)\in (\mathcal{T}_n\cap\overline{\mathcal{S}})\setminus\{(u',v')\}}\right) \\
&\geq&\sum_{u',v': (u',v')\in \mathcal{T}_n\cap\overline{\mathcal{S}}}H\left(Z_{u',v'}\big|\{Z_{u,v}\}_{(u,v)\in (\mathcal{T}_n)\setminus\{(u',v')\}}\right) \\
&\overset{(\ref{lemma3_eqX})}{\geq}&|\mathcal{T}_n\cap\overline{\mathcal{S}}|L.\label{pf_lemma4_t4}
\end{eqnarray}
The last step holds because, for each $(u',v') \in \mathcal{T}_n \cap \overline{\mathcal{S}}$, there exists $m \in [M]$ such that $(u',v') \in \mathcal{S}_m \cap \mathcal{K}_{\{u'\}}$, and we consider only one element $(u',v')$ at a time.

% \hfill\QED

We first establish that $R_X \ge 1$ and $R_Y \ge 1$, which holds under Theorems~\ref{thm:1}--\ref{thm:3}.  
By Lemma~\ref{lemma1}, we have
\begin{align}
    H(X_{u,v}) &\ge H\Big(X_{u,v} \,\big|\, \{W_{i,j},Z_{i,j}\}_{(i,j)\in \mathcal{K}_{[U]}\setminus \{(u,v)\}}\Big) 
    \overset{(\ref{lemma1_eqX})}{\ge} L, \label{lemma1_eqXX} \\
    H(Y_{u}) &\ge H\Big(Y_{u} \,\big|\, \{W_{i,j},Z_{i,j}\}_{(i,j)\in \mathcal{K}_{[U]}\setminus \{(u,v)\}}\Big)
    \overset{(\ref{lemma1_eqY})}{\ge} L. \label{lemma1_eqYY}
\end{align}
It follows that
\begin{align}
    R_X \ge \frac{L_X}{L} \ge \frac{H(X_{u,v})}{L} \ge 1, \quad
R_Y \ge \frac{L_Y}{L} \ge \frac{H(Y_{u})}{L} \ge 1.
\end{align}

second, we show that $R_{Z_\Sigma} \geq \max(a^*, d^* - 1)$ when $|\mathcal{K}_{\mathcal{U}^{(m,n)}} \cup \mathcal{T}_n| = K$, and $R_{Z_\Sigma} \geq \max(a^*, d^*)$ when $|\mathcal{K}_{\mathcal{U}^{(m,n)}} \cup \mathcal{T}_n| \leq K - 1$. From (\ref{lemma4_eqX}) in Lemma \ref{lemma4}, it directly follows that $R_{Z_\Sigma} \geq a^*$. Now consider any $\mathcal{A}_{u',m,n}$ where $\mathcal{K}_{\{u'\}}, \mathcal{S}_m, \mathcal{T}_n$ satisfy the conditions in Lemma \ref{lemma4}.
\begin{eqnarray}
    H(Z_{\Sigma})&\overset{(\ref{total rand})}{\geq}&H\left(\{Z_{u,v}\}_{(u,v)\in \mathcal{K}_{[U]}}\right)\geq H\left(\{Z_{u,v}\}_{(u,v) \in ((\mathcal{S}_m\cap\mathcal{K}_{\{u'\}})\cup\mathcal{T}_n)\cap\overline{\mathcal{S}}}\big|\{Z_{u,v}\}_{(u,v)\in\mathcal{T}_n\setminus \overline{\mathcal{S}}}\right)\\
    &\overset{(\ref{lemma4_eqX})}{\geq}& |((\mathcal{S}_m\cap\mathcal{K}_{\{u'\}})\cup\mathcal{T}_n)\cap\overline{\mathcal{S}}| L\\
\Rightarrow ~~H(Z_{\Sigma}) &\geq& \mbox{max}_{u',m,n} |\mathcal{A}_{u',m,n}| L =a^*L\label{totalrand_eq11}
\end{eqnarray}
which immediately implies
\begin{align}
    R_{Z_\Sigma} = \frac{L_{Z_\Sigma}}{L} \ge \frac{H(Z_\Sigma)}{L} \ge a^*.\label{maxXa}
\end{align}

From (\ref{lemma4_eqY}) in Lemma \ref{lemma4}, we have $R_{Z_\Sigma} \geq d^* - 1$ when $|\mathcal{K}_{\mathcal{U}^{(m,n)}} \cup \mathcal{T}_n| = K$, and $R_{Z_\Sigma} \geq d^*$ when $|\mathcal{K}_{\mathcal{U}^{(m,n)}} \cup \mathcal{T}_n| \leq K - 1$. This holds for any $\mathcal{E}_{m,n}$ such that $\mathcal{K}_{\mathcal{U}^{(m,n)}}, \mathcal{S}_m$, and $\mathcal{T}_n$ satisfy the conditions specified in Lemma \ref{lemma4}.
\begin{eqnarray}
    H(Z_{\Sigma})&\overset{(\ref{total rand})}{\geq}&H\left(\{Z_{u,v}\}_{(u,v)\in \mathcal{K}_{[U]}}\right)\geq H\left(\{Z_{u,v}\}_{(u,v) \in (\mathcal{K}_{\mathcal{U}^{(m,n)}}\cup\mathcal{T}_n)\cap\overline{\mathcal{S}}}\big|\{Z_{u,v}\}_{(u,v)\in\mathcal{T}_n\setminus\overline{\mathcal{S}}}\right)\\
    &\overset{(\ref{lemma4_eqY})}{\geq}& 
     |\mathcal{T}_n\cap\overline{\mathcal{S}}|L +\Bigg\{
    \begin{array}{cl}
        (|\mathcal{U}^{(m,n)}|-1)L, &\mbox{if}~|\mathcal{K}_{\mathcal{U}^{(m,n)}}\cup\mathcal{T}_n|= K\\
         |\mathcal{U}^{(m,n)}|L, &\mbox{otherwise}
    \end{array}\\
\Rightarrow ~~H(Z_{\Sigma}) &\geq& \mbox{max}_{m,n} \Bigg(|\mathcal{T}_n\cap\overline{\mathcal{S}}|L +\Bigg\{
    \begin{array}{cl}
        (|\mathcal{U}^{(m,n)}|-1)L, &\mbox{if}~|\mathcal{K}_{\mathcal{U}^{(m,n)}}\cup\mathcal{T}_n|= K\\
         |\mathcal{U}^{(m,n)}|L, &\mbox{otherwise}
    \end{array}\Bigg)\\
    &\overset{(\ref{eq:def d*})}{=}&\Bigg\{
    \begin{array}{cl}
        (d^*-1)L, &\mbox{if}~|\mathcal{K}_{\mathcal{U}^{(m,n)}}\cup\mathcal{T}_n|= K,\\
         d^*L, &\mbox{otherwise},
    \end{array}\label{totalrand_eq1}
\end{eqnarray}
which immediately implies
\begin{align}
    R_{Z_{\Sigma}} = \frac{L_{Z_{\Sigma}}}{L} \geq \frac{H(Z_{\Sigma})}{L} \geq\Bigg\{
    \begin{array}{cl}
        (d^*-1), &\mbox{if}~|\mathcal{K}_{\mathcal{U}^{(m,n)}}\cup\mathcal{T}_n|= K,\\
         d^*, &\mbox{otherwise}. 
    \end{array}\label{maxYd}
\end{align}

From (\ref{maxXa}) and (\ref{maxYd}), we have
\begin{eqnarray}
    R_{Z_{\Sigma}}\geq \Bigg\{
    \begin{array}{cl}
        \max(a^*,d^*-1), &\mbox{Condition 1, Case 2)},\\
         \max(a^*,d^*), &\mbox{Condition 1, Case 3), 4)}. 
    \end{array}
\end{eqnarray}

Next, we show that $R_{Z_{\Sigma}} = \max(a^, d^ - 1)$ is achievable under Condition~1, Case~2), i.e., $a^* \le K-1$ and $|\mathcal{K}_{\mathcal{U}^{(m,n)}} \cup \mathcal{T}_n| = K$.

\subsection{General Achievable Scheme under Condition 1, Case 2)}\label{sec:ach11}
Every user in the (implicit and explicit) security input set $\overline{\mathcal{S}}$ is assigned a key variable. We operate over the field $\mathbb{F}_{q}$. To satisfy the field size requirements for security, we set $q > \max(a^*,d^*-1) \binom{|\overline{\mathcal{S}}|}{\max(a^*,d^*-1)}$. Next, we consider $\max(a^*,d^*-1)$ independent and identically distributed (i.i.d.) uniform random variables, represented as a column vector $Z_\Sigma={\bf N} = (N_1; \cdots ; N_{\max(a^*,d^*-1)}) \in \mathbb{F}_{q}^{\max(a^*,d^*-1) \times 1}$, and assign these variables as the key variables.
 \begin{eqnarray}
 Z_{u,v} &=& {\bf h}_{u,v} \, {\bf N}, \forall (u,v) \in \overline{\mathcal{S}}\notag\\
 Z_{u,v} &=& 0, \forall (u,v) \in \mathcal{K}_{[U]}\setminus\overline{\mathcal{S}}\label{eq:c111}
 \end{eqnarray}
 where ${\bf h}_{u,v} \in \mathbb{F}_{q}^{1\times \max(a^*,d^*-1)}$ are chosen as follows (suppose $\overline{\mathcal{S}} = \{(u_1,v_1), \cdots, (u_{|\overline{\mathcal{S}}|},v_{|\overline{\mathcal{S}}|})\}$)
 \begin{eqnarray}
&& \mbox{each element of}~{\bf h}_{u_1,v_1}, \cdots, {\bf h}_{u_{|\overline{\mathcal{S}}|-1},v_{|\overline{\mathcal{S}}|-1}}~\mbox{is chosen uniformly and i.i.d. from}~\mathbb{F}_{q}, \notag\\
&& {\bf h}_{u_{|\overline{\mathcal{S}}|},v_{|\overline{\mathcal{S}}|}} \eqdef -\left( {\bf h}_{u_1,v_1} + \cdots + {\bf h}_{u_{|\overline{\mathcal{S}}|-1},v_{|\overline{\mathcal{S}}|-1}}\right) \label{eq:c121}
 \end{eqnarray}
There must exist a realization\footnote{Since $\max(a^*,d^*-1) < |\overline{\mathcal{S}}|$, the determinant associated with any selection of $\max(a^*,d^*-1)$ distinct vectors ${\bf h}_{u,v}$ defines a non-zero polynomial. Applying the Schwartz--Zippel lemma, the product of all such determinant polynomials has degree $\max(a^*,d^*-1) \binom{|\overline{\mathcal{S}}|}{\max(a^*,d^*-1)}$. By choosing a field $\mathbb{F}_q$ with $q$ larger than this degree, the probability that the product polynomial evaluates to a non-zero value is strictly positive. This ensures the validity of (\ref{eq:sz111}).
} of ${\bf h}_{u,v}, (u,v) \in \overline{\mathcal{S}}$ such that
\begin{eqnarray}
&&\mbox{any $\max(a^*,d^*-1)$ or fewer distinct ${\bf h}_{u,v}, (u,v) \in \overline{\mathcal{S}}$ and $\sum_{v\in[V_u]}{\bf h}_{u,v},u\in \mathcal{U}^{(m,n)}$} \notag\\
&&\mbox{vectors are linearly independent.}  \label{eq:sz111}
\end{eqnarray}
 Moreover,
 \begin{eqnarray}
 \sum_{(u,v) \in \mathcal{K}_{[U]}} Z_{u,v} \overset{(\ref{eq:c111})}{=} \sum_{(u,v) \in \overline{\mathcal{S}}} Z_{u,v} \overset{(\ref{eq:c121})}{=} 0. \label{eq:c131}
 \end{eqnarray}
 Finally, set the sent messages as
 \begin{eqnarray}
     X_{u,v} &=& W_{u,v} + Z_{u,v}, \forall (u,v) \in \mathcal{K}_{[U]}.\label{eq:achm1X1}\\
     Y_{u} &=& \sum_{v\in [V_u]}X_{u,v}, \forall u \in [U].\label{eq:achm1Y1}
 \end{eqnarray}
Thus, the achieved communication rates satisfy $R_X = L_X / L = 1$ and $R_Y = L_Y / L = 1$, while the source key rate is given by $R_{Z_\Sigma} = L_{Z_\Sigma}/L = \max(a^*, d^* - 1)$, as desired.
 The correctness can be verified through the following derivation: $\sum_{u\in[U]} Y_u \overset{(\ref{eq:achm1Y1})}{=} \sum_{u\in[U]} \sum_{v\in [V_u]} X_{u,v} = \sum_{(u,v)\in \mathcal{K}_{[U]}} X_{u,v} \overset{(\ref{eq:achm1X1})(\ref{eq:c131})}{=} \sum_{(u,v)\in \mathcal{K}_{[U]}} W_{u,v}$. The security proof is deferred to Section \ref{sec:security}, which relies on proving the existence of a realization of the randomly generated vectors ${\bf h}_{u,v}$ that satisfy a certain full-rank property.

Next, when $a^*\leq K-1$ and $|\mathcal{K}_{\mathcal{U}^{(m,n)}}\cup\mathcal{T}_n|\leq K-1$, we will show that $R_{Z_{\Sigma}}^* =\max(a^*,d^*)$ is achievable for the case $\max(a^*,e^*)<|\overline{\mathcal{S}}|$, or $\max(a^*,e^*)=|\overline{\mathcal{S}}|$ and $|\mathcal{Q}|<K$ respectively. First, consider the case $\max(a^*,e^*)<|\overline{\mathcal{S}}|$.

\subsection{General Achievable Scheme under Condition 1, Case 3)}\label{sec:ach1}
In this case, each user in the implicit and explicit security input set $\overline{\mathcal{S}}$ is assigned a key variable. Operations are performed over the field $\mathbb{F}_{q}$. To satisfy the field size requirements for security, we ensure $q > \max(a^*,d^*) \binom{|\overline{\mathcal{S}}|}{\max(a^*,d^*)}$. Next, consider $\max(a^*,d^*)$ independent and identically distributed (i.i.d.) uniform variables, organized into a column vector $ Z_\Sigma={\bf N} = (N_1; \cdots; N_{\max(a^*,d^*)}) \in \mathbb{F}_{q}^{\max(a^*,d^*) \times 1}$ and set the key variables as 
 \begin{eqnarray}
 Z_{u,v} &=& {\bf h}_{u,v} \, {\bf N}, \forall (u,v) \in \overline{\mathcal{S}}\notag\\
 Z_{u,v} &=& 0, \forall (u,v) \in \mathcal{K}_{[U]}\setminus\overline{\mathcal{S}}\label{eq:c11}
 \end{eqnarray}
 where ${\bf h}_{u,v} \in \mathbb{F}_{q}^{1\times \max(a^*,d^*)}$ are chosen as follows (suppose $\overline{\mathcal{S}} = \{(u_1,v_1), \cdots, (u_{|\overline{\mathcal{S}}|},v_{|\overline{\mathcal{S}}|})\}$)
 \begin{eqnarray}
&& \mbox{each element of}~{\bf h}_{u_1,v_1}, \cdots, {\bf h}_{u_{|\overline{\mathcal{S}}|-1},v_{|\overline{\mathcal{S}}|-1}}~\mbox{is chosen uniformly and i.i.d.} \notag\\
&& \mbox{from}~\mathbb{F}_{q}, {\bf h}_{u_{|\overline{\mathcal{S}}|},v_{|\overline{\mathcal{S}}|}} \eqdef -\left( {\bf h}_{u_1,v_1} + \cdots + {\bf h}_{u_{|\overline{\mathcal{S}}|-1},v_{|\overline{\mathcal{S}}|-1}}\right) \label{eq:c12}
 \end{eqnarray}
 Using the same argument based on the Schwartz--Zippel lemma, one can choose a realization 
of ${\bf h}_{u,v}$, $(u,v) \in \overline{\mathcal{S}}$, such that
\begin{eqnarray}
&&\mbox{any $\max(a^*,d^*)$ or fewer distinct ${\bf h}_{u,v}, (u,v) \in \overline{\mathcal{S}}$, $\sum_{v\in[V_u]}{\bf h}_{u,v},u\in \mathcal{U}^{(m,n)}$} \notag\\
&&\mbox{vectors are linearly independent.}  \label{eq:sz1}
\end{eqnarray}
Moreover,
 \begin{eqnarray}
 \sum_{(u,v) \in \mathcal{K}_{[U]}} Z_{u,v} \overset{(\ref{eq:c11})}{=} \sum_{(u,v) \in \overline{\mathcal{S}}} Z_{u,v} \overset{(\ref{eq:c12})}{=} 0. \label{eq:c13}
 \end{eqnarray}
 Finally, set the sent messages as
 \begin{eqnarray}
     X_{u,v} &=& W_{u,v} + Z_{u,v}, \forall (u,v) \in \mathcal{K}_{[U]}.\label{eq:achm1X}\\
     Y_{u} &=& \sum_{v\in [V_u]}X_{u,v}, \forall u \in [U].\label{eq:achm1Y}
 \end{eqnarray}
 The achieved communication rate is $R_X = L_X / L = 1$ and $R_Y = L_Y / L = 1$.
Note that each symbol above is from $\mathbb{F}_{q}$, so we have $L_{Z_\Sigma} = \max(a^*,d^*)L$, leading to the key rate $R_{Z_\Sigma} = L_{Z_\Sigma}/L = \max(a^*,d^*)$, as desired. The correctness can be verified by observing that $\sum_{u\in[U]} Y_u \overset{(\ref{eq:achm1Y})}{=} \sum_{u\in[U]} \sum_{v\in [V_u]} X_{u,v} = \sum_{(u,v)\in \mathcal{K}_{[U]}} X_{u,v} \overset{(\ref{eq:achm1X})(\ref{eq:c13})}{=} \sum_{(u,v)\in \mathcal{K}_{[U]}} W_{u,v}$. The security proof is deferred to Section \ref{sec:security}, where it is shown that the randomly generated vectors ${\bf h}_{u,v}$ possess a certain full-rank property necessary for security.

% \begin{remark}
% For this case and all cases presented in the following, the communication rate $L_X/L$ is 1 and is optimal (minimum, i.e., weak security does not incur additional communication cost).
% \end{remark}

Second, consider the case $\max(a^*,e^*) = \big|\overline{\mathcal{S}}\big|$ and $|\mathcal{Q}| < K$.

\subsection{General Achievable Scheme under Condition 1, Case 4)}\label{sec:ach2}

In this case, every user in the total security input set $\overline{\mathcal{S}}$, together with one additional user outside $\mathcal{Q}$, is assigned a key variable. To ensure sufficient field size, pick $q$ such that $q > \max(a^*,d^*) \binom{|\overline{\mathcal{S}}|+1}{\max(a^*,d^*)}$, and operate over the field $\mathbb{F}_{q}$. Consider $\max(a^*,d^*)$ i.i.d. uniform variables $Z_\Sigma={\bf N} = (N_1; \cdots; N_{\max(a^*,d^*)}) \in \mathbb{F}_{q}^{\max(a^*,d^*) \times 1}$. Next, find any $(u',v') \in \mathcal{K}_{[U]} \setminus \mathcal{Q}$, which must exist because $|\mathcal{Q}| < K$. Additionally, ensure $(u',v') \notin \overline{\mathcal{S}}$, as $\max(a^*,e^*) = \big|\overline{\mathcal{S}}\big|$ and $\overline{\mathcal{S}} \subset \mathcal{Q}$. Finally, set the key variables as
 \begin{eqnarray}
 Z_{u,v} &=& {\bf h}_{u,v} \, {\bf N}, \forall (u,v) \in (\overline{\mathcal{S}} \cup \{(u',v')\}) \notag\\
 Z_{u,v} &=& 0, \forall (u,v) \in \mathcal{K}_{[U]}\setminus(\overline{\mathcal{S}}\cup\{(u',v')\})\label{eq:c21}
 \end{eqnarray}
   where ${\bf h}_{u,v} \in \mathbb{F}_{q}^{1\times \max(a^*,d^*)}$ are chosen as follows (suppose $\overline{\mathcal{S}} = \{(u_1,v_1), \cdots, (u_{|\overline{\mathcal{S}}|},v_{|\overline{\mathcal{S}}|})\}$)
 \begin{eqnarray}
&& \mbox{each element of}~{\bf h}_{u_1,v_1}, \cdots, {\bf h}_{u_{|\overline{\mathcal{S}}|},v_{|\overline{\mathcal{S}}|}}~\mbox{is chosen uniformly and i.i.d. from}~\mathbb{F}_{q}, \notag\\
&& {\bf h}_{u',v'} \eqdef -\left( {\bf h}_{u_1,v_1} + \cdots + {\bf h}_{u_{|\overline{\mathcal{S}}|},v_{|\overline{\mathcal{S}}|}}\right) \label{eq:c22}
 \end{eqnarray}
 Using the same argument based on the Schwartz--Zippel lemma, one can choose a realization 
of ${\bf h}_{u,v}$, $(u,v) \in \overline{\mathcal{S}}$, such that
\begin{eqnarray}
&&\mbox{any $\max(a^*,d^*)$ or fewer distinct ${\bf h}_{u,v}, (u,v) \in \overline{\mathcal{S}} \cup \{(u',v')\}$ and   $\sum_{v\in[V_u]}{\bf h}_{u,v},u\in \mathcal{U}^{(m,n)}$} \notag\\
&&\mbox{vectors are linearly independent.} \label{eq:sz2}
\end{eqnarray}
Moreover,
 \begin{eqnarray}
 \sum_{(u,v) \in \mathcal{K}_{[U]}} Z_{u,v} \overset{(\ref{eq:c21})}{=} \sum_{(u,v) \in (\overline{\mathcal{S}}\cup\{(u',v')\}) } Z_{u,v} \overset{(\ref{eq:c22})}{=} 0. \label{eq:c23}
 \end{eqnarray}
  Finally, set the sent messages as
 \begin{eqnarray}
     X_{u,v} &=& W_{u,v} + Z_{u,v}, \forall (u,v) \in \mathcal{K}_{[U]}.\label{eq:achm2X}\\
     Y_{u} &=& \sum_{v\in[V_u]}X_{u,v}, \forall u \in [U].\label{eq:achm2Y}
 \end{eqnarray}
The achieved communication rate is $R_X = L_X / L = 1$ and $R_Y = L_Y / L = 1$.
Note that each symbol above is from $\mathbb{F}_{q}$, so we have $L_{Z_\Sigma} = \max(a^*,d^*)L$, giving the key rate $R_{Z_\Sigma} = L_{Z_\Sigma}/L = \max(a^*,d^*)$. Correctness is verified by observing that $\sum_{u\in[U]} Y_u \overset{(\ref{eq:achm2Y})}{=} \sum_{u\in[U]} \sum_{v\in[V_u]} X_{u,v} = \sum_{(u,v)\in \mathcal{K}_{[U]}} X_{u,v}$ $ \overset{(\ref{eq:achm2X})(\ref{eq:c23})}{=} \sum_{(u,v)\in \mathcal{K}_{[U]}} W_{u,v}$. The security proof, detailed in Section \ref{sec:security}, relies on showing that the randomly generated vectors ${\bf h}_{u,v}$ satisfy a required full-rank property.

\section{Proof of Theorem \ref{thm:2}} \label{sec:thm2}

Next we consider the case of Theorem \ref{thm:2}, where we need one more step to further tighten the bound $R_{Z_\Sigma} \geq \max(a^*,d^*)$ to include an additional term $b^*$. To illustrate the key idea in a simpler setting, we first consider an example.

\begin{example}\label{ex2}
Consider $U=2$ and $V_1=4, V_2=1$, with $K=5$ users: $(1,1)$, $(1,2)$, $(1,3)$, $(1,4)$, $(2,1)$. The security input sets are $(\mathcal{S}_1, \mathcal{S}_2, \mathcal{S}_3) = (\emptyset,$ $ \{(1,1)\},$ $ \{(1,2)\})$, while the colluding user sets are $(\mathcal{T}_1, \cdots, \mathcal{T}_8) = (\emptyset, $ $\{(1,2)\},$ $ \{(1,3)\},$ $ \{(1,4)\},\{(2,1)\}, \{(1,2),(1,3)\},\{(1,2),$ $(1,4)\},$ $\{(1,2),$ $(2,1)\})$.
\end{example}

For Example~\ref{ex2}, by enumerating all $\mathcal{A}_{u,m,n}$ with maximum cardinality, we obtain
$\mathcal{A}_{1,2,6} = \mathcal{A}_{1,2,7} = \mathcal{A}_{1,2,8} = \{(1,1),(1,2)\}$.
Hence, $a^* = |\overline{\mathcal{S}}| = 2 \le K-1 = 4$, and $e^* = 1 < a^*$.
Moreover, $\mathcal{Q}_1 = \cup_{m,n:|\mathcal{A}_{u,m,n}|=a^*} (\mathcal{S}_m \cup \mathcal{T}_n)
= \{(1,1),(1,2),(1,3),(1,4),(2,1)\}$, which yields $|\mathcal{Q}|=K$.
Therefore, this example falls into the setting of Theorem~\ref{thm:2}.

We consider all sets $\mathcal{A}_{u,m,n}$ such that $|\mathcal{A}_{u,m,n}| = a^*$.
For Example~\ref{ex2}, there are three such sets:
$\mathcal{A}_{1,2,6}$, $\mathcal{A}_{1,2,7}$, and $\mathcal{A}_{1,2,8}$.
Note that $(\mathcal{S}_m \cap \mathcal{K}_{\{u\}}) \cup \mathcal{T}_n
= \mathcal{A}_{u,m,n} \cup (\mathcal{T}_n \setminus \overline{\mathcal{S}}),$
where the corresponding key variables are naturally divided into two parts:
\begin{itemize}
    \item the contribution from $\mathcal{A}_{u,m,n}$, which is bounded using Lemma~\ref{lemma4} and corresponds to $a^*$;
    \item the contribution from $\mathcal{T}_n \setminus \overline{\mathcal{S}}$, which is characterized via a linear program and corresponds to $b^*$.
\end{itemize}
\begin{align}
 H(Z_\Sigma)\ge&  \max\big( H\big( \left(Z_{u,v}\right)_{(u,v) \in \mathcal{S}_2 \cup \mathcal{T}_6} \big), H\big( \left(Z_{u,v}\right)_{(u,v) \in \mathcal{S}_2 \cup \mathcal{T}_7} \big), H\big( \left(Z_{u,v}\right)_{(u,v) \in \mathcal{S}_2 \cup \mathcal{T}_8} \big) \big) \\
 =& \max \big( H(Z_{1,1}, Z_{1,2}, Z_{1,3}), H(Z_{1,1}, Z_{1,2}, Z_{1,4}),   H(Z_{1,1}, Z_{1,2}, Z_{2,1})\big) \\
 =& \max \big( H(Z_{1,3})+H(Z_{1,1},Z_{1,2}|Z_{1,3}),H(Z_{1,4})+  H(Z_{1,1}, Z_{1,2}|Z_{1,4}), H(Z_{2,1})+H(Z_{1,1}, Z_{1,2}|Z_{2,1})\big) \label{eq:tt} \\
\overset{(\ref{lemma4_eqX})}{\geq}&  \max \big( H(Z_{1,3}), H(Z_{1,4}), H(Z_{2,1}) \big) + a^*L. \label{eq:tt1}
\end{align}
In~(\ref{eq:tt}), we split the relevant $Z_{u,v}$ variables in the set
$(\mathcal{S}_m \cap \mathcal{K}_{\{u\}}) \cup \mathcal{T}_n$
into two parts: those in $\mathcal{T}_n \setminus \overline{\mathcal{S}}$
and those in $\mathcal{A}_{u,m,n}$.
In~(\ref{eq:tt1}), we apply Lemma~\ref{lemma4} to bound the contribution of
$\mathcal{A}_{u,m,n}$, conditioned on $\mathcal{T}_n \setminus \overline{\mathcal{S}}$.
Note that we only consider sets $\mathcal{A}_{u,m,n}$ with
$|\mathcal{A}_{u,m,n}| = a^*$, i.e., $\mathcal{A}_{u,m,n} = \overline{\mathcal{S}}$
in this example.

Next, we proceed to bound the term $\max\big(H(Z_{1,3}), H(Z_{1,4}), H(Z_{2,1})\big)$. It turns out that the only constraints required for this bound come from Lemma~\ref{lemma2}. From (\ref{lemma2_eqX}), we have
\begin{align}
    \mathcal{S}_2, \mathcal{T}_6: & \quad H(Z_{1,4},Z_{2,1}\mid Z_{1,3}) \geq L, \notag \\
    \mathcal{S}_2, \mathcal{T}_7: & \quad H(Z_{1,3},Z_{2,1}\mid Z_{1,4}) \geq L, \notag \\
    \mathcal{S}_2, \mathcal{T}_8: & \quad H(Z_{1,3},Z_{1,4}\mid Z_{2,1}) \geq L. \label{eq:tt33}
\end{align}

To bound (\ref{eq:tt1}) using these constraints, we formulate a linear program (LP) with the conditional entropies as variables, consistent with the chain rule. In particular, define
\begin{align}
    H(Z_{1,3}) &= b_{1,3} L, \quad H(Z_{1,4}\mid Z_{1,3}) = b_{1,4} L, \quad H(Z_{2,1}\mid Z_{1,3},Z_{1,4}) = b_{2,1} L.
\end{align}
Then the converse bound can be written as
\begin{align}
    R_{Z_\Sigma} \ge a^* + \min \max \{b_{1,3}, b_{1,4}, b_{2,1}\}, \label{eq:lpex2}
\end{align}
where the minimum is over the following linear constraints:
\begin{align}
    \mathcal{S}_2, \mathcal{T}_6: & \quad b_{1,4} + b_{2,1} \ge 1, \notag \\
    \mathcal{S}_2, \mathcal{T}_7: & \quad b_{1,3} + b_{2,1} \ge 1, \notag \\
    \mathcal{S}_2, \mathcal{T}_8: & \quad b_{1,3} + b_{1,4} \ge 1, \notag \\
    & \quad b_{1,3}, b_{1,4}, b_{2,1} \ge 0. \label{eq:tt4}
\end{align}

Intuitively, the correlation and conflict among $H(Z_{1,3})$, $H(Z_{1,4})$, and $H(Z_{2,1})$ are captured by (\ref{eq:tt33}). The goal is to determine the tightest bound on $R_{Z_\Sigma}$ subject to these constraints. To this end, the converse problem is reduced to a linear program over the non-negative variables $b_{1,3}$, $b_{1,4}$, and $b_{2,1}$, where each set $\mathcal{A}_{u,m,n}$ with maximum cardinality contributes a linear constraint. Some redundant constraints have been removed in (\ref{eq:tt4}) for clarity.
Let $b^* = \min \max \{b_{1,3}, b_{1,4}, b_{2,1}\}$ denote the optimal value of this LP. Then the desired converse bound is $R_{Z_\Sigma} \ge a^* + b^* = \max(a^*, d^*) + b^*.$
For Example~\ref{ex2}, the optimal solution is $b^* = 1/2$, achieved when $b_{1,3} = b_{1,4} = b_{2,1} = 1/2$. This value will be instrumental in the achievable scheme presented in Section~\ref{sce2:achex2}.

\subsection{Achievable Scheme of Example \ref{ex2}}\label{sce2:achex2}

We now show that the converse bound $R^*_{Z_\Sigma} = 5/2$ is achievable. Let $q=5$, i.e., all symbols are in the field $\mathbb{F}_5$.  
Consider $5$ i.i.d. uniform variables $(N_1,N_2,N_3,N_4,N_5) = Z_\Sigma \in \mathbb{F}_5$, so that $L_\Sigma = 5$. Each input contains two symbols, i.e., 
$W_{u,v} = (W^{(1)}_{u,v}, W^{(2)}_{u,v})$ for $(u,v) \in \mathcal{K}$. The individual keys of the users are chosen according to the linear program (\ref{eq:lpex2}) and constraints (\ref{eq:tt4}), i.e., $H(Z_{1,3}) = H(Z_{1,4}) = H(Z_{2,1}) = L/2 = 1, H(Z_{1,1}) = H(Z_{1,2}) = L = 2.$
Specifically, the keys are set as
\begin{align}
 &Z_{1,1} = (-(N_1+N_3+N_4+N_5);-(N_2+N_3+2N_4+3N_5)), \notag\\
 &Z_{1,2} = (N_1;N_2),Z_{1,3} = N_3, Z_{1,4} = N_4, Z_{2,1} = N_5.\label{eq:ex2key}
\end{align}

Each user $(u,v)$ sends a message $X_{u,v}=W_{u,v}+Z_{u,v}$ to relay $u$. In particular, we have
\begin{align}
& X_{1,1} = \left[
\begin{array}{c}
W^{(1)}_{1,1} -(N_1+N_3+N_4+N_5) \\
W^{(2)}_{1,1} -(N_2+N_3+2N_4+3N_5)
\end{array}
\right],
X_{1,2} = \left[
\begin{array}{c}
W^{(1)}_{1,2} + N_1 \\
W^{(2)}_{1,2} + N_2 
\end{array}
\right],~
X_{1,3} = \left[
\begin{array}{c}
W^{(1)}_{1,3} + N_3 \\
W^{(2)}_{1,3} + N_3 
\end{array}
\right],\notag\\
& 
X_{1,4} = \left[
\begin{array}{c}
W^{(1)}_{1,4} + N_4  \\
W^{(2)}_{1,4} + 2N_4 
\end{array}
\right], ~
X_{2,1} = \left[
\begin{array}{c}
W^{(1)}_{2,1} + N_5 \\
W^{(2)}_{2,1} + 3N_5 
\end{array}
\right],~\label{eq:ex2message X}
\end{align}

Upon receiving the messages from associated users, each relay $u$ forwards $Y_u=\sum_{v\in [V_u]}X_{u,v}$
to the server. Specifically,
\begin{align}
& Y_1=X_{1,1}+X_{1,2}+X_{1,3}+X_{1,4} = \left[
\begin{array}{c}
W^{(1)}_{1,1}+W^{(1)}_{1,2} +W^{(1)}_{1,3} +W^{(1)}_{1,4} -N_5 \\
W^{(2)}_{1,1}+ W^{(2)}_{1,2}+ W^{(2)}_{1,3}+ W^{(2)}_{1,4} - 3N_5
\end{array}
\right],
\\
& 
Y_2=X_{2,1} = \left[
\begin{array}{c}
W^{(1)}_{2,1}+N_5\\
W^{(2)}_{2,1}+ 3N_5 
\end{array}
\right].\label{eq:ex2messageY}
\end{align}
Since each message contains one symbol per input, the achieved communication rates are $\rx=\ry=1$.

\textbf{Input sum recovery.} The server simply adds up all received \msgs to obtain
$Y_1+Y_2= \sum_{(u,v)\in\mathcal{K}}W_{u,v}$.
This simple recovery method owes to the zero-sum property of the \indiv keys in (\ref{eq:ex2key}). 

\textbf{Proof of Relay Security.} We next provide an intuitive explanation of the relay security constraint.  
Consider the case where Relay 1 colludes with users in $\mathcal{T}_6$ and the protection of $\mathcal{S}_2 = \{(1,1)\}$.  
Even if Relay 1 gains access to the inputs and keys from users $\{(1,2),(1,3)\}$, it only obtains 3 key symbols. Since the total number of key symbols is 5, the remaining 2 symbols suffice to protect the 2 message symbols of user $(1,1)$. Hence, relay security is guaranteed for $\mathcal{S}_2$ against $\mathcal{T}_6$.
More formally, we have
\begin{align}
&I(X_{1,1}, X_{1,2}, X_{1,3}, X_{1,4}; W_{1,1} \mid W_{1,2}, W_{1,3}, Z_{1,2}, Z_{1,3}) \notag\\
=& H(X_{1,1}, X_{1,2}, X_{1,3}, X_{1,4} \mid W_{1,2}, W_{1,3}, Z_{1,2}, Z_{1,3})  - H(X_{1,1}, X_{1,2}, X_{1,3}, X_{1,4} \mid W_{1,2}, W_{1,3}, Z_{1,2}, Z_{1,3}, W_{1,1}) \notag\\
\overset{(\ref{messageX})}{=}& H(X_{1,1}, X_{1,4} \mid W_{1,2}, W_{1,3}, Z_{1,2}, Z_{1,3}) 
- H(X_{1,1}, X_{1,4} \mid W_{1,2}, W_{1,3}, Z_{1,2}, Z_{1,3}, W_{1,1}) \label{eq:ex2t1} \\
\overset{(\ref{eq:ex2key},\ref{eq:ex2message X})}{=}& H(X_{1,1}, X_{1,4} \mid W_{1,2}, W_{1,3}, Z_{1,2}, Z_{1,3}) 
- H(N_4, N_5, W^{(1)}_{1,4}, W^{(2)}_{1,4}) \label{eq:ex2t2} \\
\leq& H(X_{1,1}, X_{1,4}) - H(N_4, N_5, W^{(1)}_{1,4}, W^{(2)}_{1,4}) 
\overset{(\ref{eq:ex2message X})}{=} 4 - 4 = 0. \label{eq:ex2t4}
\end{align}

Here, equality (\ref{eq:ex2t1}) holds because $X_{1,2}$ and $X_{1,3}$ are determined by $(W_{1,2}, Z_{1,2})$ and $(W_{1,3}, Z_{1,3})$, respectively.  
In (\ref{eq:ex2t2}), we substitute the expressions for $X_{1,1}$, $X_{1,4}$, $Z_{1,2}$, and $Z_{1,3}$ from (\ref{eq:ex2key}) and (\ref{eq:ex2message X}).  
The independence of the inputs and keys justifies the simplification to (\ref{eq:ex2t4}).

\textbf{Proof of Server Security.}  
Consider the worst-case scenario where the server colludes with users $\{(1,2),(1,3)\}$.  
Even in this case, the server obtains no information about $W_{1,1}$ due to the protection provided by the key symbols, ensuring server security.  
More formally, let the input sum be $W_{\Sigma} \triangleq W_{1,1}+W_{1,2}+W_{1,3}+W_{1,4}+W_{2,1}$. Then
\begin{align}
&I(Y_1, Y_2; W_{1,1} \mid W_{\Sigma}, W_{1,2}, W_{1,3}, Z_{1,2}, Z_{1,3}) \notag \\
=& H(Y_1, Y_2 \mid W_{\Sigma}, W_{1,2}, W_{1,3}, Z_{1,2}, Z_{1,3}) 
- H(Y_1, Y_2 \mid W_{\Sigma}, W_{1,2}, W_{1,3}, Z_{1,2}, Z_{1,3}, W_{1,1}) \\
\leq& H(Y_1, Y_2 \mid W_{\Sigma}) 
- H(Y_1, Y_2 \mid W_{\Sigma}, W_{1,2}, W_{1,3}, Z_{1,1}, Z_{1,2}, Z_{1,3}, Z_{1,4}, Z_{2,1}, W_{1,1}) \\
\overset{(\ref{eq:ex2messageY})}{=}& H(Y_1, Y_2, W_{\Sigma}) - H(W_{\Sigma})
- H(W_{1,4}, W_{2,1} \mid W_{\Sigma}, W_{1,2}, W_{1,3}, Z_{1,1}, Z_{1,2}, Z_{1,3}, Z_{1,4}, Z_{2,1}, W_{1,1}) \label{eq:ex2srv1} \\
\overset{(\ref{messageY})}{=}& H(Y_1, Y_2) - H(W_{\Sigma})
- H(W_{1,4}, W_{2,1} \mid W_{1,4}+W_{2,1}) \label{eq:ex2srv2} \\
=& 4-2-2=0.
\end{align}
Here, (\ref{eq:ex2srv1}) follows by substituting the message design from (\ref{eq:ex2messageY}) and using the independence between the inputs and keys.  
Equality (\ref{eq:ex2srv2}) holds because $W_\Sigma$ is determined by $(Y_1, Y_2)$.

% \textbf{Proof of Server security.}For server security, consider the worst-case scenario where the server colludes with users $\{(1,2),(1,3)\}$. In this case, the server infers nothing about $W_{1,1}$ due to protection from the key symbols which ensures \ssec.  More formally, denoting the input sum  by $W_{\Sigma} \eqdef W_{1,1}+W_{1,2}+W_{1,3}+W_{1,4}+W_{2,1}$, we have
% \begin{align}
% & I(Y_1, Y_2; W_{1,1}| W_{\Sigma},W_{1,2},W_{1,3},Z_{1,2},Z_{1,3} )\notag \\
% =&  H(Y_1, Y_2| W_{\Sigma},W_{1,2},W_{1,3},Z_{1,2},Z_{1,3} )
% - H(Y_1, Y_2|W_{\Sigma},W_{1,2},W_{1,3},Z_{1,2},Z_{1,3},W_{1,1} )    \\
% \leq&  H(Y_1, Y_2| W_{\Sigma})
% - H(Y_1, Y_2|W_{\Sigma},W_{1,2},W_{1,3},Z_{1,1},Z_{1,2},Z_{1,3},Z_{1,4},Z_{2,1},W_{1,1} )   \\
% \overset{(\ref{eq:ex2messageY})}{=}&  H(Y_1, Y_2,W_{\Sigma})-H(W_{\Sigma})
% - H(W_{1,4},W_{2,1}|W_{\Sigma},W_{1,2},W_{1,3},Z_{1,1},Z_{1,2},Z_{1,3},Z_{1,4},Z_{2,1},W_{1,1} )  \label{eq:ex2relayt1} \\
% \overset{(\ref{messageY})}{=}&  H(Y_1, Y_2)-H(W_{\Sigma})
% - H(W_{1,4},W_{2,1}|W_{1,4}+W_{2,1} )   \label{eq:ex2relayt2} \\
% =&4-2-2=0
% \end{align}
% In (\ref{eq:ex2relayt1}), we substitute the message design in (\ref{eq:ex2messageY}) and use the independence between the inputs and the keys (see (\ref{ind})). 
% In (\ref{eq:ex2relayt2}), the first term holds because \(W_{\Sigma}\) is determined by \(Y_1\) and \(Y_2\).
 
\subsection{General Converse Proof under Condition 2}

Note that $a^* = |\overline{\mathcal{S}}|$, so for any $\mathcal{A}_{u',m,n}$ with $|\mathcal{A}_{u',m,n}| = a^*$, we must have $\mathcal{A}_{u',m,n} = \overline{\mathcal{S}}$. 
Since $d^* < |\overline{\mathcal{S}}|$ (because $e^* < |\overline{\mathcal{S}}|$ and $d^* \leq e^*$), it follows that $\max(a^*, d^*) = a^* = |\overline{\mathcal{S}}|$. 
Next, consider all $u', m, n$ such that $|\mathcal{A}_{u',m,n}| = \max(a^*, d^*)$ or $|\mathcal{U}^{(m,n)}| + |\mathcal{T}_n \cap \overline{\mathcal{S}}| = \max(a^*, d^*)$. 
Since $d^* < a^*$, there is no $m, n$ satisfying $|\mathcal{U}^{(m,n)}| + |\mathcal{T}_n \cap \overline{\mathcal{S}}| = \max(a^*, d^*)$. 
Therefore, we only need to consider all $\mathcal{A}_{u',m,n}$ sets where $|\mathcal{A}_{u',m,n}| = a^*$:
\begin{align}
H(Z_{\Sigma}) 
&\overset{(\ref{total rand})}{\ge} \max_{u',m,n: |\mathcal{A}_{u',m,n}| = a^*}  
H\Big(\{Z_{u,v}\}_{(u,v)\in (\mathcal{S}_m\cap\mathcal{K}_{\{u'\}})\cup\mathcal{T}_n}\Big) \notag\\
&\ge \max_{u',m,n: |\mathcal{A}_{u',m,n}| = a^*}  
\Big(H(\{Z_{u,v}\}_{(u,v)\in \mathcal{T}_n\setminus\overline{\mathcal{S}}}) 
+ H(\{Z_{u,v}\}_{(u,v)\in \mathcal{A}_{u',m,n}} \mid \{Z_{u,v}\}_{(u,v)\in \mathcal{T}_n\setminus\overline{\mathcal{S}}}) \Big) \notag\\
&\ge \max_{u',m,n: |\mathcal{A}_{u',m,n}| = a^*}  
\Big(H(\{Z_{u,v}\}_{(u,v)\in \mathcal{T}_n\setminus\overline{\mathcal{S}}}) + a^*\Big) \label{eq:convt1}
\end{align}
subject to the following constraints from Lemma~\ref{lemma2}:
\begin{align}
\forall u', m, n \text{ with } |\mathcal{A}_{u',m,n}| = a^*: 
H\Big(\{Z_{u,v}\}_{(u,v) \in \mathcal{K}_{[U]} \setminus ((\mathcal{S}_m\cap\mathcal{K}_{\{u'\}}) \cup \mathcal{T}_n)} \mid \{Z_{u,v}\}_{(u,v)\in \mathcal{T}_n}\Big) \ge L. \label{eq:convt2}
\end{align}

Define variables $b_{u,v}$ for $(u,v)\in \mathcal{K}_{[U]}\setminus \overline{\mathcal{S}}$ via chain-rule expansion in lexicographic order:
\begin{align}
b_{u,v} \triangleq H\Big(Z_{u,v} \mid (Z_{i,j})_{(i,j)\in \mathcal{K}_{[U]}\setminus \overline{\mathcal{S}}, (i,j) < (u,v)}\Big)/L. \label{assume_eq2}
\end{align}

Normalizing (\ref{eq:convt1}) by $L$ and expanding all entropy terms using the $b_{u,v}$ variables, we obtain the linear program:
\begin{align}
R_{Z_\Sigma} &\ge a^* + \min \max_{u',m,n: |\mathcal{A}_{u',m,n}| = a^*} \Bigg( \sum_{(u,v)\in \mathcal{T}_n \setminus \overline{\mathcal{S}}} b_{u,v} \Bigg) \label{eq:convlp}\\
\text{subject to } & \sum_{(u,v) \in \mathcal{K}_{[U]} \setminus ((\mathcal{S}_m\cap \mathcal{K}_{\{u'\}}) \cup \mathcal{T}_n)} b_{u,v} \ge 1, ~~\forall u', m, n \text{ with } |\mathcal{A}_{u',m,n}| = a^*, \label{eq:convlp1}\\
& b_{u,v} \ge 0, ~~\forall (u,v) \in \mathcal{K}_{[U]}\setminus \overline{\mathcal{S}}. \label{eq:convlp2}
\end{align}

Since $\mathcal{A}_{u',m,n} = \overline{\mathcal{S}} \subset (\mathcal{S}_m \cap \mathcal{K}_{\{u'\}} \cup \mathcal{T}_n)$, the chain-rule expansion is valid and the LP correctly bounds the conditional entropies. Denoting the optimal value of this LP by $b^*$, we obtain the converse bound:
$R_{Z_\Sigma} \ge a^* + b^* = \max(a^*, d^*) + b^*,$
establishing a tight lower bound on the achievable key rate.

\subsection{General Achievable Scheme under Condition 2}
\label{sec:ach3}

Consider the case where $a^* \leq K-1$, $|\mathcal{K}_{\mathcal{U}^{(m,n)}} \cup \mathcal{T}_n| \leq K-1$, $e^* < a^* = |\overline{\mathcal{S}}|$, and $|\mathcal{Q}| = K$. In this setting, each user is assigned a number of key variables determined by the optimal solution of the linear program (\ref{eq:convlp}). Let 
$b_{u,v},(u,v) \in \mathcal{K}_{[U]} \setminus \overline{\mathcal{S}}$
denote the optimal values achieving $b^*$ in (\ref{eq:convlp}). We write
\begin{eqnarray}
b_{u,v} = b_{u,v}^* = \frac{p_{u,v}}{\overline{q}}, \quad \forall (u,v) \in \mathcal{K}_{[U]} \setminus \overline{\mathcal{S}}, \label{eq:bk}
\end{eqnarray}
where the coefficients of the linear program are rational, so $p_{u,v}, \overline{q}$ are integers (and non-negative). Let
\begin{eqnarray}
\sum_{(u,v) \in \mathcal{K}_{[U]} \setminus \overline{\mathcal{S}}} b_{u,v}^* = \frac{\sum_{(u,v) \in \mathcal{K}_{[U]} \setminus \overline{\mathcal{S}}} p_{u,v}}{\overline{q}} \equiv \frac{\overline{p}}{\overline{q}}. \label{eq:bk22}
\end{eqnarray}

Choose the field size $q$ such that
$q > (\max(a^*,d^*) + b^*) \, \overline{q} \, \binom{K \, \overline{q}}{(\max(a^*,d^*) + b^*) \, \overline{q}}$
and operate over $\mathbb{F}_q$. Consider $\overline{p}+(\max(a^*,d^*)-1)\overline{q}$ i.i.d. uniform variables
\begin{eqnarray}
Z_\Sigma = {\bf N} = (N_1, \ldots, N_{\overline{p}+(\max(a^*,d^*)-1)\overline{q}}) \in \mathbb{F}_q^{(\overline{p}+(\max(a^*,d^*)-1)\overline{q}) \times 1}.
\end{eqnarray}

The key variables are assigned as
\begin{eqnarray}
{Z}_{u,v} &=& {\bf F}_{u,v} \, {\bf G}_{u,v} \, {\bf N}, \quad \forall (u,v) \in \mathcal{K}_{[U]} \setminus \overline{\mathcal{S}}, \notag\\
{Z}_{u,v} &=& {\bf H}_{u,v} \, {\bf N}, \quad \forall (u,v) \in \overline{\mathcal{S}}, \label{eq:achthm3t1}
\end{eqnarray}
where each element of the matrices are drawn uniformly and independently from $\mathbb{F}_q$, with dimensions
${\bf F}_{u,v} \in \mathbb{F}_q^{\overline{q}\times p_{u,v}}, 
{\bf G}_{u,v} \in \mathbb{F}_q^{p_{u,v}\times (\overline{p}+(\max(a^*,d^*)-1)\overline{q})}, 
{\bf H}_{u,v} \in \mathbb{F}_q^{\overline{q}\times (\overline{p}+(\max(a^*,d^*)-1)\overline{q})}.$

Finally, for $(u',v')\in \overline{\mathcal{S}}$, set
\begin{eqnarray}
{\bf H}_{u',v'} = - \Bigg( \sum_{(u,v)\in\mathcal{K}_{[U]}\setminus \overline{\mathcal{S}}} {\bf F}_{u,v} \, {\bf G}_{u,v} 
+ \sum_{(u,v) \in \overline{\mathcal{S}} \setminus \{(u',v')\}} {\bf H}_{u,v} \Bigg), \label{eq:achthm3t2}
\end{eqnarray}
so that
\begin{eqnarray}
\sum_{(u,v)\in \mathcal{K}_{[U]}} Z_{u,v} = 0. \label{eq:achthm3t3}
\end{eqnarray}

By the Schwartz-Zippel lemma, the matrices can be instantiated such that the row vectors of ${\bf F}_{u,v}^1 \, {\bf G}_{u,v}$ and ${\bf H}_{u,v}$ are linearly independent, as long as the total number of rows does not exceed $(a^* + b^*) \, \overline{q}$, where ${\bf F}_{u,v}^1$ denotes the first $p_{u,v}$ rows of ${\bf F}_{u,v}$:
\begin{align}
\text{Rows of } {\bf F}_{u,v}^1 \, {\bf G}_{u,v} \text{ and } {\bf H}_{u,v} \text{ are linearly independent if their total does not exceed } (a^*+b^*)\overline{q}. \label{eq:sz3}
\end{align}

Finally, let $L = \overline{q}$, i.e., each input 
$W_{u,v} = (W_{u,v}^{(1)}, \ldots, W_{u,v}^{(\overline{q})}) \in \mathbb{F}_q^{\overline{q} \times 1},$
and define the transmitted messages as
\begin{eqnarray}
X_{u,v} &=& W_{u,v} + Z_{u,v}, \quad \forall (u,v) \in \mathcal{K}_{[U]}, \label{eq:achthm3m3X} \\
Y_u &=& \sum_{v \in [V_u]} X_{u,v}, \quad \forall u \in [U]. \label{eq:achthm3m3Y}
\end{eqnarray}
This completes the achievable scheme under Condition 2.

Correctness is similarly guaranteed by taking $\sum_{u\in[U]} Y_u \overset{(\ref{eq:achthm3m3Y})}{=} \sum_{u\in[U]} \sum_{v\in [V_u]} X_{u,v} = \sum_{(u,v)\in\mathcal{K}_{[U]}} X_{u,v} \overset{(\ref{eq:achthm3m3X})(\ref{eq:achthm3t3})}{=} \sum_{(u,v)\in\mathcal{K}_{[U]}} W_{u,v}$. The achieved communication rates are $R_X = L_X/L = \overline{q}/\overline{q} = 1$ and $R_Y = L_Y/L = \overline{q}/\overline{q} = 1$, and the key rate is $R_{Z_\Sigma} = L_{Z_\Sigma}/L = (\overline{p}+(\max(a^*,d^*)-1)\overline{q})/\overline{q} \overset{(\ref{eq:bk22})}{=} \max(a^*,d^*) + \sum_{(u,v)\in\mathcal{K}_{[U]}\setminus \overline{\mathcal{S}}} b_{u,v}^* -1 \overset{(\ref{eq:ach2lp21})}{=} \max(a^*,d^*) + b^*$, where the last equality relies on a crucial property of the linear program (\ref{eq:convlp}), proved next. The security proof is deferred to Section \ref{sec:security}, where it is shown that the randomly generated vectors ${\bf h}_{u,v}$ possess a certain full-rank property necessary for security.

\begin{lemma}\label{lemma:lp}
For the linear program (\ref{eq:convlp}), its optimal value $b^*$ and optimal solution $b^*_{u,v}$ satisfy
\begin{eqnarray}
b^* = \sum_{(u,v)\in \mathcal{K}_{[U]}\setminus\overline{\mathcal{S}}} b^*_{u,v} - 1. \label{eq:ach2lp21}
\end{eqnarray}
\end{lemma}

First, we prove that $b^* \le \sum_{(u,v)\in \mathcal{K}_{[U]}\setminus\overline{\mathcal{S}}} b_{u,v}^* - 1$. 
Since $b^*$ is the optimal value of the linear program, there exist indices $m_1,n_1,u_1$ such that
$b^*=\sum_{(u,v)\in \mathcal{T}_{n_1}\setminus\overline{\mathcal{S}}} b_{u,v}^*$,
as attained in (\ref{eq:convlp})(\ref{eq:convlp1})(\ref{eq:convlp2}). Then
\begin{eqnarray}
b^* &=& \sum_{(u,v)\in \mathcal{T}_{n_1}\setminus\overline{\mathcal{S}}} b^*_{u,v} \label{eq:s3}\\
&=& \sum_{(u,v)\in \mathcal{K}_{[U]}\setminus\overline{\mathcal{S}}} b^*_{u,v}
     - \sum_{(u,v)\in \mathcal{K}_{[U]}\setminus (\mathcal{S}_{m_1}\cap\mathcal{K}_{\{u_1\}} \cup \mathcal{T}_{n_1}) } b^*_{u,v} \label{eq:lm5t1}\\
&\overset{(\ref{eq:convlp1})}{\leq}& \sum_{(u,v)\in \mathcal{K}_{[U]}\setminus\overline{\mathcal{S}}} b^*_{u,v} - 1. \label{eq:lm5t2}
\end{eqnarray}
where (\ref{eq:lm5t1}) follows from partitioning
$\mathcal{K}_{[U]}\setminus\overline{\mathcal{S}}$ into
$\mathcal{T}_{n_1}\setminus\overline{\mathcal{S}}$ and its complement
$\mathcal{K}_{[U]}\setminus (\mathcal{S}_{m_1}\cap\mathcal{K}_{\{u_1\}} \cup \mathcal{T}_{n_1})$,
and (\ref{eq:lm5t2}) holds because for all $m,n,u$, the inequality
$\sum_{(u,v)\in \mathcal{K}_{[U]}\setminus (\mathcal{S}_{m}\cap\mathcal{K}_{\{u\}} \cup \mathcal{T}_{n}) } b^*_{u,v}\ge 1$
is ensured by the feasibility condition in (\ref{eq:convlp1}).

Second, we prove that
$b^* \geq \sum_{(u,v)\in \mathcal{K}_{[U]}\setminus\overline{\mathcal{S}}} b^*_{u,v} - 1$.
For the linear program in (\ref{eq:convlp})(\ref{eq:convlp1})(\ref{eq:convlp2}), at least one of constraints (\ref{eq:convlp1}) must be tight (otherwise the binding constraints on $b_{u,v}^*$ are all from (\ref{eq:convlp2}), i.e., $b^*_{u,v} =0$, which violate (\ref{eq:convlp1})).
Suppose it is for some $m_2,n_2,u_2$ that
$\sum_{(u,v)\in \mathcal{K}_{[U]}\setminus (\mathcal{S}_{m_2} \cap\mathcal{K}_{\{u_2\}}\cup \mathcal{T}_{n_2})} b_{u,v}^*= 1,$
we have
\begin{eqnarray}
1 &=& \sum_{(u,v)\in \mathcal{K}_{[U]}\setminus (\mathcal{S}_{m_2} \cap\mathcal{K}_{\{u_2\}}\cup \mathcal{T}_{n_2})} b_{u,v}^* \\
&=& \sum_{(u,v)\in \mathcal{K}_{[U]}\setminus\overline{\mathcal{S}}} b^*_{u,v}
     -  \sum_{(u,v)\in \mathcal{T}_{n_2} \setminus\overline{\mathcal{S}}} b^*_{u,v} \label{eq:lm5t3}\\
&\overset{(\ref{eq:convlp})}{\geq}& \sum_{(u,v)\in \mathcal{K}_{[U]}\setminus\overline{\mathcal{S}}} b^*_{u,v} - b^*.\label{eq:lm5t4}
\end{eqnarray}
where (\ref{eq:lm5t3}) follows from the identity
$\mathcal{K}_{[U]}\setminus (\mathcal{S}_{m_2} \cap\mathcal{K}_{\{u_2\}}\cup \mathcal{T}_{n_2})
= (\mathcal{K}_{[U]}\setminus\overline{\mathcal{S}})
\setminus\big(\mathcal{T}_{n_2} \setminus\overline{\mathcal{S}}\big)$,
and (\ref{eq:lm5t4}) holds because $b^*$ is the optimal value of (\ref{eq:convlp}),
i.e., the maximum of
$\sum_{(u,v)\in \mathcal{T}_{n} \setminus\overline{\mathcal{S}}} b^*_{u,v}$
over all $m,n,u$.

The proof of Lemma \ref{lemma:lp} is now complete.

\section{Proof of Theorem \ref{thm:3}} \label{sec:thm3}

\subsection{Upper bound Proof of Example \ref{ex3}}
\begin{example}
\label{ex3}
Consider $U=3$ relays and $K=9$ users where $V_1=4, V_2=2, V_3=3$ with security input sets
$(\mathcal{S}_1,$ $\cdots,$ $\mathcal{S}_{16})=(\emptyset,\{(1,1)\},\{(1,2)\},\{(1,3)\},\{(2,1)\},\{(1,1),(1,2)\},\{(1,1),(1,3)\},$ $\{(1,2),(1,3)\},$ $\{(1,1),$ $(2,1)\},$ $\{(1,2),$ $(2,1)\},\{(1,3),(2,1)\},\{(1,1),(1,2),(1,3)\},\{(1,1),(1,2),(2,1)\},\{(1,1),$ 
$(1,3),$ $(2,1)\},$ $\{(1,2),$ $(1,3),(2,1)\},\{(1,1),(1,2),(1,3),(2,1)\})$ and collusion sets
$(\mathcal{T}_1, \cdots, \mathcal{T}_{20}) =(~\emptyset, ~\{(1,4)\},~\{(2,2)\},~$ $\{(3,1)\},$ $\{(3,2)\},$ $ \{(3,3)\}, \{(1,4),(2,2)\},~\{(1,4),(3,1)\},~\{(1,4),(3,2)\},\{(1,4),(3,3)\},\{(2,2),(3,1)\},\{(2,2),$ $(3,2)\},$ $\{(2,2),$ $(3,3)\},$ $\{(3,1),(3,2)\},\{(3,1),(3,3)\},$ $\{(3,2),$ $(3,3)\},$ $ \{(1,4),(2,2),(3,1)\},\{(1,4),(2,2),$ $(3,2)\},\{(1,4),$ $(2,2),$ $(3,3)\},$ $\{(3,1),(3,2),(3,3)\}).$ 
\end{example}

In this example, $a^*=3, e^*=4, d^*=2$ and $|\overline{\mathcal{S}}|=4$,
$|\mathcal{Q}|=|\mathcal{Q}_2|=|\cup_{m,n:|\mathcal{E}_{(m,n)}|=|\overline{\mathcal{S}}|}\mathcal{K}_{\mathcal{U}^{(m,n)}}\cup\mathcal{T}_n|=|(\mathcal{K}_{\mathcal{U}^{(16,16)}}\cup\mathcal{T}_{16})\cup(\mathcal{K}_{\mathcal{U}^{(16,17)}}\cup\mathcal{T}_{17})\cup(\mathcal{K}_{\mathcal{U}^{(16,18)}}\cup\mathcal{T}_{18})|=|\{(1,1),(1,2),(1,3),(1,4),(2,1),(2,2),(3,1),$ $(3,2),(3,3)\}|=K=9$.
Consider Condition 3, we have 
$a^* = 3 \leq K-1 = 8$, 
$|\mathcal{K}_{\mathcal{U}^{(m,n)}} \cup \mathcal{T}_n| \leq K-1 = 8$, 
and $a^* = 3 \leq e^* = |\overline{\mathcal{S}}| = 4$, 
which together satisfy the requirement of Condition 3.
For the lower bound $R_{Z_{\Sigma}} \geq \max\{a^*, d^*\}$, 
which has been proved in theorem \ref{thm:1}. For the rate region 
$R_{Z_{\Sigma}} \geq \max\{a^*, d^*\} + l^*$, we need give an scheme with rate
$\max\{a^*, d^*\} + l^*$ 
that satisfy the correctness constraint (\ref{corr}) and the security constraint (\ref{serversecurity})(\ref{relaysecurity}).

Before specifying the scheme, we first determine the required key sizes for each user by solving the linear program.
For any user $(u,v)\in \overline{\mathcal{S}}$, the individual key must satisfy $H(Z_{u,v}) \ge L$.
For any user $(u,v)\in \mathcal{K}\setminus \overline{\mathcal{S}}$, the key sizes are governed by the constraints in Lemma~\ref{lemma2}.

We examine all sets $\mathcal{A}_{u,m,n}$ and $\mathcal{E}_{m,n}$.
When $|\mathcal{A}_{u,m,n}| \le |\overline{\mathcal{S}}|-1$ and
$|\mathcal{E}_{m,n}| \le |\overline{\mathcal{S}}|-1$,
the constraints in Lemma~\ref{lemma2} are automatically satisfied for the keys in
$\mathcal{K}\setminus \overline{\mathcal{S}}$.
Hence, it suffices to consider the cases where
$|\mathcal{A}_{u,m,n}| = |\overline{\mathcal{S}}|$ or
$|\mathcal{E}_{m,n}| = |\overline{\mathcal{S}}|$.

In Example~3, there are exactly three such sets,
namely $\mathcal{E}_{16,16}$, $\mathcal{E}_{16,17}$, and $\mathcal{E}_{16,18}$.
Applying Lemma~\ref{lemma2} to these cases yields the following constraints:
\begin{eqnarray}
    \mathcal{S}_{16}, \mathcal{T}_{16}: && H(Z_{3,2},Z_{3,3}\mid Z_{1,4},Z_{2,2},Z_{3,1})\geq L, \notag \\
    \mathcal{S}_{16}, \mathcal{T}_{17}: && H(Z_{3,1},Z_{3,3}\mid Z_{1,4},Z_{2,2},Z_{3,2})\geq L, \notag \\
    \mathcal{S}_{16}, \mathcal{T}_{18}: && H(Z_{3,1},Z_{3,2}\mid Z_{1,4},Z_{2,2},Z_{3,3})\geq L. \label{eq:tt3}
\end{eqnarray}

Next, we translate the constraints in (\ref{eq:tt3}) into a linear program with variables $l_{u,v}$.
Specifically, we define
\begin{align}
    &H(Z_{1,4}) = l_{1,4}L,\;
H(Z_{2,2}\mid Z_{1,4}) = l_{2,2}L,\;
H(Z_{3,1}\mid Z_{1,4},Z_{2,2}) = l_{3,1}L,\notag\\
&H(Z_{3,2}\mid Z_{1,4},Z_{2,2},Z_{3,1}) = l_{3,2}L,\;
H(Z_{3,3}\mid Z_{1,4},Z_{2,2},Z_{3,1},Z_{3,2}) = l_{3,3}L.
\end{align}

We consider the worst-case scenario where all users in
$\mathcal{K}\setminus \overline{\mathcal{S}}
=\{(1,4),(2,2),(3,1),(3,2),(3,3)\}$
belong to the colluding set $\mathcal{T}_n$.
Normalizing the entropy
$H(Z_{1,4},Z_{2,2},Z_{3,1},Z_{3,2},Z_{3,3})$
by $L$ and expanding the constraints in (\ref{eq:tt3})
using the chain rule in lexicographic order,
we obtain the following linear program:
 \begin{eqnarray}
&& \min~~  l_{1,4}+l_{2,2}+l_{3,1}+l_{3,2}+l_{3,3}   
\\
 \text{subject to} 
     &&  l_{3,2}+l_{3,3}\geq (H(Z_{3,2}|Z_{1,4},Z_{2,2},Z_{3,1})+H(Z_{3,3}|Z_{1,4},Z_{2,2},Z_{3,1},Z_{3,2}))/L\geq 1, \notag \\
      &&  l_{3,1}+l_{3,3}\geq (H(Z_{3,1}|Z_{1,4},Z_{2,2},Z_{3,2})+H(Z_{3,3}|Z_{1,4},Z_{2,2},Z_{3,1},Z_{3,2}))/L\geq 1, \notag \\
     &&  l_{3,1}+l_{3,2}\geq (H(Z_{3,1}|Z_{1,4},Z_{2,2},Z_{3,3})+H(Z_{3,2}|Z_{1,4},Z_{2,2},Z_{3,1},Z_{3,3}))/L\geq 1, \notag \\
    && l_{1,4}\geq 0, l_{2,2}\geq 0, l_{3,1}\geq 0, l_{3,2}\geq 0, l_{3,3}\geq 0.
    \label{eq:upplp222}
 \end{eqnarray}

The optimal solution of (\ref{eq:upplp222}) is
$l^*_{1,4}=l^*_{2,2}=0$ and
$l^*_{3,1}=l^*_{3,2}=l^*_{3,3}=1/2$,
which yields $l^*=3/2$.
Therefore, the source key rate
$R_{Z_\Sigma}=\max(a^*,b^*)+l^*=3+3/2=9/2$ is achievable.

The optimal LP solution dictates how the key shares should be assigned to the users in $\mathcal{K}_{[U]}\setminus\overline{\mathcal{S}}$, ensuring compliance with constraint (\ref{eq:upplp222}) and fulfilling the correctness (\ref{corr}) and security (\ref{relaysecurity})(\ref{serversecurity}) requirements. Based on this allocation, we proceed to present a concrete achievable scheme.

Suppose $q = 5$, i.e., the symbols are over $\mathbb{F}_q$.
Consider $9$ i.i.d. uniform variables $(N_1,\cdots,N_9) = Z_\Sigma$ in $\mathbb{F}_{q}$, thus the total source key length is $L_\Sigma=9$,
and set $L=2$, i.e., each input contains two symbols $W_{u,v} = (W^{(1)}_{u,v}, W^{(2)}_{u,v}), (u,v) \in \mathcal{K}$. Then we have $H(Z_{1,1})=L=2, H(Z_{1,2})=L=2, H(Z_{1,3})=L=2, H(Z_{1,4})=l^*_{1,4}L=0, H(Z_{2,1})=L=2, H(Z_{2,2})=l^*_{2,2}L=0, H(Z_{3,1})=l^*_{3,1}L=1, H(Z_{3,2})=l^*_{3,2}L=1, H(Z_{3,3})=l^*_{3,3}L=1$. According to the optimal LP (\ref{eq:upplp222}) solution, the \indiv keys of  the users are chosen  as 
\begin{align}
 &Z_{1,1} = (- (N_1 + N_3 + N_5 + N_7 + N_8 + N_9),- (N_2 + N_4 + N_6 + N_7 + 2N_8 + 3N_9)), \notag\\
 & Z_{1,2} = (N_1,N_2), Z_{1,3} = (N_3,N_4), Z_{1,4} =0, Z_{2,1} =  (N_5,N_6),  Z_{2,2} =0,  Z_{3,1} =N_7, Z_{3,2} =N_8, Z_{3,3} =N_9,\label{eq:ex3key}
\end{align}

Each user $(u,v)$ sends a message $X_{u,v}=W_{u,v}+Z_{u,v}$ to relay $u$.
%{\setlength{\mathindent}{15pt}
\begin{align}
& X_{1,1} = \left[
\begin{array}{c}
W^{(1)}_{1,1} - (N_1 + N_3 + N_5 + N_7 + N_8 + N_9)\\
W^{(2)}_{1,1} - (N_2 + N_4 + N_6 + N_7 + 2N_8 + 3N_9)
\end{array}
\right],
X_{1,2} = \left[
\begin{array}{c}
W^{(1)}_{1,2} + N_1 \\
W^{(2)}_{1,2} + N_2 
\end{array}
\right],~
X_{1,3} = \left[
\begin{array}{c}
W^{(1)}_{1,3} + N_3 \\
W^{(2)}_{1,3} + N_4 
\end{array}
\right],\notag\\
& 
X_{1,4} = \left[
\begin{array}{c}
W^{(1)}_{1,4}  \\
W^{(2)}_{1,4} 
\end{array}
\right], ~
X_{2,1} = \left[
\begin{array}{c}
W^{(1)}_{2,1} + N_5 \\
W^{(2)}_{2,1} + N_6 
\end{array}
\right],~
X_{2,2} = \left[
\begin{array}{c}
W^{(1)}_{2,2}  \\
W^{(2)}_{2,2}  
\end{array}
\right],\notag\\
& 
X_{3,1} = \left[
\begin{array}{c}
W^{(1)}_{3,1} +N_7  \\
W^{(2)}_{3,1} +N_7
\end{array}
\right], ~
X_{3,2} = \left[
\begin{array}{c}
W^{(1)}_{3,2} + N_8 \\
W^{(2)}_{3,2} + 2N_8 
\end{array}
\right],~
X_{3,3} = \left[
\begin{array}{c}
W^{(1)}_{3,3} +N_9  \\
W^{(2)}_{3,3} +3N_9
\end{array}
\right].\label{eq:ex3messageX}
\end{align}
%where $Z_{u,v}$ can be easily inferred, e.g., $Z_{1,2} = (N_1;N_2), Z_{1,4} = \emptyset, Z_{3,1}=N_7$. 

Upon receiving the message from the associated users, relay $u$ sends $Y_u=\sum_{v\in [V_u]}X_{u,v}$ to the server. We have
\begin{align}
& Y_1=X_{1,1}+X_{1,2}+X_{1,3}+X_{1,4} = \left[
\begin{array}{c}
W^{(1)}_{1,1}+W^{(1)}_{1,2} +W^{(1)}_{1,3} +W^{(1)}_{1,4} -N_5 - N_7 - N_8 - N_9\\
W^{(2)}_{1,1}+ W^{(2)}_{1,2}+ W^{(2)}_{1,3}+ W^{(2)}_{1,4} - N_6 - N_7 - 2N_8 - 3N_9
\end{array}
\right],\notag
\\
& 
Y_2=X_{2,1}+X_{2,2} = \left[
\begin{array}{c}
W^{(1)}_{2,1}+W^{(1)}_{2,2} +N_5\\
W^{(2)}_{2,1}+ W^{(2)}_{2,2}+ N_6 
\end{array}
\right],\notag\\
& 
Y_3=X_{3,1}+X_{3,2}+X_{3,3} = \left[
\begin{array}{c}
W^{(1)}_{3,1}+W^{(1)}_{3,2} +W^{(1)}_{3,3}  + N_7 + N_8 + N_9\\
W^{(2)}_{3,1}+ W^{(2)}_{3,2}+ W^{(2)}_{3,3} + N_7 + 2N_8 + 3N_9
\end{array}
\right].\label{eq:ex3messageY}
\end{align}

\textbf{Input sum recovery.} The server simply adds up all received \msgs to obtain
$Y_1+Y_2+Y_3= \sum_{(u,v)\in\mathcal{K}}W_{u,v}$.
This simple recovery method owes to the zero-sum property of the \indiv keys in (\ref{eq:ex3key}).

\textbf{Proof of relay security.}
For relay security, we verify one case of $\mathcal{S}_{16} = \{(1,1),(1,2),(1,3),(2,1)\}, \mathcal{T}_{20} = \{(3,1),(3,2),(3,3)\}$, and relay 1; other cases are similar and are deferred to the general proof in Section \ref{sec:security}. 
\begin{eqnarray}
&& I\left( W_{1,1},W_{1,2},W_{1,3},W_{2,1} ; X_{1,1},X_{1,2},X_{1,3},X_{1,4} \big|   W_{3,1}, Z_{3,1}, W_{3,2}, Z_{3,2},W_{3,3}, Z_{3,3} \right) \notag\\
&\overset{(\ref{eq:ex3key})}{=}& I\left( W_{1,1},W_{1,2},W_{1,3},W_{2,1} ; X_{1,1},X_{1,2},X_{1,3},X_{1,4} \big|   W_{3,1}, N_7, W_{3,2}, N_8,W_{3,3}, N_9 \right) \\
&=& H\left(X_{1,1},X_{1,2},X_{1,3},X_{1,4} \big|   W_{3,1}, N_7, W_{3,2}, N_8,W_{3,3}, N_9 \right) \notag\\
&&- H(  X_{1,1},X_{1,2},X_{1,3},X_{1,4} \big|   W_{3,1}, N_7, W_{3,2}, N_8,W_{3,3}, N_9,W_{1,1},W_{1,2},W_{1,3},W_{2,1} ) \\
&\leq& H\left(X_{1,1},X_{1,2},X_{1,3},X_{1,4} \right) \notag\\
&&- H\left(  X_{1,1},X_{1,2},X_{1,3} \big|   W_{3,1}, N_7, W_{3,2}, N_8,W_{3,3}, N_9, W_{1,1},W_{1,2},W_{1,3},W_{2,1} \right) \notag\\
&&- H\left(  X_{1,4} \big|   W_{3,1}, N_7, W_{3,2}, N_8,W_{3,3}, N_9, W_{1,1},W_{1,2},W_{1,3},W_{2,1},X_{1,1},X_{1,2},X_{1,3} \right) \\
&\leq& 4L- H( W^{(1)}_{1,1} - N_1 - N_3 - N_5 - N_7 - N_8 - N_9,
W^{(2)}_{1,2} - N_2 - N_4 - N_6 - N_7 - 2N_8 - 3N_9, \notag\\
&&W^{(1)}_{1,2} + N_1, W^{(2)}_{1,2} + N_2,W^{(1)}_{1,3} + N_3 ,
W^{(2)}_{1,3} + N_4\big|   W^{(1)}_{3,1}, W^{(2)}_{3,1}, N_7, W^{(1)}_{3,2}, W^{(2)}_{3,2}, N_8,W^{(1)}_{3,3},W^{(2)}_{3,3}, N_9,\notag\\
&&W^{(1)}_{1,1}, W^{(2)}_{1,1}, W^{(1)}_{1,2} , W^{(2)}_{1,2}, W^{(1)}_{1,3}, W^{(2)}_{1,3}, W^{(1)}_{2,1}, W^{(2)}_{2,1} ) - H(  W^{(1)}_{1,4},W^{(2)}_{1,4} \big|    W^{(1)}_{3,1}, W^{(2)}_{3,1}, W^{(1)}_{3,2}, W^{(2)}_{3,2}, W^{(1)}_{3,3},W^{(2)}_{3,3},\notag\\
&&W^{(1)}_{1,1}, W^{(2)}_{1,1}, W^{(1)}_{1,2} , W^{(2)}_{1,2}, W^{(1)}_{1,3}, W^{(2)}_{1,3}, W^{(1)}_{2,1}, W^{(2)}_{2,1},N_1,N_2,N_3,N_4,N_5,N_6,N_7,N_8,N_9 )\label{eq:ex3t1} \\
&\overset{(\ref{ind})}{=}&4L-H(N_1,N_2,N_3,N_4,N_5,N_6)- H(W^{(1)}_{1,4},W^{(2)}_{1,4})\label{eq:ex3t2} \\
&=&4\times 2-6-2=0
\end{eqnarray}
The first term in (\ref{eq:ex3t1}) follows from the fact that $X_{1,1}, X_{1,2}, X_{1,3}, X_{1,4}$ jointly contain $4L$ variables. 
Equation (\ref{eq:ex3t2}) holds because the randomness variables are independent of the input variables. 
Hence, the security constraint (\ref{relaysecurity}) is satisfied.

\textbf{Proof of server security.}
For server security, we verify one case of $\mathcal{S}_{16} = \{(1,1),(1,2),(1,3),(2,1)\}, \mathcal{T}_{20} = \{(3,1),(3,2),(3,3)\}$; other cases are similar and are deferred to the general proof in Section \ref{sec:security}. 
\begin{eqnarray}
&&I\Big( W_{1,1},W_{1,2},W_{1,3},W_{2,1} ;Y_1,Y_2,Y_3 \Big| \sum_{(u,v)\in \mathcal{K}}W_{u,v},  W_{3,1}, Z_{3,1}, W_{3,2}, Z_{3,2},W_{3,3}, Z_{3,3} \Big) \notag\\
&\overset{(\ref{eq:ex3key})}{=}& H\Big( Y_1,Y_2,Y_3 \Big| \sum_{(u,v)\in \mathcal{K}}W_{u,v},  W_{3,1}, N_7, W_{3,2}, N_8,W_{3,3}, N_9 \Big) \notag\\
&&-H\Big(Y_1,Y_2,Y_3 \Big| \sum_{(u,v)\in \mathcal{K}}W_{u,v},  W_{3,1}, N_7, W_{3,2}, N_8,W_{3,3}, N_9, W_{1,1},W_{1,2},W_{1,3},W_{2,1}  \Big) \notag\\
&=& H\Big( Y_1,Y_2 \Big| \sum_{(u,v)\in \mathcal{K}}W_{u,v},  W_{3,1}, N_7, W_{3,2}, N_8,W_{3,3}, N_9 \Big) \notag\\
&&+\underbrace{H\Big( Y_3 \Big| \sum_{(u,v)\in \mathcal{K}}W_{u,v},  W_{3,1}, N_7, W_{3,2}, N_8,W_{3,3}, N_9,Y_1,Y_2 \Big)}_{\overset{(\ref{messageY})}{=}0} \notag\\
&&-H\Big(Y_1,Y_2 \Big| \sum_{(u,v)\in \mathcal{K}}W_{u,v},  W_{3,1}, N_7, W_{3,2}, N_8,W_{3,3}, N_9, W_{1,1},W_{1,2},W_{1,3},W_{2,1}  \Big) \notag\\
&&-\underbrace{H\Big(Y_3 \Big| \sum_{(u,v)\in \mathcal{K}}W_{u,v}  W_{3,1}, N_7, W_{3,2}, N_8,W_{3,3}, N_9, W_{1,1},W_{1,2},W_{1,3},W_{2,1},Y_1,Y_2  \Big)}_{\overset{(\ref{messageY})}{=}0} \label{eq:ex3t11}\\
&\overset{(\ref{eq:ex3messageY})}{=}& H(W^{(1)}_{2,1}+W^{(1)}_{2,2} +N_5,W^{(2)}_{2,1}+ W^{(2)}_{2,2}+ N_6,W^{(1)}_{1,1}+W^{(1)}_{1,2} +W^{(1)}_{1,3} +W^{(1)}_{1,4} -N_5 - N_7 - N_8 - N_9,\notag\\
&&W^{(2)}_{1,1}+ W^{(2)}_{1,2}+ W^{(2)}_{1,3}+ W^{(2)}_{1,4} - N_6 - N_7 - 2N_8 - 3N_9\big|W^{(1)}_{1,1}+W^{(1)}_{1,2} +W^{(1)}_{1,3}+W^{(1)}_{2,1}+ W^{(1)}_{2,2},\notag\\
&&W^{(2)}_{1,1}+W^{(2)}_{1,2} +W^{(2)}_{1,3}+W^{(2)}_{2,1}+ W^{(2)}_{2,2},W^{(1)}_{3,1},W^{(2)}_{3,1}, N_7, W^{(1)}_{3,2},W^{(2)}_{3,2}, N_8,W^{(1)}_{3,3},W^{(2)}_{3,3}, N_9 ) \notag\\
&& -H(W^{(1)}_{2,1}+W^{(1)}_{2,2} +N_5,W^{(2)}_{2,1}+ W^{(2)}_{2,2}+ N_6,W^{(1)}_{1,1}+W^{(1)}_{1,2} +W^{(1)}_{1,3} +W^{(1)}_{1,4} -N_5 - N_7 - N_8 - N_9,\notag\\
&&W^{(2)}_{1,1}+ W^{(2)}_{1,2}+ W^{(2)}_{1,3}+ W^{(2)}_{1,4} - N_6 - N_7 - 2N_8 - 3N_9,\big| W^{(1)}_{1,4}+W^{(1)}_{2,2},W^{(2)}_{1,4}+W^{(2)}_{2,2}, W^{(1)}_{3,1},W^{(2)}_{3,1}, N_7,\notag\\
&&  W^{(1)}_{3,2},W^{(2)}_{3,2}, N_8,W^{(1)}_{3,3},W^{(2)}_{3,3}, N_9, W^{(1)}_{1,1},W^{(2)}_{1,1},W^{(1)}_{1,2}, W^{(2)}_{1,2},W^{(1)}_{1,3}, W^{(2)}_{1,3},W^{(1)}_{2,1},W^{(2)}_{2,1} ) \\
&\overset{(\ref{ind})}{=}& H(W^{(1)}_{2,1}+W^{(1)}_{2,2} +N_5,W^{(1)}_{1,1}+W^{(1)}_{1,2} +W^{(1)}_{1,3} +W^{(1)}_{1,4} -N_5,W^{(2)}_{1,1}+ W^{(2)}_{1,2}+ W^{(2)}_{1,3}+ W^{(2)}_{1,4} - N_6,\notag\\
&&W^{(2)}_{2,1}+ W^{(2)}_{2,2}+ N_6 \big|W^{(1)}_{1,1}+W^{(1)}_{1,2} +W^{(1)}_{1,3}+W^{(1)}_{2,1}+ W^{(1)}_{2,2},W^{(2)}_{1,1}+W^{(2)}_{1,2} +W^{(2)}_{1,3}+W^{(2)}_{2,1}+ W^{(2)}_{2,2} ) \notag\\
&& -H(W^{(1)}_{2,2} +N_5,W^{(2)}_{2,2}+ N_6,W^{(1)}_{1,4} -N_5,W^{(2)}_{1,4} - N_6,\big| W^{(1)}_{1,4}+W^{(1)}_{2,2},W^{(2)}_{1,4}+W^{(2)}_{2,2} ) \label{eq:ex3t12}\\
&=& H(W^{(1)}_{2,1}+W^{(1)}_{2,2} +N_5,W^{(1)}_{1,1}+W^{(1)}_{1,2} +W^{(1)}_{1,3} +W^{(1)}_{1,4} -N_5,W^{(2)}_{1,1}+ W^{(2)}_{1,2}+ W^{(2)}_{1,3}+ W^{(2)}_{1,4} - N_6,\notag\\
&&W^{(2)}_{2,1}+ W^{(2)}_{2,2}+ N_6 ,W^{(1)}_{1,1}+W^{(1)}_{1,2} +W^{(1)}_{1,3}+W^{(1)}_{2,1}+ W^{(1)}_{2,2},W^{(2)}_{1,1}+W^{(2)}_{1,2} +W^{(2)}_{1,3}+W^{(2)}_{2,1}+ W^{(2)}_{2,2} ) \notag\\
&&-H(W^{(1)}_{1,1}+W^{(1)}_{1,2} +W^{(1)}_{1,3}+W^{(1)}_{2,1}+ W^{(1)}_{2,2},W^{(2)}_{1,1}+W^{(2)}_{1,2} +W^{(2)}_{1,3}+W^{(2)}_{2,1}+ W^{(2)}_{2,2} )\notag\\
&& -H(W^{(1)}_{2,2} +N_5,W^{(2)}_{2,2}+ N_6,W^{(1)}_{1,4} -N_5,W^{(2)}_{1,4} - N_6, W^{(1)}_{1,4}+W^{(1)}_{2,2},W^{(2)}_{1,4}+W^{(2)}_{2,2} ) \notag\\
&&+H(W^{(1)}_{1,4}+W^{(1)}_{2,2},W^{(2)}_{1,4}+W^{(2)}_{2,2} )\label{eq:ex3t13}\\
&=&4-2-4+2=0
\end{eqnarray}
The second and fourth terms in (\ref{eq:ex3t11}) are zero because $Y_3 = W_{\Sigma} - Y_1 - Y_2$. Equation (\ref{eq:ex3t12}) follows from the independence between the random variables and the input variables. For (\ref{eq:ex3t13}), the first term equals 4 because the sum $W^{(1)}_{1,1}+W^{(1)}_{1,2}+W^{(1)}_{1,3}+W^{(1)}_{2,1}+W^{(1)}_{2,2}$ is determined by $W^{(1)}_{2,1}+W^{(1)}_{2,2}+N_5$ and $W^{(1)}_{1,1}+W^{(1)}_{1,2}+W^{(1)}_{1,3}+W^{(1)}_{1,4}-N_5$, while the sum $W^{(2)}_{1,1}+W^{(2)}_{1,2}+W^{(2)}_{1,3}+W^{(2)}_{2,1}+W^{(2)}_{2,2}$ is determined by $W^{(2)}_{1,1}+W^{(2)}_{1,2}+W^{(2)}_{1,3}+W^{(2)}_{1,4}-N_6$ and $W^{(2)}_{2,1}+W^{(2)}_{2,2}+N_6$, so the total contribution is 4. The third term of (\ref{eq:ex3t13}) also equals 4 because the quantities $W^{(1)}_{1,4}+W^{(1)}_{2,2}$ can be obtained from $W^{(1)}_{2,2}+N_5$ and $W^{(1)}_{1,4}-N_5$, and the quantities $W^{(2)}_{1,4}+W^{(2)}_{2,2}$ can be obtained from $W^{(2)}_{2,2}+N_6$ and $W^{(2)}_{1,4}-N_6$, which again gives a total contribution of 4.

we note that the achieved communication rate is $R_X = L_X / L=2/2 = 1$ and $R_Y = L_Y / L=2/2 = 1$. The achieved key rate is $R_{Z_\Sigma} = L_{Z_\Sigma}/L = 9/2 =3+3/2= \max(a^*,d^*) + l^*$, as desired.

\subsection{General Proof of Theorem \ref{thm:3}}

The lower bound is proved in Section~\ref{sec:convpfthm1}. 
When $a^*\le K-1$ and $|\mathcal{K}_{\mathcal{U}^{(m,n)}}\cup \mathcal{T}_n| \le K-1$, 
the rate region
$R_X\geq 1, R_Y\geq 1, R_{Z_\Sigma}\geq \max(a^*,d^*)+l^*$
is achievable. 
To determine the optimal $l^*$, we assign key of size $L$ to each user in the total security input set $\overline{\mathcal{S}}$. 
Additional keys are assigned to users in $\mathcal{K}_{[U]}\setminus \overline{\mathcal{S}}$, subject to the following constraints from Lemma~\ref{lemma2}:
\begin{align}
&\forall u,m,n \text{ such that } |\mathcal{A}_{u,m,n}| = |\overline{\mathcal{S}}|: 
H\Big(\{Z_{u,v}\}_{(u,v) \in \mathcal{K}_{[U]} \setminus ((\mathcal{S}_m\cap \mathcal{K}_{\{u\}})\cup \mathcal{T}_n)} 
\;\Big|\; \{Z_{u,v}\}_{(u,v)\in\mathcal{T}_n}\Big) \ge L, \label{eq:upp1}\\
&\forall m,n \text{ such that } |\mathcal{E}_{m,n}| = |\overline{\mathcal{S}}|: 
H\Big(\{Z_{u,v}\}_{(u,v) \in \mathcal{K}_{[U]} \setminus (\mathcal{K}_{\mathcal{U}^{(m,n)}} \cup \mathcal{T}_n)} 
\;\Big|\; \{Z_{u,v}\}_{(u,v)\in\mathcal{T}_n}\Big) \ge L. \label{eq:upp2}
\end{align}

%\subsection{Linear Program Formulation for $l^*$}

Next, we translate the constraints (\ref{eq:upp1}) and (\ref{eq:upp2}) into a linear program over variables $l_{u,v}$, defined as
\begin{equation}
l_{u,v} \triangleq H\Big(Z_{u,v} \,\big|\, (Z_{i,j})_{(i,j)\in\mathcal{K}_{[U]}\setminus\overline{\mathcal{S}},\,(i,j)<(u,v)}\Big)/L, \quad \forall (u,v)\in \mathcal{K}_{[U]}\setminus \overline{\mathcal{S}}. \label{assume_eq1}
\end{equation}

For any $u,m,n$ with $|\mathcal{A}_{u,m,n}|<|\overline{\mathcal{S}}|$ and $|\mathcal{E}_{m,n}|<|\overline{\mathcal{S}}|$, all users in $\mathcal{K}_{[U]}\setminus\overline{\mathcal{S}}$ may belong to the colluding set $\mathcal{T}_n$. By normalizing the entropy $H(\{Z_{u,v}\}_{(u,v)\in\mathcal{K}_{[U]}\setminus\overline{\mathcal{S}}})$ by $L$ and expanding the terms in (\ref{eq:upp1}) and (\ref{eq:upp2}) via the chain rule in lexicographic order, we obtain the following linear program:
 \begin{eqnarray}
&& \min  \sum_{(u,v)\in \mathcal{K}_{[U]}\setminus\overline{\mathcal{S}}} l_{u,v} \label{eq:uppmin}
\\
 \text{subject to} 
    && \sum_{(u,v)\in \mathcal{K}_{[U]}\setminus ((\mathcal{S}_m\cap\mathcal{K}_{\{u\}})\cup\mathcal{T}_n)} l_{u,v} 
    \geq 1,~~~\forall u,m,n ~\mbox{such that}~ |\mathcal{A}_{u,m,n}| = |\overline{\mathcal{S}}| \label{eq:upplp20}
    \\
    && \sum_{(u,v)\in \mathcal{K}_{[U]}\setminus (\mathcal{K}_{\mathcal{U}^{(m,n)}}\cup\mathcal{T}_n)} l_{u,v} 
    \geq 1,~~~~~~\forall m,n ~\mbox{such that}~|\mathcal{E}_{m,n}|=|\overline{\mathcal{S}}| \label{eq:upplp21}\\
    && ~~~~~~~~~~~l_{u,v}\geq 0, ~~~~~~~~~~~~~~~~~~~~~\forall (u,v)\in \mathcal{K}_{[U]}\setminus\overline{\mathcal{S}}.
    %\end{array}
    %\right.  
    \label{eq:upplp22}
 \end{eqnarray}

Returning to the key assignment for users in $\mathcal{K}_{[U]}\setminus\overline{\mathcal{S}}$, 
the amount of key for each user is determined by the optimal solution of the linear program (\ref{eq:upplp22}). 
Let the optimal values be
\begin{eqnarray}
    l_{u,v}=l_{u,v}^*=\frac{p_{u,v}}{\overline{q}}, \quad \forall (u,v)\in \mathcal{K}_{[U]}\setminus\overline{\mathcal{S}}, 
\end{eqnarray}
where $p_{u,v},\overline{q}$ are non-negative integers. Then the total key contribution is
\begin{eqnarray}
   l^* = \sum_{(u,v)\in \mathcal{K}_{[U]}\setminus\overline{\mathcal{S}}} l_{u,v}^* = \frac{\overline{p}}{\overline{q}}. \label{eq:pq111}
\end{eqnarray}

Pick a field $\mathbb{F}_q$ with 
$q > (\max(a^*,d^*) + l^*)\overline{q} \binom{K \overline{q}}{(\max(a^*,d^*) + l^*)\overline{q}}$.
Consider $\overline{p}+\max(a^*,d^*)\overline{q}$ i.i.d. uniform variables
$Z_\Sigma={\bf N}=(N_1;\cdots;N_{\overline{p}+\max(a^*,d^*)\overline{q}}) \in \mathbb{F}_q^{(\overline{p}+\max(a^*,d^*)\overline{q})\times 1}$.
Set the key variables as
\begin{align}
Z_{u,v} &= {\bf F}_{u,v} {\bf G}_{u,v} {\bf N}, \quad (u,v) \in \mathcal{K}_{[U]}\setminus \overline{\mathcal{S}},\\
Z_{u,v} &= {\bf H}_{u,v} {\bf N}, \quad (u,v) \in \overline{\mathcal{S}},
\end{align}
where the entries of ${\bf F}_{u,v}\in \mathbb{F}_q^{\overline{q}\times p_{u,v}}$, 
${\bf G}_{u,v}\in \mathbb{F}_q^{p_{u,v}\times (\overline{p}+\max(a^*,d^*)\overline{q})}$, 
${\bf H}_{u,v}\in \mathbb{F}_q^{\overline{q}\times (\overline{p}+\max(a^*,d^*)\overline{q})}$ 
are drawn uniformly and independently. 
For a chosen $(u',v')\in \overline{\mathcal{S}}$, let
\begin{eqnarray}
{\bf H}_{u',v'} = - \Bigg( \sum_{(u,v)\in \mathcal{K}_{[U]}\setminus \overline{\mathcal{S}}} {\bf F}_{u,v} {\bf G}_{u,v} 
+ \sum_{(u,v)\in \overline{\mathcal{S}}\setminus \{(u',v')\}} {\bf H}_{u,v} \Bigg). \label{eq:uppttt22}
\end{eqnarray}

Consequently,
\begin{eqnarray}
\sum_{(u,v)\in \mathcal{K}_{[U]}\setminus \overline{\mathcal{S}}} {\bf F}_{u,v} {\bf G}_{u,v} + \sum_{(u,v)\in \overline{\mathcal{S}}} {\bf H}_{u,v} = {\bf 0} \quad \Rightarrow \quad \sum_{(u,v)\in\mathcal{K}_{[U]}} Z_{u,v} = 0. \label{eq:achthm3t33311}
\end{eqnarray}

By a similar argument using the Schwartz-Zippel lemma, there exists an instantiation of the matrices 
${\bf H}_{u,v}, {\bf F}_{u,v}, {\bf G}_{u,v}$ such that (here, ${\bf F}_{u,v}^1$ denotes the first $p_{u,v}$ rows of ${\bf F}_{u,v}$):
\begin{align}
    &\mbox{If } |\overline{\mathcal{S}}| \ge \max(a^*,d^*)+1, 
\mbox{ the rows of } {\bf F}_{u,v}^1 {\bf G}_{u,v}, {\bf H}_{u,v}, 
\mbox{ and  $\sum_{v\in[V_u]} ({\bf F}_{u,v}^1 {\bf G}_{u,v} + {\bf H}_{u,v}),~ u \in \mathcal{U}^{(m,n)},$} \notag\\
&\mbox{are linearly independent whenever their total number of rows does not exceed } (\max(a^*,d^*)+l^*)\overline{q}, \label{eq:sz32} \\
&\mbox{If } |\overline{\mathcal{S}}| = \max(a^*,d^*), 
\mbox{ the rows of } {\bf F}_{u,v}^1 {\bf G}_{u,v}, {\bf H}_{u,v}, 
\mbox{ and  $\sum_{v\in[V_u]} ({\bf F}_{u,v}^1 {\bf G}_{u,v} + {\bf H}_{u,v}),~ u \in \mathcal{U}^{(m,n)},$} \notag\\
&\mbox{are linearly independent whenever their total number of rows does not exceed } (\max(a^*,d^*)-1+l^*)\overline{q}. \label{eq:sz321}
\end{align}

Finally, set $L=\overline{q}$, i.e., ${W_{u,v}}=(W^{(1)}_{u,v}; \cdots; W^{(\overline{q})}_{u,v}) \in \mathbb{F}_{q}^{\overline{q} \times 1}$ and the sent messages as
 \begin{eqnarray}
     {X}_{u,v} &=& {W}_{u,v} + {Z}_{u,v}, \forall (u,v) \in \mathcal{K}_{[U]}.\label{eq:uppm3X}\\
     {Y}_{u}&=&\sum_{v\in [V_u]}X_{u,v},\forall u \in [U].\label{eq:uppm3Y}
 \end{eqnarray}
 
Correctness is guaranteed by the following derivation: $\sum_{u \in [U]} Y_u \overset{(\ref{eq:uppm3Y})}{=} \sum_{u \in [U]} \sum_{v \in [V_u]} X_{u,v} = \sum_{(u,v) \in \mathcal{K}_{[U]}} X_{u,v} $ $\overset{(\ref{eq:uppm3X})(\ref{eq:achthm3t33311})}{=} \sum_{(u,v) \in \mathcal{K}_{[U]}} W_{u,v}$. This derivation shows that the total sum of $Y_u$ is equal to the total sum of $W_{u,v}$, thereby ensuring correctness. The achieved communication rates are $R_X = L_X / L=\overline{q}/\overline{q} = 1$ and $R_Y = L_Y / L=\overline{q}/\overline{q} = 1$. The achieved key rate is $R_{Z_\Sigma} = L_{Z_\Sigma}/L = (\overline{p} +\max(a^*,d^*)\overline{q})/\overline{q} \overset{(\ref{eq:pq111})}{=} \max(a^*,d^*) + \sum_{(u,v)\in\mathcal{K}_{[U]} \setminus \overline{\mathcal{S}}} l_{u,v}^* \overset{(\ref{eq:pq111})}{=} \max(a^*,d^*) + l^*$. The security proof is presented in a unified manner in Section \ref{sec:security}.

Next, we prove a lemma that will be used later.

\begin{lemma}\label{lemma:lp23}
For the linear program (\ref{eq:upplp22}), its optimal value $l^*$ and optimal solution $l^*_{u,v}$ satisfy
\begin{eqnarray}
l^* \geq \sum_{(u,v)\in \mathcal{T}_{n}\setminus\overline{\mathcal{S}}} l^*_{u,v} + 1. \label{eq:ach2lp233}
\end{eqnarray}
\end{lemma}

\emph{Proof:} First, suppose the optimal value $l^* = \min \sum_{(u,v)\in \mathcal{K}_{[U]}\setminus\overline{\mathcal{S}}} l^*_{u,v}$ in (\ref{eq:uppmin}) is attained for some $u_1, m_1, n_1$ corresponding to constraint (\ref{eq:upplp20}), i.e., $|\mathcal{A}_{u_1,m_1,n_1}| = |\overline{\mathcal{S}}|$, and without loss of generality, assume $|\mathcal{S}_{m_1} \cap\mathcal{K}_{\{u_1\}}\cup \mathcal{T}_{n_1}| \leq K-1$. Then, by partitioning $\mathcal{K}_{[U]}\setminus\overline{\mathcal{S}}$ into $\mathcal{T}_{n_1}\setminus\overline{\mathcal{S}}$ and its complement, we have
\begin{eqnarray}
l^* &=& \sum_{(u,v)\in \mathcal{K}_{[U]}\setminus\overline{\mathcal{S}}} l^*_{u,v} \label{eq:sss3}\\
&=& \sum_{(u,v)\in \mathcal{T}_{n_1}\setminus\overline{\mathcal{S}}} l^*_{u,v} + \sum_{(u,v)\in \mathcal{K}_{[U]}\setminus (\mathcal{S}_{m_1}\cap\mathcal{K}_{\{u_1\}} \cup \mathcal{T}_{n_1}) } l^*_{u,v} \label{eq:lm6t1}\\
&\overset{(\ref{eq:upplp20})}{\geq}&  \sum_{(u,v)\in \mathcal{T}_{n_1}\setminus\overline{\mathcal{S}}} l^*_{u,v} + 1. \label{eq:lm6t2}
\end{eqnarray}
Here, (\ref{eq:lm6t1}) follows from the partitioning, and (\ref{eq:lm6t2}) holds because the feasibility condition (\ref{eq:upplp20}) guarantees $\sum_{(u,v)\in \mathcal{K}_{[U]}\setminus (\mathcal{S}_{m_1}\cap\mathcal{K}_{\{u_1\}} \cup \mathcal{T}_{n_1}) } l^*_{u,v}\ge 1$.

Second, suppose the optimal value $l^*$ is attained for some $m_2, n_2$ corresponding to constraint (\ref{eq:upplp21}), i.e., $|\mathcal{E}_{m_2,n_2}| = |\overline{\mathcal{S}}|$, and without loss of generality, assume $|\mathcal{K}_{\mathcal{U}^{(m_2,n_2)}}\cup\mathcal{T}_{n_2}| \leq K-1$. Then, by partitioning $\mathcal{K}_{[U]}\setminus\overline{\mathcal{S}}$ into $\mathcal{T}_{n_2}\setminus\overline{\mathcal{S}}$ and its complement, we have
\begin{eqnarray}
l^* &=& \sum_{(u,v)\in \mathcal{K}_{[U]}\setminus\overline{\mathcal{S}}} l^*_{u,v} \label{eq:ss3}\\
&=& \sum_{(u,v)\in \mathcal{T}_{n_2}\setminus\overline{\mathcal{S}}} l^*_{u,v} + \sum_{(u,v)\in \mathcal{K}_{[U]}\setminus (\mathcal{K}_{\mathcal{U}^{(m_2,n_2)}}\cup\mathcal{T}_{n_2}) } l^*_{u,v} \label{eq:lm6t3}\\
&\overset{(\ref{eq:upplp21})}{\geq}&  \sum_{(u,v)\in \mathcal{T}_{n_2}\setminus\overline{\mathcal{S}}} l^*_{u,v} + 1. \label{eq:lm6t4}
\end{eqnarray}
Here, (\ref{eq:lm6t3}) follows from the partitioning, and (\ref{eq:lm6t4}) holds because the feasibility condition (\ref{eq:upplp21}) guarantees $\sum_{(u,v)\in \mathcal{K}_{[U]}\setminus (\mathcal{K}_{\mathcal{U}^{(m_2,n_2)}}\cup\mathcal{T}_{n_2}) } l^*_{u,v}\ge 1$.

This completes the proof of Lemma~\ref{lemma:lp23}.

 % \hfill\QED

\section{Proof of Security}
\label{sec:security}

\textbf{Proof of relay security.} 
Consider any sets $\mathcal{K}_{\{u\}}$, $\mathcal{S}_m$, and $\mathcal{T}_n$, where $u\in[U]$, $m\in[M]$, and $n\in[N]$, such that
$\bigl| \mathcal{S}_m \cap \bigl( \mathcal{K}_{\{u\}} \cup \mathcal{T}_n \bigr) \bigr| \le K-1$.
Under this condition, we show that the relay security constraint in \eqref{relaysecurity} is satisfied by all the schemes described above. 
\begin{align}
& I\left(\left\{W_{u,v}\right\}_{(u,v) \in \mathcal{S}_m}; \left\{X_{u,v}\right\}_{(u,v) \in \mathcal{K}_{\{u\}}} \big| \left\{W_{u,v}, Z_{u,v}\right\}_{(u,v) \in \mathcal{T}_n} \right)\notag\\
=& H\left(\left\{W_{u,v}+Z_{u,v}\right\}_{(u,v) \in \mathcal{K}_{\{u\}}} \big| \left\{W_{u,v}, Z_{u,v}\right\}_{(u,v) \in \mathcal{T}_n} \right)\notag\\
&-~H\left(\left\{W_{u,v}+Z_{u,v}\right\}_{(u,v) \in \mathcal{K}_{\{u\}}} \big| \left\{W_{u,v}, Z_{u,v}\right\}_{(u,v) \in \mathcal{T}_n},\left\{W_{u,v}\right\}_{(u,v) \in \mathcal{S}_m} \right)\label{eq_thm1_a4a5}\\
=& H\left(\left\{W_{u,v}+Z_{u,v}\right\}_{(u,v) \in \mathcal{K}_{\{u\}}} \big|  \left\{W_{u,v}, Z_{u,v}\right\}_{(u,v) \in \mathcal{T}_n} \right)\notag\\
&~-H\left(\left\{W_{u,v}+Z_{u,v}\right\}_{(u,v) \in \mathcal{K}_{\{u\}}\setminus\mathcal{T}_n} \big|  \left\{W_{u,v}, Z_{u,v}\right\}_{(u,v) \in \mathcal{T}_n},\left\{W_{u,v}\right\}_{(u,v) \in \mathcal{S}_m} \right)\label{eq_thm1_a5}\\
=& H\left(\left\{W_{u,v}+Z_{u,v}\right\}_{(u,v) \in \mathcal{K}_{\{u\}}} \big|  \left\{W_{u,v}, Z_{u,v}\right\}_{(u,v) \in \mathcal{T}_n} \right)\notag\\
&-~H\left(\left\{W_{u,v}+Z_{u,v}\right\}_{(u,v) \in (\mathcal{S}_m\cap\mathcal{K}_{\{u\}})\setminus\mathcal{T}_n} \big|  \left\{W_{u,v}\right\}_{(u,v) \in (\mathcal{S}_m\cup\mathcal{T}_n)},\left\{Z_{u,v}\right\}_{(u,v) \in \mathcal{T}_n}  \right)\notag\\
&-~H\left(\left\{W_{u,v}+Z_{u,v}\right\}_{(u,v) \in \mathcal{K}_{\{u\}}\setminus((\mathcal{S}_m\cap\mathcal{K}_{\{u\}})\cup\mathcal{T}_n)} \big|\left\{W_{u,v}\right\}_{(u,v) \in (\mathcal{S}_m\cup\mathcal{T}_n)},\left\{Z_{u,v}\right\}_{(u,v) \in ((\mathcal{S}_m\cap\mathcal{K}_{\{u\}})\cup\mathcal{T}_n)}\right)\\
%&&\left.\left\{W_{u,v}\right\}_{(u,v) \in (\mathcal{S}_m\cup\mathcal{T}_n)},\left\{Z_{u,v}\right\}_{(u,v) \in ((\mathcal{S}_m\cap\mathcal{K}_{\{u\}})\cup\mathcal{T}_n)}\right)\\
\overset{(\ref{ind})}{=}& H\left(\left\{W_{u,v}+Z_{u,v}\right\}_{(u,v) \in \mathcal{K}_{\{u\}}} \big| \left\{W_{u,v},Z_{u,v}\right\}_{(u,v) \in \mathcal{T}_n} \right)-H\left(\left\{Z_{u,v}\right\}_{(u,v) \in (\mathcal{S}_m\cap\mathcal{K}_{\{u\}})\setminus\mathcal{T}_n} \big|\left\{Z_{u,v}\right\}_{(u,v) \in \mathcal{T}_n}  \right) \notag\\
&-~H\left(\left\{W_{u,v}+Z_{u,v}\right\}_{(u,v) \in \mathcal{K}_{\{u\}}\setminus((\mathcal{S}_m\cap\mathcal{K}_{\{u\}})\cup\mathcal{T}_n)} \big|\left\{W_{u,v}\right\}_{(u,v) \in (\mathcal{S}_m\cup\mathcal{T}_n)},\left\{Z_{u,v}\right\}_{(u,v) \in ((\mathcal{S}_m\cap\mathcal{K}_{\{u\}})\cup\mathcal{T}_n)}\right)\label{relaysecurity_eq1}\\
\leq&|(\mathcal{S}_m\cap\mathcal{K}_{\{u\}})\setminus\mathcal{T}_n|L-|(\mathcal{S}_m\cap\mathcal{K}_{\{u\}})\setminus\mathcal{T}_n|L\\
=&0
\end{align}
where in (\ref{relaysecurity_eq1}), the difference of the first term and the third term is no greater than $|(\mathcal{S}_m\cap\mathcal{K}_{\{u\}})\setminus\mathcal{T}_n|L$, derived below and the second term is also $|(\mathcal{S}_m\cap\mathcal{K}_{\{u\}})\setminus\mathcal{T}_n|L$, proved in Lemma \ref{lemma:independent} below.
\begin{align}
& H\left(\left\{W_{u,v}+Z_{u,v}\right\}_{(u,v) \in \mathcal{K}_{\{u\}}} \big| \left\{W_{u,v},Z_{u,v}\right\}_{(u,v) \in \mathcal{T}_n} \right)\notag\\
&-~H\left(\left\{W_{u,v}+Z_{u,v}\right\}_{(u,v) \in \mathcal{K}_{\{u\}}\setminus((\mathcal{S}_m\cap\mathcal{K}_{\{u\}})\cup\mathcal{T}_n)} \big|\left\{W_{u,v}\right\}_{(u,v) \in (\mathcal{S}_m\cup\mathcal{T}_n)},\left\{Z_{u,v}\right\}_{(u,v) \in ((\mathcal{S}_m\cap\mathcal{K}_{\{u\}})\cup\mathcal{T}_n)}\right)\label{security_eq12}\\
=& H\left(\left\{W_{u,v}+Z_{u,v}\right\}_{(u,v) \in ((\mathcal{S}_m\cap\mathcal{K}_{\{u\}})\cup\mathcal{T}_n)} \big| \left\{W_{u,v}, Z_{u,v}\right\}_{(u,v) \in \mathcal{T}_n} \right)\notag\\
&+~H\left(\left\{W_{u,v}+Z_{u,v}\right\}_{(u,v) \in \mathcal{K}_{\{u\}}\setminus((\mathcal{S}_m\cap\mathcal{K}_{\{u\}})\cup\mathcal{T}_n)} \big| \left\{W_{u,v}, Z_{u,v}\right\}_{(u,v) \in \mathcal{T}_n},\left\{W_{u,v}+Z_{u,v}\right\}_{(u,v) \in ((\mathcal{S}_m\cap\mathcal{K}_{\{u\}})\cup\mathcal{T}_n)} \right)\notag\\
&-~H\left(\left\{W_{u,v}+Z_{u,v}\right\}_{(u,v) \in \mathcal{K}_{\{u\}}\setminus((\mathcal{S}_m\cap\mathcal{K}_{\{u\}})\cup\mathcal{T}_n)} \big|\left\{W_{u,v}\right\}_{(u,v) \in (\mathcal{S}_m\cup\mathcal{T}_n)},\left\{Z_{u,v}\right\}_{(u,v) \in ((\mathcal{S}_m\cap\mathcal{K}_{\{u\}})\cup\mathcal{T}_n)}\right)\label{security_eq1}\\
\leq& H\left(\left\{W_{u,v}+Z_{u,v}\right\}_{(u,v) \in ((\mathcal{S}_m\cap\mathcal{K}_{\{u\}})\setminus\mathcal{T}_n)} \right)+H\left(\left\{W_{u,v}+Z_{u,v}\right\}_{(u,v) \in \mathcal{K}_{\{u\}}\setminus((\mathcal{S}_m\cap\mathcal{K}_{\{u\}})\cup\mathcal{T}_n)}  \right)\notag\\
&-~H\left(\left\{W_{u,v}+Z_{u,v}\right\}_{(u,v) \in \mathcal{K}_{\{u\}}\setminus((\mathcal{S}_m\cap\mathcal{K}_{\{u\}})\cup\mathcal{T}_n)} \big|\left\{W_{u,v}\right\}_{(u,v) \in (\mathcal{S}_m\cup\mathcal{T}_n)},\left\{Z_{u,v}\right\}_{(u,v) \in \mathcal{K}_{\{u\}}\cup\mathcal{T}_n}\right)\label{security_eq11}\\
=& H\left(\left\{W_{u,v}+Z_{u,v}\right\}_{(u,v) \in ((\mathcal{S}_m\cap\mathcal{K}_{\{u\}})\setminus\mathcal{T}_n)} \right)+H\left(\left\{W_{u,v}+Z_{u,v}\right\}_{(u,v) \in \mathcal{K}_{\{u\}}\setminus(\mathcal{S}_m\cup\mathcal{T}_n)}  \right)\notag\\
&-~H\left(\left\{W_{u,v}\right\}_{(u,v) \in \mathcal{K}_{\{u\}}\setminus(\mathcal{S}_m\cup\mathcal{T}_n)} \big|\left\{W_{u,v}\right\}_{(u,v) \in (\mathcal{S}_m\cup\mathcal{T}_n)},\left\{Z_{u,v}\right\}_{(u,v) \in \mathcal{K}_{\{u\}}}\right)\\
\leq&|(\mathcal{S}_m\cap\mathcal{K}_{\{u\}})\setminus\mathcal{T}_n|L+|\mathcal{K}_{\{u\}}\setminus(\mathcal{S}_m\cup\mathcal{T}_n)|L-|\mathcal{K}_{\{u\}}\setminus(\mathcal{S}_m\cup\mathcal{T}_n)|L\label{relaysecurity_eq2}\\
=&|(\mathcal{S}_m\cap\mathcal{K}_{\{u\}})\setminus\mathcal{T}_n|L.
\end{align}
In (\ref{relaysecurity_eq2}), we bound the first two terms with the number of elements and the third term is due to the independence of $\{W_{u,v}\}_{(u,v)\in\mathcal{K}_{[U]}}$ (which is further uniform) and $\{Z_{u,v}\}_{(u,v)\in\mathcal{K}_{[U]}}$.

\textbf{Proof of server security.} 
When $a^* \leq K-1$, consider any sets $\mathcal{K}_{\mathcal{U}^{(m,n)}}, \mathcal{S}_m,$ and $\mathcal{T}_n$, where $u \in [U]$, $m \in [M]$, and $n \in [N]$, such that
$\left|\mathcal{K}_{\mathcal{U}^{(m,n)}} \cap \mathcal{T}_n\right| = K$,
equivalently, $\left|\mathcal{S}_m \cup \mathcal{T}_n\right| = K$.
We show that, under this condition, the server security constraint~$\eqref{serversecurity}$ is satisfied by the scheme described in Section~$\ref{sec:ach11}$.
\begin{eqnarray}
&& I\Big(\left\{W_{u,v}\right\}_{(u,v) \in \mathcal{S}_m}; \left\{Y_u\right\}_{u \in [U]} \Big| \sum_{(u,v) \in \mathcal{K}_{[U]}} W_{u,v}, \left\{W_{u,v}, Z_{u,v}\right\}_{(u,v) \in \mathcal{T}_n} \Big)\notag\\
&=& H\Big(\Big\{\sum_{v\in[V_u]}\left(W_{u,v}+Z_{u,v}\right)\Big\}_{u \in [U]} \Big| \sum_{(u,v) \in \mathcal{K}_{[U]}} W_{u,v}, \left\{W_{u,v}, Z_{u,v}\right\}_{(u,v) \in \mathcal{T}_n} \Big)\notag\\
&&-~H\Big(\Big\{\sum_{v\in[V_u]}\left(W_{u,v}+Z_{u,v}\right)\Big\}_{u \in [U]} \Big| \sum_{(u,v) \in \mathcal{K}_{[U]}} W_{u,v}, \left\{ Z_{u,v}\right\}_{(u,v) \in \mathcal{T}_n},\left\{W_{u,v}\right\}_{(u,v) \in \mathcal{T}_n\cup\mathcal{S}_m} \Big)\label{eq_thm1_a4a6}~~~~\\
&=& H\Big(\Big\{\sum_{v\in[V_u]}\left(W_{u,v}+Z_{u,v}\right)\Big\}_{u \in \mathcal{U}^{(m,n)}} \Big| \sum_{(u,v) \in \mathcal{K}_{[U]}} W_{u,v}, \left\{W_{u,v}, Z_{u,v}\right\}_{(u,v) \in \mathcal{T}_n} \Big)\notag\\
&&+~H\Big(\Big\{\sum_{v\in[V_u]}\left(W_{u,v}+Z_{u,v}\right)\Big\}_{u \in [U]\setminus\mathcal{U}^{(m,n)}} \Big|\notag\\
&&\sum_{(u,v) \in \mathcal{K}_{[U]}} W_{u,v}, \left\{W_{u,v}, Z_{u,v}\right\}_{(u,v) \in \mathcal{T}_n},\Big\{\sum_{v\in[V_u]}\left(W_{u,v}+Z_{u,v}\right)\Big\}_{u \in \mathcal{U}^{(m,n)}} \Big)\notag\\
&&-~H\Big(\Big\{\sum_{v\in[V_u]}\left(W_{u,v}+Z_{u,v}\right)\Big\}_{u \in \mathcal{U}^{(m,n)}} \Big| \sum_{(u,v) \in \mathcal{K}_{[U]}} W_{u,v}, \left\{W_{u,v}, Z_{u,v}\right\}_{(u,v) \in \mathcal{T}_n},\left\{W_{u,v}\right\}_{(u,v) \in \mathcal{S}_m} \Big)\notag\\
&&-~H\Big(\Big\{\sum_{v\in[V_u]}\left(W_{u,v}+Z_{u,v}\right)\Big\}_{u \in [U]\setminus\mathcal{U}^{(m,n)}} \Big| \notag\\
&&\sum_{(u,v) \in \mathcal{K}_{[U]}} W_{u,v}, \{W_{u,v}, Z_{u,v}\}_{(u,v) \in \mathcal{T}_n},\left\{W_{u,v}\right\}_{(u,v) \in \mathcal{S}_m},\Big\{\sum_{v\in[V_u]}\left(W_{u,v}+Z_{u,v}\Big)\right\}_{u \in \mathcal{U}^{(m,n)}} \Big)\label{eq_thm1_a44}\\
&=& H\Big(\Big\{\sum_{v\in[V_u]}\left(W_{u,v}+Z_{u,v}\right)\Big\}_{u \in \mathcal{U}^{(m,n)}} \Big| \sum_{(u,v) \in \mathcal{K}_{\mathcal{U}^{(m,n)}}} (W_{u,v}+Z_{u,v}), \left\{W_{u,v}, Z_{u,v}\right\}_{(u,v) \in \mathcal{T}_n} \Big)\notag\\
&&-~H\Big(\Big\{\sum_{v\in[V_u]}\left(Z_{u,v}\right)\Big\}_{u \in \mathcal{U}^{(m,n)}} \Big| \left\{Z_{u,v}\right\}_{(u,v) \in \mathcal{T}_n}\Big)\label{eq_secupoof_t2}\\
&=&(|\mathcal{U}^{(m,n)}|-1)L-(|\mathcal{U}^{(m,n)}|-1)L\\
&=&0,
\end{eqnarray}
where in (\ref{eq_thm1_a44}), the second term and the fourth term are both zero because $|\mathcal{K}_{\mathcal{U}^{(m,n)}}\cup\mathcal{T}_n|= K$, and $\{\sum_{v\in[V_u]}(W_{u,v}+$ $Z_{u,v})\}_{u \in [U]\setminus\mathcal{U}^{(m,n)}}$ is determined by $\{W_{u,v}, Z_{u,v}\}_{(u,v) \in \mathcal{T}_n}$. Where in (\ref{eq_secupoof_t2}), the first term follows the fact that: (1), $\sum_{(u,v) \in \mathcal{K}_{[U]}} (W_{u,v}+Z_{u,v})=\sum_{(u,v) \in \mathcal{K}_{[U]}} W_{u,v}$ because $\sum_{(u,v) \in \mathcal{K}_{[U]}} Z_{u,v}=0$; (2), $\sum_{(u,v) \in \mathcal{U}^{(m,n)}} (W_{u,v}+Z_{u,v})$ can be derived from $\sum_{(u,v) \in \mathcal{K}_{[U]}} (W_{u,v}+Z_{u,v})$ and $\{W_{u,v}, Z_{u,v}\}_{(u,v) \in \mathcal{T}_n}$. The second term is also $(|\mathcal{U}^{(m,n)}|-1)L$, proved in Lemma \ref{lemma:independent} below.

Note that when $|\mathcal{K}_{\mathcal{U}^{(m,n)}} \cup \mathcal{T}_n| \le K-1$, 
all sums involved below remain well defined and the corresponding partial sums 
are either independent or fully determined by 
$\sum_{(u,v)\in\mathcal{K}_{[U]}} W_{u,v}$ and 
$\{W_{u,v},Z_{u,v}\}_{(u,v)\in\mathcal{T}_n}$ under the proposed schemes.

Consider any sets $\mathcal{K}_{\mathcal{U}^{(m,n)}}$, $\mathcal{S}_m$, and $\mathcal{T}_n$, with $u\in[U]$, $m\in[M]$, and $n\in[N]$, such that $|\mathcal{K}_{\mathcal{U}^{(m,n)}} \cup \mathcal{T}_n| \le K-1$. We show that the server security constraint in~\eqref{serversecurity} is satisfied by all the schemes proposed in Sections~\ref{sec:ach1}, \ref{sec:ach2}, \ref{sec:ach3}, and~\ref{sec:thm3}.
\begin{eqnarray}
&& I\Big(\left\{W_{u,v}\right\}_{(u,v) \in \mathcal{S}_m}; \left\{Y_u\right\}_{u \in [U]} \Big| \sum_{(u,v) \in \mathcal{K}_{[U]}} W_{u,v}, \left\{W_{u,v}, Z_{u,v}\right\}_{(u,v) \in \mathcal{T}_n} \Big)\notag\\
&=& H\Big(\Big\{\sum_{v\in[V_u]}\left(W_{u,v}+Z_{u,v}\right)\Big\}_{u \in [U]} \Big| \sum_{(u,v) \in \mathcal{K}_{[U]}} W_{u,v}, \left\{W_{u,v}, Z_{u,v}\right\}_{(u,v) \in \mathcal{T}_n} \Big)\notag\\
&&-~H\Big(\Big\{\sum_{v\in[V_u]}\left(W_{u,v}+Z_{u,v}\right)\Big\}_{u \in [U]} \Big| \sum_{(u,v) \in \mathcal{K}_{[U]}} W_{u,v}, \left\{ Z_{u,v}\right\}_{(u,v) \in \mathcal{T}_n},\left\{W_{u,v}\right\}_{(u,v) \in \mathcal{T}_n\cup\mathcal{S}_m} \Big)\label{eq_thm1_a4a7}\\
&=& H\Big(\Big\{\sum_{v\in[V_u]}\left(W_{u,v}+Z_{u,v}\right)\Big\}_{u \in [U]} \Big| \sum_{(u,v) \in \mathcal{K}_{[U]}} W_{u,v}, \left\{W_{u,v}, Z_{u,v}\right\}_{(u,v) \in \mathcal{T}_n} \Big)\notag\\
&&-~H\Big(\Big\{\sum_{v\in[V_u]}\left(W_{u,v}+Z_{u,v}\right)\Big\}_{u \in \mathcal{U}^{(m,n)}} \Big| \sum_{(u,v) \in \mathcal{K}_{[U]}} W_{u,v}, \left\{W_{u,v}, Z_{u,v}\right\}_{(u,v) \in \mathcal{T}_n},\left\{W_{u,v}\right\}_{(u,v) \in \mathcal{S}_m} \Big)\notag\\
&&-~H\Big(\Big\{\sum_{v\in[V_u]}\left(W_{u,v}+Z_{u,v}\right)\Big\}_{u \in [U]\setminus\mathcal{U}^{(m,n)}} \Big| \notag\\
&&\sum_{(u,v) \in \mathcal{K}_{[U]}} W_{u,v}, \left\{W_{u,v}, Z_{u,v}\right\}_{(u,v) \in \mathcal{T}_n},\left\{W_{u,v}\right\}_{(u,v) \in \mathcal{S}_m},\Big\{\sum_{v\in[V_u]}\left(W_{u,v}+Z_{u,v}\right)\Big\}_{u \in \mathcal{U}^{(m,n)}} \Big)\label{eq_thm1_a4}\\
&=& H\Big(\Big\{\sum_{v\in[V_u]}\left(W_{u,v}+Z_{u,v}\right)\Big\}_{u \in [U]} \Big| \sum_{(u,v) \in \mathcal{K}_{[U]}} W_{u,v}, \left\{W_{u,v}, Z_{u,v}\right\}_{(u,v) \in \mathcal{T}_n} \Big)\notag\\
&&-~H\Big(\Big\{\sum_{v\in[V_u]}\left(Z_{u,v}\right)\Big\}_{u \in \mathcal{U}^{(m,n)}} \Big| \left\{Z_{u,v}\right\}_{(u,v) \in \mathcal{T}_n}\Big)\label{eq_thm1_a4a1}\\
&&-~H\Big(\Big\{\sum_{v\in[V_u]}\left(W_{u,v}+Z_{u,v}\right)\Big\}_{u \in [U]\setminus\mathcal{U}^{(m,n)}} \Big| \notag\\
&&\sum_{(u,v) \in \mathcal{K}_{[U]}} W_{u,v}, \left\{W_{u,v}, Z_{u,v}\right\}_{(u,v) \in \mathcal{T}_n},\left\{W_{u,v}\right\}_{(u,v) \in \mathcal{S}_m},\Big\{\sum_{v\in[V_u]}\left(W_{u,v}+Z_{u,v}\right)\Big\}_{u \in \mathcal{U}^{(m,n)}} \Big)\label{serversecurity_t1}\\
&\leq&|\mathcal{U}^{(m,n)}|L-|\mathcal{U}^{(m,n)}|L\\
&=&0
\end{eqnarray}
where in (\ref{serversecurity_t1}), the difference of the first term and the third term is no greater than $|\mathcal{U}^{(m,n)}|L$, derived below and the second term is also $|\mathcal{U}^{(m,n)}|L$, proved in Lemma \ref{lemma:independent} below.
\begin{eqnarray}
&&H\Big(\Big\{\sum_{v\in[V_u]}\left(W_{u,v}+Z_{u,v}\right)\Big\}_{u \in [U]} \Big| \sum_{(u,v) \in \mathcal{K}_{[U]}} W_{u,v}, \left\{W_{u,v}, Z_{u,v}\right\}_{(u,v) \in \mathcal{T}_n} \Big)\notag\\
&&-~H\Big(\Big\{\sum_{v\in[V_u]}\left(W_{u,v}+Z_{u,v}\right)\Big\}_{u \in [U]\setminus\mathcal{U}^{(m,n)}} \Big| \notag\\
&&\sum_{(u,v) \in \mathcal{K}_{[U]}} W_{u,v}, \left\{W_{u,v}, Z_{u,v}\right\}_{(u,v) \in \mathcal{T}_n},\left\{W_{u,v}\right\}_{(u,v) \in \mathcal{S}_m},\Big\{\sum_{v\in[V_u]}\left(W_{u,v}+Z_{u,v}\right)\Big\}_{u \in \mathcal{U}^{(m,n)}} \Big)\label{eq_thm1_a4a2}\\
&=&H\Big(\Big\{\sum_{v\in[V_u]}\left(W_{u,v}+Z_{u,v}\right)\Big\}_{u \in \mathcal{U}^{(m,n)}} \Big| \sum_{(u,v) \in \mathcal{K}_{[U]}} W_{u,v}, \left\{W_{u,v}, Z_{u,v}\right\}_{(u,v) \in \mathcal{T}_n} \Big)\notag\\
&&+H\Big(\Big\{\sum_{v\in[V_u]}\left(W_{u,v}+Z_{u,v}\right)\Big\}_{u \in [U]\setminus\mathcal{U}^{(m,n)}} \Big| \notag\\
&&\sum_{(u,v) \in \mathcal{K}_{[U]}} W_{u,v}, \left\{W_{u,v}, Z_{u,v}\right\}_{(u,v) \in \mathcal{T}_n},\Big\{\sum_{v\in[V_u]}\left(W_{u,v}+Z_{u,v}\right)\Big\}_{u \in \mathcal{U}^{(m,n)}} \Big)\notag\\
&&-~H\Big(\Big\{\sum_{v\in[V_u]}\left(W_{u,v}+Z_{u,v}\right)\Big\}_{u \in [U]\setminus\mathcal{U}^{(m,n)}} \Big| \notag\\
&&\sum_{(u,v) \in \mathcal{K}_{[U]}} W_{u,v}, \left\{W_{u,v}, Z_{u,v}\right\}_{(u,v) \in \mathcal{T}_n},\left\{W_{u,v}\right\}_{(u,v) \in \mathcal{S}_m},\Big\{\sum_{v\in[V_u]}\left(W_{u,v}+Z_{u,v}\right)\Big\}_{u \in \mathcal{U}^{(m,n)}} \Big)\label{eq_thm1_a4a3}\\
&\leq&H\Big(\Big\{\sum_{v\in[V_u]}\left(W_{u,v}+Z_{u,v}\right)\Big\}_{u \in \mathcal{U}^{(m,n)}}\Big)\notag\\
&&+H\Big(\Big\{\sum_{v\in[V_u]}\left(W_{u,v}+Z_{u,v}\right)\Big\}_{u \in [U]\setminus\mathcal{U}^{(m,n)}} \Big|\sum_{(u,v) \in \mathcal{K}_{[U]\setminus \mathcal{U}^{(m,n)}}} (W_{u,v}+Z_{u,v}), \left\{W_{u,v}, Z_{u,v}\right\}_{(u,v) \in \mathcal{T}_n}\Big)\notag\\
&&-~H\Big(\Big\{\sum_{v\in[V_u]}\left(W_{u,v}+Z_{u,v}\right)\Big\}_{u \in [U]\setminus\mathcal{U}^{(m,n)}} \Big| \sum_{(u,v) \in \mathcal{K}_{[U]}} W_{u,v}, \left\{W_{u,v}\right\}_{(u,v) \in \mathcal{T}_n},\left\{Z_{u,v}\right\}_{(u,v) \in \mathcal{K}_{[U]}}, \notag\\
&&\left\{W_{u,v}\right\}_{(u,v) \in \mathcal{S}_m},\Big\{\sum_{v\in[V_u]}\left(W_{u,v}+Z_{u,v}\right)\Big\}_{u \in \mathcal{U}^{(m,n)}} \Big)\label{eq_thm1_a4a4}\\
&\leq&H\Big(\Big\{\sum_{v\in[V_u]}\left(W_{u,v}+Z_{u,v}\right)\Big\}_{u \in \mathcal{U}^{(m,n)}} \Big)\notag\\
&&+H\Big(\Big\{\sum_{v\in[V_u]}\left(W_{u,v}+Z_{u,v}\right)\Big\}_{u \in [U]\setminus(\mathcal{U}^{(m,n)}\cup\mathcal{F}^{(m,n)})} \Big| \sum_{(u,v)\in\mathcal{K}_{[U]\setminus(\mathcal{U}^{(m,n)}\cup\mathcal{F}^{(m,n)})}} \left(W_{u,v}+Z_{u,v}\right) \Big)\notag\\
&&-~H\Big(\Big\{\sum_{v\in[V_u]}\left(W_{u,v}\right)\Big\}_{u \in [U]\setminus(\mathcal{U}^{(m,n)}\cup\mathcal{F}^{(m,n)})} \Big|\sum_{(u,v) \in \mathcal{K}_{[U]\setminus(\mathcal{U}^{(m,n)}\cup\mathcal{F}^{(m,n)})}} W_{u,v}, \notag\\
&& \left\{W_{u,v}\right\}_{(u,v) \in \mathcal{T}_n},\left\{W_{u,v}\right\}_{(u,v) \in \mathcal{S}_m},\Big\{\sum_{v\in[V_u]}\left(W_{u,v}\right)\Big\}_{u \in \mathcal{U}^{(m,n)}} \Big)\label{serversecurity_t2}\\
&\leq&|\mathcal{U}^{(m,n)}|L+(|[U]\setminus(\mathcal{U}^{(m,n)}\cup\mathcal{F}^{(m,n)})|-1)L-(|[U]\setminus(\mathcal{U}^{(m,n)}\cup\mathcal{F}^{(m,n)})|-1)L\\
&=&|\mathcal{U}^{(m,n)}|L.
\end{eqnarray}
where in (\ref{serversecurity_t2}), the second term holds because $\mathcal{K}_{\mathcal{F}^{(m,n)}}\subset \mathcal{T}_n$ (see Definition \ref{def:secrelay11}) and $\sum_{(u,v)\in\mathcal{F}^{(m,n)}} \left(W_{u,v}+Z_{u,v}\right)$ is determined by $\{W_{u,v}+Z_{u,v}\}_{(u,v)\in \mathcal{T}_n}$; The third term holds because $\mathcal{K}_{\mathcal{F}^{(m,n)}}\subset \mathcal{T}_n$ and $W_{u,v}$ is independent.

% where in (\ref{serversecurity_t2}), the first term is $U-1$ if $|\mathcal{U}^{(m,n)}\setminus\mathcal{F}^{(m,n)}|=U$; else if $|\mathcal{U}^{(m,n)}\setminus\mathcal{F}^{(m,n)}|\leq U-1$, the first term is $|\mathcal{U}^{(m,n)}\setminus\mathcal{F}^{(m,n)}|L$.

To complete the proof, we are left to prove the following lemma.
\begin{lemma}\label{lemma:independent}
For all cases of the achievable scheme presented in section \ref{sec:ach11}, \ref{sec:ach1}, \ref{sec:ach2},  \ref{sec:ach3}, and section \ref{sec:thm3}, we have
\begin{eqnarray}
H\left(\{Z_{u,v}\}_{(u,v) \in (\mathcal{S}_m\cap\mathcal{K}_{\{u\}})\setminus\mathcal{T}_n} \big|\{Z_{u,v}\}_{(u,v) \in \mathcal{T}_n}  \right) &=& |(\mathcal{S}_m\cap\mathcal{K}_{\{u\}})\setminus\mathcal{T}_n|L, \label{eq:independent}\\
%\forall u\in[U],m \in [M], n \in [N]. 
H\Big(\Big\{\sum_{v\in[V_u]}\left(Z_{u,v}\right)\Big\}_{u \in \mathcal{U}^{(m,n)}} \Big| \left\{Z_{u,v}\right\}_{(u,v) \in \mathcal{T}_n}\Big)&=&(|\mathcal{U}^{(m,n)}|-1)L,~\mbox{if}~|\mathcal{K}_{\mathcal{U}^{(m,n)}}\cup\mathcal{T}_n|= K, ~~\label{eq:independent1}\\
H\Big(\Big\{\sum_{v\in[V_u]}\left(Z_{u,v}\right)\Big\}_{u \in \mathcal{U}^{(m,n)}} \Big| \left\{Z_{u,v}\right\}_{(u,v) \in \mathcal{T}_n}\Big)&=&|\mathcal{U}^{(m,n)}|L,~\mbox{else}~|\mathcal{K}_{\mathcal{U}^{(m,n)}}\cup\mathcal{T}_n|< K. \label{eq:independent2}
\end{eqnarray}
\end{lemma}

\interfootnotelinepenalty=10000

 We prove each case one by one. First, consider the scheme in Section \ref{sec:ach11}, we are now ready to prove (\ref{eq:independent}).
\begin{eqnarray}
&&H\left(\{Z_{u,v}\}_{(u,v) \in (\mathcal{S}_m\cap\mathcal{K}_{\{u\}})\setminus\mathcal{T}_n} |\{Z_{u,v}\}_{(u,v) \in \mathcal{T}_n}  \right) \notag\\
&=& H\big(\{Z_{u,v}\}_{(u,v) \in \mathcal{S}_m\cap\mathcal{K}_{\{u\}} \cup \mathcal{T}_n}\big) - H\left(\{Z_{u,v}\}_{(u,v) \in \mathcal{T}_n}  \right) \\
&\overset{(\ref{eq:c111})}{=}& H\left(\{Z_{u,v}\}_{(u,v) \in (\mathcal{S}_m\cap\mathcal{K}_{\{u\}} \cup \mathcal{T}_n) \cap \overline{\mathcal{S}} } \right) - H\left(\{Z_{u,v}\}_{(u,v) \in \mathcal{T}_n \cap \overline{\mathcal{S}} }  \right) \label{eq:st11}\\
&\overset{(\ref{eq:sz111})}{=}&  \big|(\mathcal{S}_m\cap\mathcal{K}_{\{u\}} \cup \mathcal{T}_n) \cap \overline{\mathcal{S}}\big| L - \big| \mathcal{T}_n \cap \overline{\mathcal{S}} \big| L \label{eq:st31} \\
&=& \big|((\mathcal{S}_m\cap\mathcal{K}_{\{u\}})\setminus\mathcal{T}_n) \cap \overline{\mathcal{S}} \big| L= |(\mathcal{S}_m\cap\mathcal{K}_{\{u\}})\setminus\mathcal{T}_n|L,
\end{eqnarray}
where (\ref{eq:st11}) uses the fact that $Z_{u,v} = 0, (u,v) \in \mathcal{K}_{[U]}\setminus\overline{\mathcal{S}}$ (see (\ref{eq:c111})). 
To obtain (\ref{eq:st31}), we use the property that $|\mathcal{A}_{u,m,n}| = \big|(\mathcal{S}_m\cap\mathcal{K}_{\{u\}} \cup \mathcal{T}_n) \cap \overline{\mathcal{S}}\big| \leq a^*\leq \max(a^*,d^*-1)$ for Section \ref{sec:ach11} so that the ${\bf h}_{u,v}$ vectors are linearly independent (see (\ref{eq:sz111})). The last step follows from the fact that $\mathcal{S}_m \cap \mathcal{K}_{\{u\}} \subset \overline{\mathcal{S}}$.

Consider the scheme in section \ref{sec:ach11} where $|\mathcal{K}_{\mathcal{U}^{(m,n)}}\cup\mathcal{T}_n|= K$. We are ready to prove (\ref{eq:independent1}).
\begin{eqnarray}
&&H\Big(\Big\{\sum_{v\in[V_u]}\left(Z_{u,v}\right)\Big\}_{u \in \mathcal{U}^{(m,n)}} \Big| \left\{Z_{u,v}\right\}_{(u,v) \in \mathcal{T}_n}\Big)\\
&=&H\Big(\Big\{\sum_{v\in[V_u]}\left(Z_{u,v}\right)\Big\}_{u \in \mathcal{U}^{(m,n)}} , \left\{Z_{u,v}\right\}_{(u,v) \in \mathcal{T}_n}\Big)-H\left(\left\{Z_{u,v}\right\}_{(u,v) \in \mathcal{T}_n}\right)\\
&\overset{(\ref{eq:c111})}{=}&H\Big(\Big\{\sum_{v\in[V_u]}\left(Z_{u,v}\right)\Big\}_{u \in \mathcal{U}^{(m,n)}} , \left\{Z_{u,v}\right\}_{(u,v) \in \mathcal{T}_n\cap\overline{\mathcal{S}}}\Big)-H\left(\left\{Z_{u,v}\right\}_{(u,v) \in \mathcal{T}_n\cap\overline{\mathcal{S}}}\right)\label{eq:st12t}\\
&\overset{(\ref{eq:sz111})}{=}&(|\mathcal{U}^{(m,n)}|+|\mathcal{T}_n \cap \overline{\mathcal{S}}|-1)L-|\mathcal{T}_n \cap \overline{\mathcal{S}}|L\label{eq:st32t} \\
&=&(|\mathcal{U}^{(m,n)}|-1)L,
\end{eqnarray}
where (\ref{eq:st12t}) uses the fact that $Z_{u,v} = 0, (u,v) \in \mathcal{K}_{[U]}\setminus\overline{\mathcal{S}}$. For any set $\{(u,v)\}_{v\in [V_u], (u,v)\in \mathcal{K}_{\mathcal{U}^{(m,n)}}}$, there exist at least an $(u,v)$ such that $Z_{u,v} \neq 0$ (see Defination \ref{def:secrelay11}). 
To obtain (\ref{eq:st32t}), we follow the fact that $|\mathcal{K}_{\mathcal{U}^{(m,n)}}\cup \mathcal{T}_n|=K$ and according to the key assignment in (\ref{eq:c121}), one of the key is the linear combination of other $|\overline{\mathcal{S}}|-1$ keys.

Second, consider the scheme in Section \ref{sec:ach1}, 
we are now ready to prove (\ref{eq:independent}). 
\begin{eqnarray}
&&H\left(\{Z_{u,v}\}_{(u,v) \in (\mathcal{S}_m\cap\mathcal{K}_{\{u\}})\setminus\mathcal{T}_n} |\{Z_{u,v}\}_{(u,v) \in \mathcal{T}_n}  \right) \notag\\
&=& H\big(\{Z_{u,v}\}_{(u,v) \in \mathcal{S}_m\cap\mathcal{K}_{\{u\}} \cup \mathcal{T}_n}\big) - H\left(\{Z_{u,v}\}_{(u,v) \in \mathcal{T}_n}  \right) \\
&\overset{(\ref{eq:c11})}{=}& H\left(\{Z_{u,v}\}_{(u,v) \in (\mathcal{S}_m\cap\mathcal{K}_{\{u\}} \cup \mathcal{T}_n) \cap \overline{\mathcal{S}} } \right) - H\left(\{Z_{u,v}\}_{(u,v) \in \mathcal{T}_n \cap \overline{\mathcal{S}} }  \right) \label{eq:st1}\\
&\overset{(\ref{eq:sz1})}{=}&  \big|(\mathcal{S}_m\cap\mathcal{K}_{\{u\}} \cup \mathcal{T}_n) \cap \overline{\mathcal{S}}\big| L - \big| \mathcal{T}_n \cap \overline{\mathcal{S}} \big| L \label{eq:st3} \\
&=& \big|((\mathcal{S}_m\cap\mathcal{K}_{\{u\}})\setminus\mathcal{T}_n) \cap \overline{\mathcal{S}} \big| L= |(\mathcal{S}_m\cap\mathcal{K}_{\{u\}})\setminus\mathcal{T}_n|L,
\end{eqnarray}
where (\ref{eq:st1}) uses the fact that $Z_{u,v} = 0, k \in \mathcal{K}_{[U]}\setminus\overline{\mathcal{S}}$ (see (\ref{eq:c11})).
To obtain (\ref{eq:st3}), we use the property that $|\mathcal{A}_{u,m,n}| = \big|(\mathcal{S}_m\cap\mathcal{K}_{\{u\}} \cup \mathcal{T}_n) \cap \overline{\mathcal{S}}\big| \leq a^*\leq \max(a^*,d^*)$ for Section \ref{sec:ach1} so that the ${\bf h}_{u,v}$ vectors are linearly independent (see (\ref{eq:sz1})). The last step follows from the fact that $\mathcal{S}_m\cap\mathcal{K}_{\{u\}} \subset \overline{\mathcal{S}}$.

Consider the scheme in section \ref{sec:ach1} where $|\mathcal{K}_{\mathcal{U}^{(m,n)}}\cup\mathcal{T}_n|\leq K-1$. We are ready to prove (\ref{eq:independent2}).
\begin{eqnarray}
&&H\Big(\Big\{\sum_{v\in[V_u]}\left(Z_{u,v}\right)\Big\}_{u \in \mathcal{U}^{(m,n)}} \Big| \left\{Z_{u,v}\right\}_{(u,v) \in \mathcal{T}_n}\Big)\\
&=&H\Big(\Big\{\sum_{v\in[V_u]}\left(Z_{u,v}\right)\Big\}_{u \in \mathcal{U}^{(m,n)}} , \left\{Z_{u,v}\right\}_{(u,v) \in \mathcal{T}_n}\Big)-H\left(\left\{Z_{u,v}\right\}_{(u,v) \in \mathcal{T}_n}\right)\\
&\overset{(\ref{eq:c11})}{=}&H\Big(\Big\{\sum_{v\in[V_u]}\left(Z_{u,v}\right)\Big\}_{u \in \mathcal{U}^{(m,n)}} , \left\{Z_{u,v}\right\}_{(u,v) \in \mathcal{T}_n\cap\overline{\mathcal{S}}}\Big)-H\left(\left\{Z_{u,v}\right\}_{(u,v) \in \mathcal{T}_n\cap\overline{\mathcal{S}}}\right)\label{eq:st12}\\
&\overset{(\ref{eq:sz1})}{=}&(|\mathcal{U}^{(m,n)}|+|\mathcal{T}_n \cap \overline{\mathcal{S}}|)L-|\mathcal{T}_n \cap \overline{\mathcal{S}}|L\label{eq:st32} \\
&=&|\mathcal{U}^{(m,n)}|L,
\end{eqnarray}
where (\ref{eq:st12}) uses the fact that $Z_{u,v} = 0, (u,v) \in \mathcal{K}_{[U]}\setminus\overline{\mathcal{S}}$ and for any set $\{(u,v)\}_{v\in [V_u], (u,v)\in \mathcal{K}_{\mathcal{U}^{(m,n)}}}$, there exist at least a $(u,v)$ such that $Z_{u,v} \neq 0$ (see definition \ref{def:secrelay11}). 
To obtain (\ref{eq:st32}), 
we use property that $(|\mathcal{U}^{(m,n)}|+|\mathcal{T}_n \cap \overline{\mathcal{S}}|) \leq d^*\leq \max(a^*,d^*)$ for Section \ref{sec:ach1} so that the ${\bf h}_{u,v}$ vectors are linearly independent (see (\ref{eq:sz1})). The last step follows from the fact that $\mathcal{S}_m \cap \mathcal{K}_{\{u\}} \subset \overline{\mathcal{S}}$.

Third, consider the scheme in Section \ref{sec:ach2}, 
we now proceed to prove (\ref{eq:independent}). We have two sub-cases. 
For the first sub-case, $(u',v') \in \mathcal{T}_n$. Then $\big|(\mathcal{S}_m\cap \mathcal{K}_{\{u\}} \cup \mathcal{T}_n) \cap \overline{\mathcal{S}}\big| < \big|\overline{\mathcal{S}}\big|= a^*\leq \max(a^*,d^*)$ because otherwise $\big|(\mathcal{S}_m \cap \mathcal{K}_{\{u\}}\cup \mathcal{T}_n) \cap \overline{\mathcal{S}}\big| = \big|\overline{\mathcal{S}}\big| = a^*$ and according to Definition \ref{def:totset1} and \ref{def:totset2}, we have $(u',v') \in (\mathcal{S}_m \cap \mathcal{K}_{\{u\}}\cup \mathcal{T}_n) \subset \mathcal{Q}$, which violates our choice of $(u',v')$ to be outside $\mathcal{Q}$ in Section \ref{sec:ach2}.
\begin{eqnarray}
&& H\left(\{Z_{u,v}\}_{(u,v) \in \mathcal{S}_m\cap\mathcal{K}_{\{u\}} \setminus\mathcal{T}_n} |\{Z_{u,v}\}_{(u,v) \in \mathcal{T}_n}  \right) 
\notag\\
&\overset{(\ref{eq:c21})}{=}& H\left(\{Z_{u,v}\}_{(u,v) \in (\mathcal{S}_m\cap\mathcal{K}_{\{u\}} \cup \mathcal{T}_n) \cap (\overline{\mathcal{S}} \cup \{(u',v')\})} \right) - H\left(\{Z_{u,v}\}_{(u,v) \in \mathcal{T}_n \cap (\overline{\mathcal{S}} \cup \{(u',v')\}) }  \right) \\
&\overset{(\ref{eq:sz2})}{=}&  \big|(\mathcal{S}_m\cap\mathcal{K}_{\{u\}} \cup \mathcal{T}_n) \cap (\overline{\mathcal{S}} \cup \{(u',v')\}) \big| L - \big| \mathcal{T}_n \cap ( \overline{\mathcal{S}} \cup \{(u',v')\}) \big| L \label{eq:st5} \\
&=& \big|(\mathcal{S}_m\cap\mathcal{K}_{\{u\}} \setminus\mathcal{T}_n) \cap (\overline{\mathcal{S}} \cup \{(u',v')\})\big| L= |\mathcal{S}_m\cap\mathcal{K}_{\{u\}} \setminus\mathcal{T}_n|L
\end{eqnarray}
whereas $\big|(\mathcal{S}_m \cap \mathcal{K}_{\{u\}}\cup \mathcal{T}_n) \cap \overline{\mathcal{S}}\big| < \max(a^*,d^*)$ , the first term of (\ref{eq:st5}) is no greater than $\max(a^*,d^*)$ enabling us to use (\ref{eq:sz2}) to reduce the rank to the set cardinality. The last step uses the fact that $(u',v') \in \mathcal{T}_n$.
For the second sub-case, $(u',v') \notin \mathcal{T}_n$ (recall that $(u',v') \notin \overline{\mathcal{S}}$ and $\mathcal{S}_m \subset \overline{\mathcal{S}}$).
\begin{eqnarray}
&&H\left(\{Z_{u,v}\}_{(u,v) \in (\mathcal{S}_m\cap\mathcal{K}_{\{u\}})\setminus\mathcal{T}_n} |\{Z_{u,v}\}_{(u,v) \in \mathcal{T}_n}  \right)\notag\\
&\overset{(\ref{eq:c21})}{=}& H\left(\{Z_{u,v}\}_{(u,v) \in (\mathcal{S}_m\cap\mathcal{K}_{\{u\}} \cup \mathcal{T}_n) \cap (\overline{\mathcal{S}} \cup \{(u',v')\})} \right) - H\left(\{Z_{u,v}\}_{(u,v) \in \mathcal{T}_n \cap (\overline{\mathcal{S}} \cup \{(u',v')\}) }  \right) \\
&\overset{}{=}& H\left(\{Z_{u,v}\}_{(u,v) \in (\mathcal{S}_m\cap\mathcal{K}_{\{u\}} \cup \mathcal{T}_n) \cap \overline{\mathcal{S}} } \right) - H\left(\{Z_{u,v}\}_{(u,v) \in \mathcal{T}_n \cap \overline{\mathcal{S}} }  \right) \label{eq:st4}\\
&\overset{(\ref{eq:sz2})}{=}&  \big|(\mathcal{S}_m\cap\mathcal{K}_{\{u\}} \cup \mathcal{T}_n) \cap \overline{\mathcal{S}}\big| L - \big| \mathcal{T}_n \cap \overline{\mathcal{S}} \big| L 
= |(\mathcal{S}_m\cap\mathcal{K}_{\{u\}})\setminus\mathcal{T}_n|L.\label{eq:stst1}
\end{eqnarray}
where (\ref{eq:st4}) uses the assumption that $(u',v') \notin \mathcal{T}_n$. The first term of (\ref{eq:stst1}) holds because $\big|(\mathcal{S}_m\cap\mathcal{K}_{\{u\}} \cup \mathcal{T}_n) \cap \overline{\mathcal{S}}\big|\leq a^*\leq \max(a^*,d^*)$ so that the ${\bf h}_{u,v}$ vectors are linearly independent (see (\ref{eq:sz2})).

Consider the scheme in section \ref{sec:ach2} where $|\mathcal{K}_{\mathcal{U}^{(m,n)}}\cup\mathcal{T}_n|\leq K-1$. We are ready to prove (\ref{eq:independent2}). Similarly,  We have two sub-cases. 
For the first sub-case, $(u',v') \in \mathcal{T}_n$. Then 
$|\mathcal{U}^{(m,n)}|+| \mathcal{T}_n \cap \overline{\mathcal{S}}|< d^*\leq \max(a^*,d^*)$ because otherwise $|\mathcal{U}^{(m,n)}|+| \mathcal{T}_n \cap \overline{\mathcal{S}}|  = d^*$ and according to Definition \ref{def:totset1} and \ref{def:totset2}, we have $(u',v') \in (\mathcal{S}_m \cap \mathcal{K}_{\{u\}}\cup \mathcal{T}_n) \subset \mathcal{Q}$, which violates our choice of $(u',v')$ to be outside $\mathcal{Q}$ in Section \ref{sec:ach2}. 
\begin{eqnarray}
&&H\Big(\Big\{\sum_{v\in[V_u]}\left(Z_{u,v}\right)\Big\}_{u \in \mathcal{U}^{(m,n)}} \Big| \left\{Z_{u,v}\right\}_{(u,v) \in \mathcal{T}_n}\Big)\\
&=&H\Big(\Big\{\sum_{v\in[V_u]}\left(Z_{u,v}\right)\Big\}_{u \in \mathcal{U}^{(m,n)}} , \left\{Z_{u,v}\right\}_{(u,v) \in \mathcal{T}_n}\Big)-H\left(\left\{Z_{u,v}\right\}_{(u,v) \in \mathcal{T}_n}\right)\\
&\overset{(\ref{eq:c21})}{=}&H\Big(\Big\{\sum_{v\in[V_u]}\left(Z_{u,v}\right)\Big\}_{u \in \mathcal{U}^{(m,n)}} , \left\{Z_{u,v}\right\}_{(u,v) \in \mathcal{T}_n\cap(\overline{\mathcal{S}} \cup \{(u',v')\})}\Big)-H\left(\left\{Z_{u,v}\right\}_{(u,v) \in \mathcal{T}_n\cap(\overline{\mathcal{S}} \cup \{(u',v')\})}\right)\\
&\overset{(\ref{eq:sz2})}{=}&(|\mathcal{U}^{(m,n)}|+|\mathcal{T}_n \cap (\overline{\mathcal{S}} \cup \{(u',v')\})|)L-|\mathcal{T}_n \cap (\overline{\mathcal{S}} \cup \{(u',v')\})|L\label{eq:st2221}\\
&=&|\mathcal{U}^{(m,n)}|L,
\end{eqnarray}
whereas $|\mathcal{U}^{(m,n)}|+| \mathcal{T}_n \cap \overline{\mathcal{S}}|< d^*\leq \max(a^*,d^*)$, then $|\mathcal{U}^{(m,n)}|+|\mathcal{T}_n \cap (\overline{\mathcal{S}} \cup \{(u',v')\})|\leq  d^*\leq \max(a^*,d^*)$. The first term of (\ref{eq:st2221}) is no greater than $\max(a^*,d^*)$ enabling us to use (\ref{eq:sz2}) to reduce the rank to the set cardinality. The last step uses the fact that $(u',v') \in \mathcal{T}_n$.
For the second sub-case, $(u',v') \notin \mathcal{T}_n$ (recall that $(u',v') \notin \overline{\mathcal{S}}$ and $\mathcal{S}_m \subset \overline{\mathcal{S}}$).
\begin{eqnarray}
&&H\Big(\Big\{\sum_{v\in[V_u]}\left(Z_{u,v}\right)\Big\}_{u \in \mathcal{U}^{(m,n)}} \Big| \left\{Z_{u,v}\right\}_{(u,v) \in \mathcal{T}_n}\Big)\\
&=&H\Big(\Big\{\sum_{v\in[V_u]}\left(Z_{u,v}\right)\Big\}_{u \in \mathcal{U}^{(m,n)}} , \left\{Z_{u,v}\right\}_{(u,v) \in \mathcal{T}_n}\Big)-H\left(\left\{Z_{u,v}\right\}_{(u,v) \in \mathcal{T}_n}\right)\\
&\overset{(\ref{eq:c21})}{=}&H\Big(\Big\{\sum_{v\in[V_u]}\left(Z_{u,v}\right)\Big\}_{u \in \mathcal{U}^{(m,n)}} , \left\{Z_{u,v}\right\}_{(u,v) \in \mathcal{T}_n\cap(\overline{\mathcal{S}} \cup \{(u',v')\})}\Big)-H\left(\left\{Z_{u,v}\right\}_{(u,v) \in \mathcal{T}_n\cap(\overline{\mathcal{S}} \cup \{(u',v')\})}\right)\\
&\overset{(\ref{eq:sz2})}{=}&(|\mathcal{U}^{(m,n)}|+|\mathcal{T}_n \cap \overline{\mathcal{S}}|)L-|\mathcal{T}_n \cap \overline{\mathcal{S}} |L\label{eq:st222}\\
&=&|\mathcal{U}^{(m,n)}|L,
\end{eqnarray}
whereas $|\mathcal{U}^{(m,n)}|+| \mathcal{T}_n \cap \overline{\mathcal{S}}|\leq d^*\leq \max(a^*,d^*)$, the first term of (\ref{eq:st222}) is no greater than $\max(a^*,d^*)$ enabling us to use (\ref{eq:sz2}) to reduce the rank to the set cardinality.

Fourth, consider the scheme in Section \ref{sec:ach3}, 
we now proceed to prove (\ref{eq:independent}).
\begin{eqnarray}
&&H\left(\{Z_{u,v}\}_{(u,v) \in (\mathcal{S}_m\cap\mathcal{K}_{\{u\}})\setminus\mathcal{T}_n} \big|\{Z_{u,v}\}_{(u,v) \in \mathcal{T}_n}  \right) \notag\\
&=& H\big(\{Z_{u,v}\}_{(u,v) \in (\mathcal{S}_m\cap\mathcal{K}_{\{u\}})\cup\mathcal{T}_n} \big) - H\big( \{Z_{u,v}\}_{(u,v) \in \mathcal{T}_n}  \big)\\
&=& H\left(\{Z_{u,v}\}_{(u,v) \in \left( ((\mathcal{S}_m\cap\mathcal{K}_{\{u\}})\cup\mathcal{T}_n) \cap \overline{\mathcal{S}}\right) \cup \left( ((\mathcal{S}_m\cap\mathcal{K}_{\{u\}})\cup\mathcal{T}_n) \setminus \overline{\mathcal{S}}\right) } \right) - H\left( \{Z_{u,v}\}_{(u,v) \in (\mathcal{T}_n \cap \overline{\mathcal{S}}) \cup (\mathcal{T}_n \setminus \overline{\mathcal{S}})}  \right) ~~~\\
&=& H\left(\{Z_{u,v}\}_{(u,v) \in \left( ((\mathcal{S}_m\cap\mathcal{K}_{\{u\}})\cup\mathcal{T}_n) \cap \overline{\mathcal{S}}\right) \cup ( \mathcal{T}_n \setminus \overline{\mathcal{S}}) } \right) - H\left( \{Z_{u,v}\}_{(u,v) \in (\mathcal{T}_n \cap \overline{\mathcal{S}}) \cup (\mathcal{T}_n \setminus \overline{\mathcal{S}})}  \right) \\
&\overset{(\ref{eq:achthm3t1})(\ref{eq:sz3})}{=}& \Big(\big|((\mathcal{S}_m\cap\mathcal{K}_{\{u\}})\cup\mathcal{T}_n) \cap \overline{\mathcal{S}}\big| \overline{q} + \sum_{(u,v) \in \mathcal{T}_n \setminus \overline{\mathcal{S}}} p_{u,v} \Big)  - \Big( \big|\mathcal{T}_n  \cap \overline{\mathcal{S}}\big| \overline{q} + \sum_{(u,v) \in \mathcal{T}_n \setminus \overline{\mathcal{S}}} p_{u,v} \Big)  \label{eq:sf11} \\
&=& |(\mathcal{S}_m\cap\mathcal{K}_{\{u\}})\setminus\mathcal{T}_n| \overline{q} = |(\mathcal{S}_m\cap\mathcal{K}_{\{u\}})\setminus\mathcal{T}_n| L
\end{eqnarray}
where $L = \overline{q}$ for the scheme in Section \ref{sec:ach3} is used in the last step and in order to apply (\ref{eq:sz3}) to obtain (\ref{eq:sf11}), we note that $H\left((Z_{u,v})_{(u,v)\in\mathcal{T}_n\setminus\overline{\mathcal{S}}}\right)$ is captured by the first $p_{u,v}$ rows (as ${\bf F}_{u,v} \, {\bf G}_{u,v}$ has rank $p_{u,v}$, refer to (\ref{eq:achthm3t1})) and need to show that the first term is no greater than $(\max(a^*,d^*) + b^*)\overline{q}$, the proof of which is provided for two sub-cases. For the first sub-case, $\big|(\mathcal{S}_{m}\cap\mathcal{K}_{\{u\}}\cup\mathcal{T}_{n})\cap\overline{\mathcal{S}}\big|=a^*$, then $\big|((\mathcal{S}_m\cap\mathcal{K}_{\{u\}})\cup\mathcal{T}_n) \cap \overline{\mathcal{S}}\big| \overline{q} + \sum_{(u,v) \in \mathcal{T}_n \setminus \overline{\mathcal{S}}} p_{u,v} = a^* \overline{q} + \overline{q} \sum_{(u,v) \in \mathcal{T}_n \setminus \overline{\mathcal{S}}} b_{u,v}^* \leq (a^* + b^*)\overline{q}$, where the last step follows from the `max' objective function of the linear program (\ref{eq:convlp})(\ref{eq:convlp1})(\ref{eq:convlp2}).
For the second sub-case, $\big|(\mathcal{S}_{m}\cap\mathcal{K}_{\{u\}}\cup\mathcal{T}_{n})\cap\overline{\mathcal{S}}\big| < a^*$,
then $\big|((\mathcal{S}_m\cap\mathcal{K}_{\{u\}})\cup\mathcal{T}_n) \cap \overline{\mathcal{S}}\big| \overline{q} + \sum_{(u,v) \in \mathcal{T}_n \setminus \overline{\mathcal{S}}} p_{u,v} \leq (a^*-1) \overline{q} + \overline{q} \sum_{(u,v) \in \mathcal{K}_{[U]} \setminus \overline{\mathcal{S}}} b_{u,v}^* \overset{(\ref{eq:ach2lp21})}{=} (a^* -1 + b^* + 1)\overline{q}$ = $(a^*+b^*)\overline{q}$.

Consider the scheme in section \ref{sec:ach3} where $|\mathcal{K}_{\mathcal{U}^{(m,n)}}\cup\mathcal{T}_n|\leq K-1$. We are ready to prove (\ref{eq:independent2}).
\begin{eqnarray}
&&H\Big(\Big\{\sum_{v\in[V_u]}\left(Z_{u,v}\right)\Big\}_{u \in \mathcal{U}^{(m,n)}} \Big| \left\{Z_{u,v}\right\}_{(u,v) \in \mathcal{T}_n}\Big)\\
&=&H\Big(\Big\{\sum_{v\in[V_u]}\left(Z_{u,v}\right)\Big\}_{u \in \mathcal{U}^{(m,n)}} , \left\{Z_{u,v}\right\}_{(u,v) \in \mathcal{T}_n}\Big)-H\left(\left\{Z_{u,v}\right\}_{(u,v) \in \mathcal{T}_n}\right)\\
&=&H\Big(\Big\{\sum_{v\in[V_u]}\left(Z_{u,v}\right)\Big\}_{u \in \mathcal{U}^{(m,n)}} , \left\{Z_{u,v}\right\}_{(u,v) \in (\mathcal{T}_n \cap \overline{\mathcal{S}}) \cup (\mathcal{T}_n \setminus \overline{\mathcal{S}})}\Big)-H\left(\left\{Z_{u,v}\right\}_{(u,v) \in (\mathcal{T}_n \cap \overline{\mathcal{S}}) \cup (\mathcal{T}_n \setminus \overline{\mathcal{S}})}\right)~~\\
&\overset{(\ref{eq:achthm3t1})(\ref{eq:sz3})}{=}& \Big((|\mathcal{U}^{(m,n)}|+\big|\mathcal{T}_n  \cap \overline{\mathcal{S}}\big|) \overline{q} + \sum_{(u,v) \in \mathcal{T}_n \setminus \overline{\mathcal{S}}} p_{u,v} \Big)  - \Big( \big|\mathcal{T}_n  \cap \overline{\mathcal{S}}\big| \overline{q} + \sum_{(u,v) \in \mathcal{T}_n \setminus \overline{\mathcal{S}}} p_{u,v} \Big)  \label{eq:sf133} \\
&=&|\mathcal{U}^{(m,n)}|L
\end{eqnarray}
where $L = \overline{q}$ for the scheme in Section \ref{sec:ach3} is used in the last step and in order to apply (\ref{eq:sz3}) to obtain (\ref{eq:sf133}), we note that $H\left((Z_{u,v})_{(u,v)\in\mathcal{T}_n\setminus\overline{\mathcal{S}}}\right)$ is captured by the first $p_{u,v}$ rows (as ${\bf F}_{u,v} \, {\bf G}_{u,v}$ has rank $p_{u,v}$, refer to (\ref{eq:achthm3t1})) and need to show that the first term is no greater than $(a^* + b^*)\overline{q}$. To see this, note that 
$|\mathcal{U}^{(m,n)}|+\big|\mathcal{T}_n  \cap \overline{\mathcal{S}}\big| \leq d^*\leq a^*-1$,
then $(|\mathcal{U}^{(m,n)}|+\big|\mathcal{T}_n  \cap \overline{\mathcal{S}}\big|) \overline{q} + \sum_{(u,v) \in \mathcal{T}_n \setminus \overline{\mathcal{S}}} p_{u,v} \leq (a^*-1) \overline{q} + \overline{q} \sum_{(u,v) \in \mathcal{K}_{[U]} \setminus \overline{\mathcal{S}}} b_{u,v}^* \overset{(\ref{eq:ach2lp21})}{=} (a^* -1 + b^* + 1)\overline{q}$ = $(a^*+b^*)\overline{q}$.

Fifth, consider the scheme in Section \ref{sec:thm3}, 
we now proceed to prove (\ref{eq:independent}).
\begin{eqnarray}
&&H\left(\{Z_{u,v}\}_{(u,v) \in (\mathcal{S}_m\cap\mathcal{K}_{\{u\}})\setminus\mathcal{T}_n} \big|\{Z_{u,v}\}_{(u,v) \in \mathcal{T}_n}  \right) \notag\\
&=& H\big(\{Z_{u,v}\}_{(u,v) \in (\mathcal{S}_m\cap\mathcal{K}_{\{u\}})\cup\mathcal{T}_n} \big) - H\big( \{Z_{u,v}\}_{(u,v) \in \mathcal{T}_n}  \big)\\
%&=& H\left(\{Z_{u,v}\}_{(u,v) \in \left( ((\mathcal{S}_m\cap\mathcal{K}_{\{u\}})\cup\mathcal{T}_n) \cap \overline{\mathcal{S}}\right) \cup \left( ((\mathcal{S}_m\cap\mathcal{K}_{\{u\}})\cup\mathcal{T}_n) \setminus \overline{\mathcal{S}}\right) } \right) - H\left( \{Z_{u,v}\}_{(u,v) \in (\mathcal{T}_n \cap \overline{\mathcal{S}}) \cup (\mathcal{T}_n \setminus \overline{\mathcal{S}})}  \right) \\
&=& H\left(\{Z_{u,v}\}_{(u,v) \in \left( ((\mathcal{S}_m\cap\mathcal{K}_{\{u\}})\cup\mathcal{T}_n) \cap \overline{\mathcal{S}}\right) \cup ( \mathcal{T}_n \setminus \overline{\mathcal{S}}) } \right) - H\left( \{Z_{u,v}\}_{(u,v) \in (\mathcal{T}_n \cap \overline{\mathcal{S}}) \cup (\mathcal{T}_n \setminus \overline{\mathcal{S}})}  \right) \\
&\overset{(\ref{eq:sz32})(\ref{eq:sz321})}{=}& \Big(\big|((\mathcal{S}_m\cap\mathcal{K}_{\{u\}})\cup\mathcal{T}_n) \cap \overline{\mathcal{S}}\big| \overline{q} + \sum_{(u,v) \in \mathcal{T}_n \setminus \overline{\mathcal{S}}} p_{u,v} \Big)  - \Big( \big|\mathcal{T}_n  \cap \overline{\mathcal{S}}\big| \overline{q} + \sum_{(u,v) \in \mathcal{T}_n \setminus \overline{\mathcal{S}}} p_{u,v} \Big)  \label{eq:sf112} \\
&=& |(\mathcal{S}_m\cap\mathcal{K}_{\{u\}})\setminus\mathcal{T}_n| \overline{q} = |(\mathcal{S}_m\cap\mathcal{K}_{\{u\}})\setminus\mathcal{T}_n| L
\end{eqnarray}
where $L = \overline{q}$ for the scheme in Section \ref{sec:thm3} is used in the last step. The first term of (\ref{eq:sf112}) holds because of (\ref{eq:sz32}) and (\ref{eq:sz321}). Let us consider two cases one by one.

When $a^*\leq e^*=|\overline{\mathcal{S}}|$ and $\max(a^*,d^*)\leq |\overline{\mathcal{S}}|-1$, we note that $H\left((Z_{u,v})_{(u,v)\in\mathcal{T}_n\setminus\overline{\mathcal{S}}}\right)$ is captured by the first $p_{u,v}$ rows (as ${\bf F}_{u,v} \, {\bf G}_{u,v}$ has rank $p_{u,v}$, refer to (\ref{eq:uppttt22})) and need to show that the first term is no greater than $(\max(a^*,d^*) + l^*)\overline{q}$. Here is the proof: $\big|((\mathcal{S}_m\cap\mathcal{K}_{\{u\}})\cup\mathcal{T}_n) \cap \overline{\mathcal{S}}\big| \overline{q} + \sum_{(u,v) \in \mathcal{T}_n \setminus \overline{\mathcal{S}}} p_{u,v} \leq  (\max(a^*,d^*) + l^*)\overline{q}$.

When $a^*\leq e^*=|\overline{\mathcal{S}}|$ and $\max(a^*,d^*)=|\overline{\mathcal{S}}|$, we note that $H\left((Z_{u,v})_{(u,v)\in\mathcal{T}_n\setminus\overline{\mathcal{S}}}\right)$ is captured by the first $p_{u,v}$ rows (as ${\bf F}_{u,v} \, {\bf G}_{u,v}$ has rank $p_{u,v}$, refer to (\ref{eq:uppttt22})) and need to show that the first term is no greater than $(\max(a^*,d^*)-1 + l^*)\overline{q}$, the proof of which is provided for two sub-cases. 
For the first sub-case, $\big|(\mathcal{S}_{m}\cap\mathcal{K}_{\{u\}}\cup\mathcal{T}_{n})\cap\overline{\mathcal{S}}\big|=|\overline{\mathcal{S}}|$, then $\big|((\mathcal{S}_m\cap\mathcal{K}_{\{u\}})\cup\mathcal{T}_n) \cap \overline{\mathcal{S}}\big| \overline{q} + \sum_{(u,v) \in \mathcal{T}_n \setminus \overline{\mathcal{S}}} p_{u,v} = \max(a^*,d^*) \overline{q} + \overline{q} \sum_{(u,v) \in \mathcal{T}_n \setminus \overline{\mathcal{S}}} l_{u,v}^* \overset{(\ref{eq:ach2lp233})}{\leq} (\max(a^*,d^*) + l^*-1)\overline{q}$, where the last step follows from (\ref{eq:ach2lp233}) in lemma \ref{lemma:lp23}.
For the second sub-case, $\big|(\mathcal{S}_{m}\cap\mathcal{K}_{\{u\}}\cup\mathcal{T}_{n})\cap\overline{\mathcal{S}}\big| \leq|\overline{\mathcal{S}}|-1$,
then $\big|((\mathcal{S}_m\cap\mathcal{K}_{\{u\}})\cup\mathcal{T}_n) \cap \overline{\mathcal{S}}\big| \overline{q} + \sum_{(u,v) \in \mathcal{T}_n \setminus \overline{\mathcal{S}}} p_{u,v} \leq (\max(a^*,d^*)-1) \overline{q}  + l^*\overline{q}$ = $(\max(a^*,d^*)+l^*-1)\overline{q}$.

Consider the scheme in section \ref{sec:thm3} where $|\mathcal{K}_{\mathcal{U}^{(m,n)}}\cup\mathcal{T}_n|\leq K-1$. We are ready to prove (\ref{eq:independent2}).
\begin{eqnarray}
&&H\Big(\Big\{\sum_{v\in[V_u]}\left(Z_{u,v}\right)\Big\}_{u \in \mathcal{U}^{(m,n)}} \Big| \left\{Z_{u,v}\right\}_{(u,v) \in \mathcal{T}_n}\Big)\\
&=&H\Big(\Big\{\sum_{v\in[V_u]}\left(Z_{u,v}\right)\Big\}_{u \in \mathcal{U}^{(m,n)}} , \left\{Z_{u,v}\right\}_{(u,v) \in \mathcal{T}_n}\Big)-H\left(\left\{Z_{u,v}\right\}_{(u,v) \in \mathcal{T}_n}\right)\\
&=&H\Big(\Big\{\sum_{v\in[V_u]}\left(Z_{u,v}\right)\Big\}_{u \in \mathcal{U}^{(m,n)}} , \left\{Z_{u,v}\Big\}_{(u,v) \in (\mathcal{T}_n \cap \overline{\mathcal{S}}) \cup (\mathcal{T}_n \setminus \overline{\mathcal{S}})}\right)-H\left(\left\{Z_{u,v}\right\}_{(u,v) \in (\mathcal{T}_n \cap \overline{\mathcal{S}}) \cup (\mathcal{T}_n \setminus \overline{\mathcal{S}})}\right)~~\\
&\overset{(\ref{eq:sz32})(\ref{eq:sz321})}{=}& \Big((|\mathcal{U}^{(m,n)}|+\big|\mathcal{T}_n  \cap \overline{\mathcal{S}}\big|) \overline{q} + \sum_{(u,v) \in \mathcal{T}_n \setminus \overline{\mathcal{S}}} p_{u,v} \Big)  - \Big( \big|\mathcal{T}_n  \cap \overline{\mathcal{S}}\big| \overline{q} + \sum_{(u,v) \in \mathcal{T}_n \setminus \overline{\mathcal{S}}} p_{u,v} \Big)  \label{eq:sf1333} \\
&=&|\mathcal{U}^{(m,n)}|L
\end{eqnarray}
where $L = \overline{q}$ for the scheme in Section \ref{sec:thm3} is used in the last step. The first term of (\ref{eq:sf1333}) holds because of (\ref{eq:sz32}) and (\ref{eq:sz321}). Let us consider two cases one by one.

When $a^*\leq e^*=|\overline{\mathcal{S}}|$ and $\max(a^*,d^*)\leq |\overline{\mathcal{S}}|-1$, we note that $H\left((Z_{u,v})_{(u,v)\in\mathcal{T}_n\setminus\overline{\mathcal{S}}}\right)$ is captured by the first $p_{u,v}$ rows (as ${\bf F}_{u,v} \, {\bf G}_{u,v}$ has rank $p_{u,v}$, refer to (\ref{eq:uppttt22})) and need to show that the first term is no greater than $(\max(a^*,d^*) + l^*)\overline{q}$. Here is the proof: $(|\mathcal{U}^{(m,n)}|+ |\mathcal{T}_n\cap\overline{\mathcal{S}}|) \overline{q} + \sum_{(u,v) \in \mathcal{T}_n \setminus \overline{\mathcal{S}}} p_{u,v} \leq  (\max(a^*,d^*) + l^*)\overline{q}$.

When $a^*\leq e^*=|\overline{\mathcal{S}}|$ and $\max(a^*,d^*)=|\overline{\mathcal{S}}|$, we note that $H\left((Z_{u,v})_{(u,v)\in\mathcal{T}_n\setminus\overline{\mathcal{S}}}\right)$ is captured by the first $p_{u,v}$ rows (as ${\bf F}_{u,v} \, {\bf G}_{u,v}$ has rank $p_{u,v}$, refer to (\ref{eq:uppttt22})) and need to show that the first term is no greater than $(\max(a^*,d^*)-1 + l^*)\overline{q}$, the proof of which is provided for two sub-cases. 
For the first sub-case, $|\mathcal{U}^{(m,n)}|+ |\mathcal{T}_n\cap\overline{\mathcal{S}}|=|\overline{\mathcal{S}}|$, then $(|\mathcal{U}^{(m,n)}|+ |\mathcal{T}_n\cap\overline{\mathcal{S}}|) \overline{q} + \sum_{(u,v) \in \mathcal{T}_n \setminus \overline{\mathcal{S}}} p_{u,v}=\max(a^*,d^*) \overline{q} + \overline{q} \sum_{(u,v) \in \mathcal{T}_n \setminus \overline{\mathcal{S}}} l_{u,v}^* \overset{(\ref{eq:ach2lp233})}{\leq} (\max(a^*,d^*) + l^*-1)\overline{q}$, where the last step follows from (\ref{eq:ach2lp233}) in lemma \ref{lemma:lp23}.
For the second sub-case, $|\mathcal{U}^{(m,n)}|+ |\mathcal{T}_n\cap\overline{\mathcal{S}}|\leq |\overline{\mathcal{S}}|-1$,
then $(|\mathcal{U}^{(m,n)}|+ |\mathcal{T}_n\cap\overline{\mathcal{S}}|) \overline{q} + \sum_{(u,v) \in \mathcal{T}_n \setminus \overline{\mathcal{S}}} p_{u,v} \leq (\max(a^*,d^*)-1) \overline{q} + \overline{q} \sum_{(u,v) \in \mathcal{T}_{n} \setminus \overline{\mathcal{S}}} l_{u,v}^* \leq (\max(a^*,d^*) -1 + l^*)\overline{q}$.

\section{Conclusion}
\label{sec:discussion and conclusion}

In this paper, we studied hierarchical secure aggregation with heterogeneous security constraints and arbitrary user collusion from an information-theoretic perspective. Under an honest-but-curious threat model, we characterized the fundamental communication efficiency and randomness consumption in a three-layer aggregation architecture. In particular, we established tight characterizations of the optimal source key rate in two broad regimes of security and collusion constraints, and derived a general lower bound together with a bounded-gap achievable scheme for the remaining regime. These results reveal how heterogeneous security requirements and hierarchical network structures fundamentally affect the required randomness, and provide principled insights for the design of randomness-efficient secure aggregation protocols in federated learning systems.

\subsection{Future Work}

Despite the progress made in this work, several problems remain open. First, while the first two regimes admit tight characterizations of the optimal source key rate, there exists a remaining regime for which we only derive a general information-theoretic lower bound on \( R_{Z_\Sigma}^\star \). Although an explicit achievable scheme is provided and shown to approach this bound within a bounded gap, the tightness of the converse remains unresolved. Closing this gap likely requires either a sharper converse or new coding strategies that more effectively exploit heterogeneous security constraints.

Second, unlike prior studies on hierarchical secure aggregation~\cite{10806947, zhang2024optimal} and its cyclic variants~\cite{zhang2025fundamental}, which also characterize individual key rates, this work focuses solely on the source key rate. Under heterogeneous security constraints, users face highly asymmetric security requirements, and the minimum individual key rate generally varies across users, even in star-network settings. A complete treatment would therefore require characterizing an individual key rate profile, whose explicit determination appears challenging and is left for future work.

Finally, our model enforces the server and relay security constraints in (\ref{serversecurity}) and (\ref{relaysecurity}) for all security input sets and all colluding user sets, corresponding to a product-form security model. A natural extension is to allow the protected security input set to depend on the realized collusion pattern, thereby restricting attention to a prescribed subset of collusion--security pairs rather than their full Cartesian product. This more general formulation lies beyond the scope of the current techniques and remains an interesting direction for future investigation.

\bibliographystyle{IEEEtran}
\bibliography{references_secagg.bib}
\end{document}